\documentclass[12pt,a4paper]{article}
\pdfoutput=1
\usepackage{lineno}  
\usepackage{graphicx}  
\usepackage{xspace} 

\usepackage[usenames,dvipsnames]{color}
\usepackage{colortbl}
\usepackage{amsmath} 
\usepackage{ifthen} 
\usepackage[normalem]{ulem} 

\graphicspath{{./figs/}}

\newboolean{pdflatex}
\setboolean{pdflatex}{true} 
\newboolean{articletitles}
\setboolean{articletitles}{true} 

\newboolean{uprightparticles}
\setboolean{uprightparticles}{true} 

\usepackage{amssymb}
\usepackage{amsfonts}
\usepackage{upgreek} 

\usepackage{hyperref}    
\usepackage[all]{hypcap} 

\usepackage{rotating}
\usepackage{multirow}
\usepackage{graphpap}
\usepackage{subfigure} 
\usepackage{wasysym} 
\usepackage{array} 
\usepackage{pifont} 
\definecolor{RootOne}  {rgb}{0,0,0}
\definecolor{RootTwo}  {rgb}{1,0,0}
\definecolor{RootThree}{rgb}{0,1,0}
\definecolor{RootFour} {rgb}{0,0,1}
\definecolor{RootFive} {rgb}{1,1,0}
\definecolor{RootSix}  {rgb}{1,0,1}
\definecolor{RootSeven}{rgb}{0,1,1}

\def\lhcb {LHCb\xspace}
\def\ux85 {UX85\xspace}

\ifthenelse{\boolean{uprightparticles}}%
{

 \def\Pmu         {\ensuremath{\upmu}\xspace}

 \def\Ppi         {\ensuremath{\uppi}\xspace}

 \def\Pphi        {\ensuremath{\upphi}\xspace}

 \def\Ppsi        {\ensuremath{\uppsi}\xspace}

 \def\PDelta      {\ensuremath{\Delta}\xspace}                 
 \def\PXi      {\ensuremath{\Xi}\xspace}                 
 \def\PLambda      {\ensuremath{\Lambda}\xspace}                 
 \def\PSigma      {\ensuremath{\Sigma}\xspace}                 
 \def\POmega      {\ensuremath{\Omega}\xspace}                 
 \def\PUpsilon      {\ensuremath{\Upsilon}\xspace}                 
 
 \def\PB      {\ensuremath{\mathrm{B}}\xspace}                 
                  
 \def\PD      {\ensuremath{\mathrm{D}}\xspace}

 \def\PJ      {\ensuremath{\mathrm{J}}\xspace}                 
 \def\PK      {\ensuremath{\mathrm{K}}\xspace}

 \def\Pb      {\ensuremath{\mathrm{b}}\xspace}                 
 \def\Pc      {\ensuremath{\mathrm{c}}\xspace}

 \def\Pi      {\ensuremath{\mathrm{i}}\xspace}

 \def\Pp      {\ensuremath{\mathrm{p}}\xspace}

 \def\Ps      {\ensuremath{\mathrm{s}}\xspace}

}
{

 \def\Pmu         {\ensuremath{\mu}\xspace}

 \def\Ppi         {\ensuremath{\pi}\xspace}

 \def\Pphi        {\ensuremath{\phi}\xspace}

 \def\Ppsi        {\ensuremath{\psi}\xspace}                 
                  
 \mathchardef\PDelta="7101
 \mathchardef\PXi="7104
 \mathchardef\PLambda="7103
 \mathchardef\PSigma="7106
 \mathchardef\POmega="710A
 \mathchardef\PUpsilon="7107
                  
 \def\PB      {\ensuremath{B}\xspace}                 
                  
 \def\PD      {\ensuremath{D}\xspace}

 \def\PJ      {\ensuremath{J}\xspace}                 
 \def\PK      {\ensuremath{K}\xspace}

 \def\Pb      {\ensuremath{b}\xspace}                 
 \def\Pc      {\ensuremath{c}\xspace}

 \def\Pi      {\ensuremath{i}\xspace}

 \def\Pp      {\ensuremath{p}\xspace}

 \def\Ps      {\ensuremath{s}\xspace}

}

\def\mup        {\ensuremath{\Pmu^+}\xspace}
\def\mumu       {\ensuremath{\Pmu^+\Pmu^-}\xspace}

\def\squark    {\ensuremath{\Ps}\xspace}

\def\cquark    {\ensuremath{\Pc}\xspace}

\def\bquark    {\ensuremath{\Pb}\xspace}

\def\pion  {\ensuremath{\Ppi}\xspace}

\def\pip   {\ensuremath{\pion^+}\xspace}

\def\pipm  {\ensuremath{\pion^\pm}\xspace}

\def\kaon  {\ensuremath{\PK}\xspace}
  \def\Kbar  {\kern 0.2em\overline{\kern -0.2em \PK}{}\xspace}

\def\Kz    {\ensuremath{\kaon^0}\xspace}
\def\Kzb   {\ensuremath{\Kbar^0}\xspace}
\def\KzKzb {\ensuremath{\Kz \kern -0.16em \Kzb}\xspace}
\def\Kp    {\ensuremath{\kaon^+}\xspace}
\def\Km    {\ensuremath{\kaon^-}\xspace}
\def\Kpm   {\ensuremath{\kaon^\pm}\xspace}

\def\KpKm  {\ensuremath{\Kp \kern -0.16em \Km}\xspace}

\def\Dbar    {\ensuremath{\overline{\PD}}\xspace}
\def\D       {\ensuremath{\PD}\xspace}

\def\Dz      {\ensuremath{\D^0}\xspace}
\def\Dzb     {\ensuremath{\Dbar^0}\xspace}
\def\DzDzb   {\ensuremath{\Dz {\kern -0.16em \Dzb}}\xspace}
\def\Dp      {\ensuremath{\D^+}\xspace}
\def\Dm      {\ensuremath{\D^-}\xspace}

\def\DpDm    {\ensuremath{\Dp {\kern -0.16em \Dm}}\xspace}

\def\Ds      {\ensuremath{\D^+_\squark}\xspace}

\def\Dsm     {\ensuremath{\D^-_\squark}\xspace}

  \def\Bbar    {\kern 0.18em\overline{\kern -0.18em \PB}{}\xspace}

\def\jpsi     {\ensuremath{{\PJ\mskip -3mu/\mskip -2mu\Ppsi\mskip 2mu}}\xspace}

  \def\Y#1S{\ensuremath{\PUpsilon{(#1S)}}\xspace}

\def\proton      {\ensuremath{\Pp}\xspace}

\def\L {\ensuremath{\PLambda}\xspace}
\def\Lbar{\ensuremath{\bar \L}\xspace}

\def\Lc      {\ensuremath{\L_\cquark^+}\xspace}
\def\Lcbar   {\ensuremath{\Lbar_\cquark^-}\xspace}

\def\BF         {{\ensuremath{\cal B}\xspace}}

\def\BR         {\BF}

\def\to                 {\ensuremath{\rightarrow}\xspace}

\def\AT#1     {\ensuremath{A_{\mathrm{T}}^{#1}}\xspace}           

\def\C#1      {\ensuremath{\mathcal{C}_{#1}}\xspace}                       
\def\Cp#1     {\ensuremath{\mathcal{C}_{#1}^{'}}\xspace}                    
\def\Ceff#1   {\ensuremath{\mathcal{C}_{#1}^{\mathrm{(eff)}}}\xspace}        
\def\Cpeff#1  {\ensuremath{\mathcal{C}_{#1}^{'\mathrm{(eff)}}}\xspace}       
\def\Ope#1    {\ensuremath{\mathcal{O}_{#1}}\xspace}                       
\def\Opep#1   {\ensuremath{\mathcal{O}_{#1}^{'}}\xspace}                    

\newcommand{\tev}{\ensuremath{\mathrm{\,Te\kern -0.1em V}}\xspace}
\newcommand{\gev}{\ensuremath{\mathrm{\,Ge\kern -0.1em V}}\xspace}
\newcommand{\mev}{\ensuremath{\mathrm{\,Me\kern -0.1em V}}\xspace}
\newcommand{\kev}{\ensuremath{\mathrm{\,ke\kern -0.1em V}}\xspace}
\newcommand{\ev}{\ensuremath{\mathrm{\,e\kern -0.1em V}}\xspace}
\newcommand{\gevc}{\ensuremath{{\mathrm{\,Ge\kern -0.1em V\!/}c}}\xspace}
\newcommand{\mevc}{\ensuremath{{\mathrm{\,Me\kern -0.1em V\!/}c}}\xspace}
\newcommand{\gevcc}{\ensuremath{{\mathrm{\,Ge\kern -0.1em V\!/}c^2}}\xspace}
\newcommand{\gevgevcccc}{\ensuremath{{\mathrm{\,Ge\kern -0.1em V^2\!/}c^4}}\xspace}
\newcommand{\mevcc}{\ensuremath{{\mathrm{\,Me\kern -0.1em V\!/}c^2}}\xspace}

\def\mum  {\ensuremath{\,\upmu\rm m}\xspace}

\def\mub{\ensuremath{\rm \,\upmu b}\xspace}

\def\gsim{{~\raise.15em\hbox{$>$}\kern-.85em
          \lower.35em\hbox{$\sim$}~}\xspace}
\def\lsim{{~\raise.15em\hbox{$<$}\kern-.85em
          \lower.35em\hbox{$\sim$}~}\xspace}

\def\tell1  {TELL1\xspace}
\def\ukl1   {UKL1\xspace}

\def\CC      {\ensuremath{\mathrm{C}\mathrm{C}}\xspace}
\def\psiC    {\ensuremath{\jpsi{}\mathrm{C}}\xspace}

\def\CCbar   {\ensuremath{\mathrm{C}\overline{\mathrm{C}}}\xspace}
\def\CCbb    {\ensuremath{{\overline{\mathrm{C}}}{\overline{\mathrm{C}}}}\xspace}
\def\psiCb   {\ensuremath{\jpsi{}\overline{\mathrm{C}}}\xspace}

\def\DzDz    {\ensuremath{\Dz{}\Dz}\xspace}
\def\DzDp    {\ensuremath{\Dz{}\Dp}\xspace}
\def\DzDs    {\ensuremath{\Dz{}\Ds}\xspace}
\def\DzLc    {\ensuremath{\Dz{}\Lc}\xspace}
\def\DpDp    {\ensuremath{\Dp{}\Dp}\xspace}
\def\DpDs    {\ensuremath{\Dp{}\Ds}\xspace}
\def\DpLc    {\ensuremath{\Dp{}\Lc}\xspace}

\def\DzDzb  {\ensuremath{\Dz{}\Dzb}\xspace}
\def\DzDpb  {\ensuremath{\Dz{}\Dm}\xspace}
\def\DzDsb  {\ensuremath{\Dz{}\Dsm}\xspace}
\def\DzLcb  {\ensuremath{\Dz{}\Lcbar}\xspace}
\def\DpDpb  {\ensuremath{\Dp{}\Dm}\xspace}
\def\DpDsb  {\ensuremath{\Dp{}\Dsm}\xspace}
\def\DpLcb  {\ensuremath{\Dp{}\Lcbar}\xspace}

\def\DpDm   {\ensuremath{\Dp{}\Dm}\xspace}

\def\psiDz   {\ensuremath{\jpsi{}\PD^0}\xspace}
\def\psiDp   {\ensuremath{\jpsi{}\PD^+}\xspace}
\def\psiDs   {\ensuremath{\jpsi{}\PD^+_{\mathrm{s}}}\xspace}
\def\psiLc   {\ensuremath{\jpsi{}\PLambda^+_{\mathrm{c}}}\xspace}

\def\sPlot{\ensuremath{_s{\mathcal{P}}lot}}

\usepackage{cite}
\usepackage{mciteplus}

\begin{document}

\renewcommand{\thefootnote}{\fnsymbol{footnote}}
\setcounter{footnote}{1}



\begin{titlepage}
\pagenumbering{roman}

\vspace*{-1.5cm}
\centerline{\large EUROPEAN ORGANIZATION FOR NUCLEAR RESEARCH (CERN)}
\vspace*{1.5cm}
\hspace*{-0.5cm}
\begin{tabular*}{\linewidth}{lc@{\extracolsep{\fill}}r}
\ifthenelse{\boolean{pdflatex}}
{\vspace*{-2.7cm}\mbox{\!\!\!\includegraphics[width=.14\textwidth]{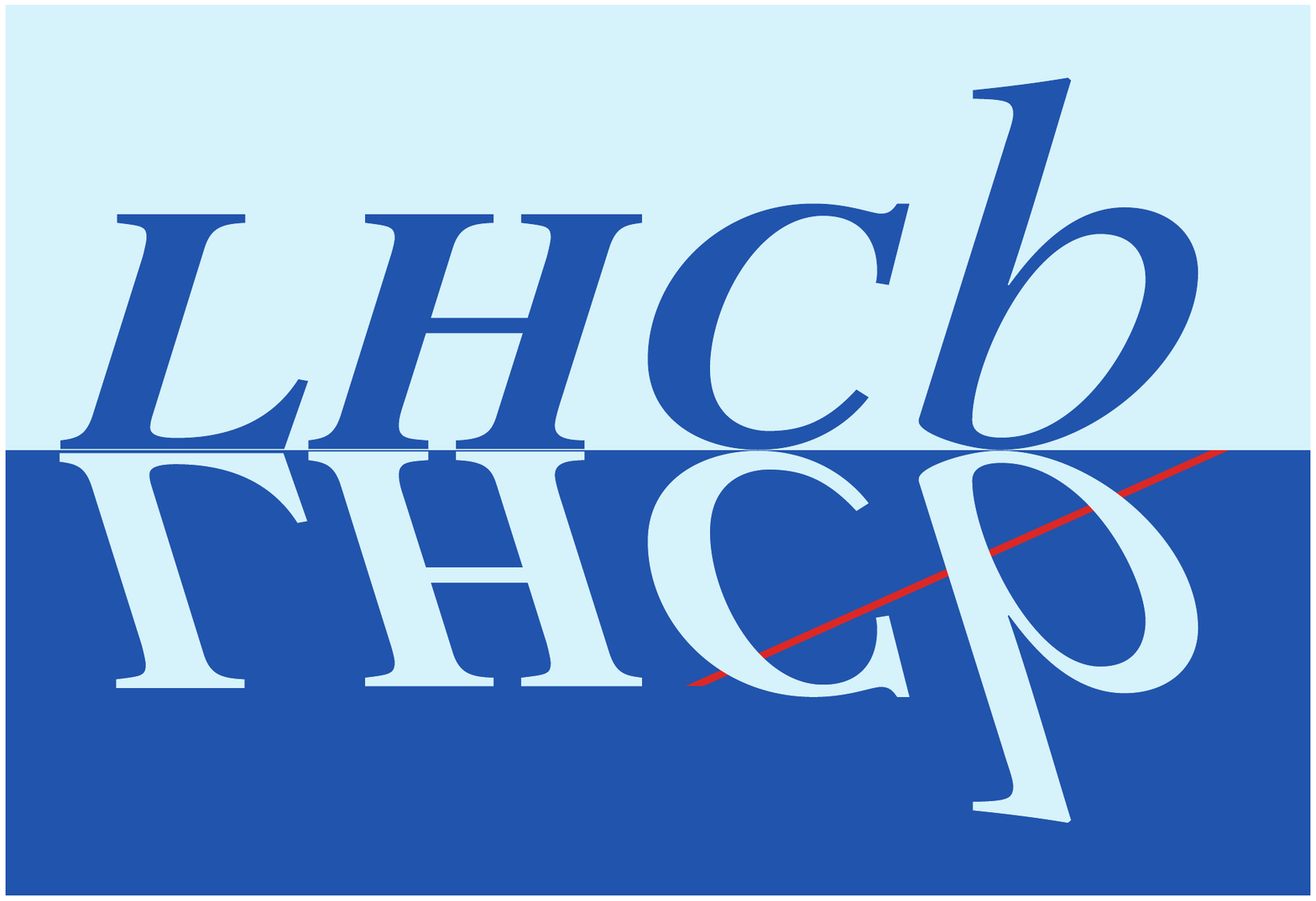}} & &}%
{\vspace*{-1.2cm}\mbox{\!\!\!\includegraphics[width=.12\textwidth]{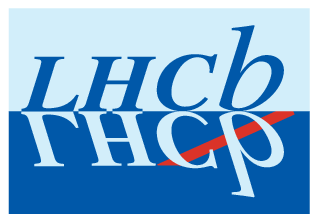}} & &}%
\\
 & & CERN-PH-EP-2012-109 \\  
 & & LHCb-PAPER-2012-003 \\  
 & & January 27, 2014 \\ 
& & \\
\end{tabular*}

\vspace*{3.0cm}

{\bf\boldmath\huge
\begin{center}
  Observation of double charm production 
  involving open charm 
  in pp~collisions at $\sqrt{s}=7~\mathrm{TeV}$
\end{center}
}

\vspace*{1.0cm}

\begin{center}
The LHCb collaboration\footnote{Authors are listed on the following pages.}
\end{center}

\vspace{\fill}

\begin{abstract}
  \noindent
The production of \jpsi~mesons
accompanied by open charm,  
and of pairs of open charm hadrons
are observed in pp~collisions at a centre-of-mass  
energy of 7~TeV using an integrated luminosity of  
$355~\mathrm{pb}^{-1}$~collected with  the LHCb detector. 
Model independent measurements of absolute
cross-sections are given together with ratios to the measured 
\jpsi~and open charm cross-sections. 
The properties of these events are studied and compared to theoretical predictions.
\end{abstract}

\vspace*{2.0cm}

\begin{center}
(Published in the Journal of High Energy Physics.)
\end{center}

\vspace{\fill}

\end{titlepage}


\newpage
\setcounter{page}{2}
\mbox{~}
\newpage

\begin{center}
{\bf LHCb collaboration}
\end{center}
\begin{flushleft}
R.~Aaij$^{38}$, 
C.~Abellan~Beteta$^{33,n}$, 
B.~Adeva$^{34}$, 
M.~Adinolfi$^{43}$, 
C.~Adrover$^{6}$, 
A.~Affolder$^{49}$, 
Z.~Ajaltouni$^{5}$, 
J.~Albrecht$^{35}$, 
F.~Alessio$^{35}$, 
M.~Alexander$^{48}$
% ...

{\footnotesize \it
% ...
$ ^{5}$Clermont Universit\'{e}, Universit\'{e} Blaise Pascal, CNRS/IN2P3, LPC, Clermont-Ferrand, France\\
$ ^{6}$CPPM, Aix-Marseille Universit\'{e}, CNRS/IN2P3, Marseille, France\\
% ...
$ ^{33}$Universitat de Barcelona, Barcelona, Spain\\
$ ^{34}$Universidad de Santiago de Compostela, Santiago de Compostela, Spain\\
$ ^{35}$European Organization for Nuclear Research (CERN), Geneva, Switzerland\\
% ...
$ ^{38}$Nikhef National Institute for Subatomic Physics, Amsterdam, The Netherlands\\
% ...
$ ^{43}$H.H. Wills Physics Laboratory, University of Bristol, Bristol, United Kingdom\\
% ...
$ ^{48}$School of Physics and Astronomy, University of Glasgow, Glasgow, United Kingdom\\
$ ^{49}$Oliver Lodge Laboratory, University of Liverpool, Liverpool, United Kingdom\\
% ...
$ ^{n}$LIFAELS, La Salle, Universitat Ramon Llull, Barcelona, Spain\\
% ...
}
\bigskip
\end{flushleft}

\cleardoublepage


\renewcommand{\thefootnote}{\arabic{footnote}}
\setcounter{footnote}{0}



\pagestyle{plain} 
\setcounter{page}{1}
\pagenumbering{arabic}


%


%

\section{Introduction}
\label{sec:Introduction}
Due to the high energy and luminosity of the LHC,
charm production studies 
can be carried out in a new kinematic domain with
unprecedented precision. As the cross-sections of 
open charm~\cite{LHCb-CONF-2010-013} 
and charmonium~\cite{Aaij:2011jh} production 
are large, 
the question of multiple production 
of these states in a single proton-proton collision naturally
arises. Recently, studies of double charmonium and charmonium 
with associated open charm production have been proposed as probes of
the quarkonium production mechanism~\cite{lansbergetc}.  
In pp~collisions, additional contributions from other mechanisms, 
such as Double Parton Scattering
(DPS)~\cite{Kom:2011bd, Baranov:2011ch, Novoselov:2011ff,Luszczak:2011zp}
or the intrinsic charm content of the proton~\cite{Brodsky1980451}
to the total cross-section, 
are possible, 
though these constributions may not be mutually exclusive. 

In this paper, both the production of $\jpsi$~mesons
together with an associated open charm hadron (either 
a \Dz, \Dp, \Ds~or \Lc)\footnote{The inclusion 
of charge-conjugate modes  is implied throughout this paper, 
unless explicitly stated otherwise.}
and double open charm hadron
production are studied in pp~collisions at a centre-of-mass
energy of 7~TeV.  
We denote the former process as \psiC
and the latter as \CC. In addition, as a control channel, 
$\mathrm{c}\mathrm{\bar{c}}$~events where two open charm hadrons are 
reconstructed in the LHCb fiducial
volume (denoted \CCbar) are studied. 
While the production of \psiC~events have not been observed 
before in hadron interactions, evidence for
the production of four charmed particles in pion-nuclear 
interactions has been reported 
by the WA75~collaboration~\cite{Aoki:1986uq}.

Leading order (LO) calculations for the $\mathrm{gg} \to \jpsi{}\jpsi$~process 
in perturbative QCD exist and give consistent 
results~\cite{Kartvelishvili:1984ur,Humpert:1983yj,Berezhnoy:2011xy}. 
In the LHCb fiducial region 
($2<y_{\jpsi}<4.5$, $p^{\mathrm{T}}_{\jpsi}<10~\mathrm{GeV}/c$), 
where $y_{\jpsi}$ and $p^{\mathrm{T}}_{\jpsi}$~stand for rapidity and
transverse momentum respectively, the calculated 
$\jpsi{}\jpsi$~production 
cross-section is
$4.1\pm1.2~\mathrm{nb}$~\cite{Berezhnoy:2011xy} in agreement with the
measured value of $5.1\pm1.0\pm1.1~\mathrm{nb}$~\cite{Aaij:2011yc}. 
Similar calculations for 
the $\mathrm{gg} \to \jpsi \mathrm{c}\bar{\mathrm{c}}$
and $\mathrm{gg}\to \mathrm{c}\bar{\mathrm{c}}\mathrm{c}\bar{\mathrm{c}}$~matrix elements 
exist~\cite{Berezhnoy:1998aa,Baranov:2006dh}. 
The calculated cross-sections for these processes in the acceptance
region considered here ($2<y_{\jpsi},y_{\mathrm{C}}<4$, 
$p^{\mathrm{T}}_{\jpsi}<12~\mathrm{GeV}/c$,
$3<p^{\mathrm{T}}_{\mathrm{C}}<12~\mathrm{GeV}/c$) 
are 
$\sigma\left(\psiC+\psiCb\right)\sim 18~\mathrm{nb}$ and 
$\sigma\left(\CC+\CCbb\right) \sim 100~\mathrm{nb}$, 
where $\mathrm{C}$ stands for the open charm hadron. 
The predictions are summarized in Table~\ref{tab:intro_theo}.
These LO~$\alpha_s^4$ perturbative QCD results are affected by
uncertainties  originating from the selection of the scale for the $\alpha_s$
calculation that can amount to a factor of two.

The DPS contribution can be estimated, 
neglecting partonic correlations in the proton,
as the product of the cross-sections of the sub-processes involved 
divided by an effective 
cross-section~\cite{Kom:2011bd, Baranov:2011ch, Novoselov:2011ff,Luszczak:2011zp}
%
\begin{equation}
\sigma^{\mathrm{DPS}}_{\mathrm{C}_1\mathrm{C}_2}= 
 \upalpha 
          \dfrac { \sigma_{\mathrm{C}_1}\times\sigma_{\mathrm{C}_2}}
                       {\sigma^{\mathrm{DPS}}_{\mathrm{eff}} },
\label{eq:dps}
\end{equation}
where 
$\upalpha=\tfrac{1}{4}$~if $\mathrm{C}_1$~and $\mathrm{C}_2$~are identical and non-self-conjugate (e.g. $\Dz\Dz$),
$\upalpha=1$~if $\mathrm{C}_1$~and $\mathrm{C}_2$~are different and either $\mathrm{C}_1$~or 
$\mathrm{C}_2$~is self-conjugate (e.g. $\psiDz$), and $\upalpha=\tfrac{1}{2}$~otherwise.
Using this equation and the measured single charm
cross-sections given in~\cite{LHCb-CONF-2010-013,Aaij:2011jh} 
together with the effective cross-section measured
in multi-jet events at the Tevatron $\sigma^{\mathrm{DPS}}_{\mathrm{eff}}=
14.5\pm1.7^{+1.7}_{-2.3}~\mathrm{mb}$~\cite{Abe:1997xk}, the size of this
contribution is estimated (see Table~\ref{tab:intro_theo}). 
However, this approach
has been criticized as being too naive~\cite{Blok:2011bu}. 

Extra charm particles in the event can originate from
the sea charm quarks of the interacting 
protons themselves. Estimates for
the possible contribution in the fiducial volume used here are given in
the Appendix and summarized in
Table~\ref{tab:intro_theo}. It should be stressed that the charm
parton density functions are not well known, 
nor are the $p^{\mathrm{T}}$~distributions of 
the resulting charm particles,
so these calculations should be considered as upper estimates.

\begin{table}[htb]
  \centering
  \caption{ \small
    Estimates for the production cross-sections of the \psiC~and
    \CC~modes in the LHCb fiducial range
    given by the leading order
    $\mathrm{gg}\to\jpsi\mathrm{c}\bar{\mathrm{c}}$ matrix 
    element,
    $\sigma^{\mathrm{gg}}$~\cite{Berezhnoy:1998aa,Baranov:2006dh,Lansberg:2008gk},
    the double parton scattering approach, 
    $\sigma^{\mathrm{DPS}}$
    and the sea charm quarks 
   from the interacting protons,
    $\sigma^{\mathrm{sea}}$.
  } \label{tab:intro_theo}
  \vspace*{3mm}
  \setlength{\extrarowheight}{1.5pt}
  \begin{tabular*}{0.75\textwidth}{@{\hspace{5mm}}c@{\extracolsep{\fill}}cccc@{\hspace{5mm}}}
    \multirow{2}{*}{Mode}  
    & \multicolumn{2}{c}{$\sigma^{\mathrm{gg}}$} 
    & \multirow{2}{*}{$\sigma^{\mathrm{DPS}}$} 
    & \multirow{2}{*}{$\sigma^{\mathrm{sea}}$}
    \\
    & \cite{Berezhnoy:1998aa,Baranov:2006dh}
    & \cite{Lansberg:2008gk}
    & 
    & 
    \\
    \hline 
      & \multicolumn{4}{c}{$\left[\mathrm{nb}\right]$}
    \\
    \hline
    \psiDz  
    & $10 \pm6\phantom{0}$
    & $7.4\pm3.7$
    & $146\pm39\phantom{0}$   
    & $220$   
    \\
    \psiDp  
    & $5\pm3$
    & $2.6\pm1.3$
    & $60\pm17$   
    & $100$   
    \\
    \psiDs  
    & $1.0\pm0.8$
    & $1.5\pm0.7$
    & $24\pm7\phantom{0}$   
    & $30$   
    \\
    \psiLc  
    & $0.8\pm0.5$
    & $0.9\pm0.5$
    & $56\pm22$   
    & 
    \\
    \hline 
      & \multicolumn{4}{c}{$\left[\mub\right]$}
    \\
    \hline 
    \DzDz  
    & 
    &
    & $1.0\phantom{0} \pm 0.25$
    & $1.5 $
    \\
    \DzDp  
    &
    &
    & $0.85           \pm 0.2\phantom{0}$
    & $1.4 $
    \\
    \DzDs  
    &
    &
    & $0.33 \pm 0.07$
    & $0.4$
    \\
    \DzLc 
    &
    &
    & $0.75\pm 0.25$
    & 
    \\
    \DpDp 
    &
    &
    & $0.17 \pm 0.05$
    & $0.3$
    \\
    \DpDs 
    &
    &
    & $0.14 \pm 0.03$
    & $0.2$
    \\
    \DpLc 
    &
    &
    & $0.32 \pm 0.12$
    &
    \\
 \end{tabular*}   
\end{table}


\section{The LHCb detector and dataset}
\label{sec:Detector}

The \lhcb detector~\cite{Alves:2008zz} is a single-arm forward
spectrometer covering the pseudorapidity range $2<\eta <5$, 
and is designed
for the study of particles containing \bquark or \cquark quarks. The
detector includes a high precision tracking system consisting of a
silicon-strip vertex detector surrounding the proton-proton 
interaction region,
a large-area silicon-strip detector located upstream of a dipole
magnet with a bending power of about $4{\rm\,Tm}$, and three stations
of silicon-strip detectors and straw drift tubes placed
downstream. The combined tracking system has a momentum resolution
$\Delta p/p$ that varies from 0.4\% at 5\gevc to 0.6\% at 100\gevc,
and an impact parameter resolution of 20\mum for tracks with high
transverse momentum. Charged hadrons are identified using two
ring-imaging Cherenkov (RICH) detectors. 
Photon, electron and hadron
candidates are identified by a calorimeter system consisting of
scintillating-pad and pre-shower detectors, and electromagnetic
and hadronic calorimeters. Muons are identified by a muon
system composed of alternating layers of iron and multiwire
proportional chambers. The trigger consists of a hardware stage
based on information from the calorimeter and muon systems, 
followed by a
software stage which applies a full event reconstruction.

Events with a~$\jpsi\to\mumu$~final state are triggered using 
two hardware trigger
decisions: the single-muon decision, which requires one muon
candidate with a transverse momentum $p^{\mathrm{T}}$~larger 
than 1.5~$\mathrm{GeV}/c$,
and the di-muon decision, which requires two muon candidates
with transverse momenta 
$p^{\mathrm{T}}_1$ and 
$p^{\mathrm{T}}_2$ 
satisfying the relation
$\sqrt{p^{\mathrm{T}}_1 \cdot p^{\mathrm{T}}_2}> 1.3~\mathrm{GeV}/c$.
The di-muon trigger decision in the software trigger
requires muon pairs of opposite charge with $p^{\mathrm{T}}>500~\mathrm{MeV}/c$, 
forming a common vertex and with an invariant mass 
$2.97 < m_{\mumu} < 3.21~\mathrm{GeV}/c^2$. 
Events with purely hadronic final states are accepted by the
hardware trigger if there is a calorimeter cluster with transverse energy
$E^{\mathrm{T}} > 3.6~\mathrm{GeV}$.  The software trigger decisions 
select generic displaced vertices from tracks  with large $\chi^2$ of 
impact parameter with respect to all  primary pp~interaction 
vertices in the event, providing high efficiency for 
purely hadronic decays~\cite{LHCb-PUB-2011-016}.

To prevent a few events with high occupancy 
from dominating the CPU time in the software trigger, 
a set of global event cuts 
is applied on the hit multiplicities of each sub-detector used by the
pattern recognition algorithms. These cuts were chosen to reject 
events with a large number of 
pile-up interactions with minimal loss of data.

The data used for this analysis comprises $355\pm13~\mathrm{pb}^{-1}$ 
of $\mathrm{pp}$~collisions at a centre-of-mass energy of $\sqrt{s}=7~\tev$
collected by the LHCb experiment in the first half of the 2011 data-taking
period. Simulation samples used are based on the 
{\sc{Pythia}}~6.4 generator~\cite{Sjostrand:2006za}
configured with the parameters detailed in Ref.~\cite{LHCb-PROC-2011-005}.
The {\sc{EvtGen}}~\cite{Lange:2001uf} and  
{\sc{Geant4}}~\cite{Agostinelli:2002hh} packages are used
to describe hadron decays and for the detector simulation, respectively.
The prompt charmonium production is simulated in
{\sc{Pythia}}~according to the  leading-order colour-singlet
and colour-octet mechanisms.


\section{Event selection}
\label{sec:EventSelection}
To select events containing multiple charm hadrons,
first \jpsi,
\Dz, \Dp, \Ds~and \Lc~candidates are formed from charged tracks 
reconstructed in the
spectrometer. Subsequently, these candidates are combined to form
\psiC, \CC and \CCbar candidates.

Well reconstructed tracks are selected for these studies by requiring
that the $\chi^2_{\rm{tr}}$  provided by the track fit 
satisfy $\chi^2_{\mathrm{tr}}/{\mathrm{ndf}}<5$, 
where ndf represents the number of degrees of freedom in the fit, 
and that the transverse momentum is greater than 
$650~(250)~\mathrm{MeV}/c$ for each muon~(hadron) candidate. 
For each track, the global likelihoods of the muon and hadron hypotheses
provided by reconstruction of the muon system are evaluated, 
and well identified 
muons  are selected by a requirement on the difference in likelihoods
\mbox{$\Delta \ln \mathcal{L}_{\Pmu/\mathrm{h}}>0$}.

Good quality particle identification by the ring-imaging Cherenkov detectors is ensured
by requiring the momentum of  the hadron candidate to be between    
$3.2~\mathrm{GeV}/c$ ($10~\mathrm{GeV}/c$ for protons) and 
$100~\mathrm{GeV}/c$, and  the pseudorapidity to be in the
range $2<\eta<5$. To select 
kaons (pions) the corresponding 
difference in logarithms of 
the global likelihood of the kaon (pion) hypothesis provided by the RICH system with 
respect to the pion (kaon) hypothesis, $\Delta \ln \mathcal{L} _{\mathrm{K}/\Ppi}$ 
($\Delta \ln \mathcal{L}_{\Ppi/\mathrm{K}}$), is required to be greater than 2. 
For protons,  the differences in logarithms of the global likelihood of the proton 
hypothesis provided by the RICH system with respect to the pion and kaon hypotheses, 
are required to be $\Delta \ln \mathcal{L}_{\proton/\Ppi}>10$ and 
$\Delta \ln \mathcal{L}_{\proton/\mathrm{K}}>10$, respectively.  

Pions, kaons and protons, used for the reconstruction of long-lived charm particles, 
are required  to be inconsistent with being produced in a pp~interaction vertex.  
Only particles with a minimal value of impact parameter $\chi^2$ 
with respect to any reconstructed proton-proton collision vertex 
$\chi^{2}_{\mathrm{IP}}>9$,
are considered for subsequent analysis. These selection criteria are summarized 
in Table~\ref{tab:basic}.
\begin{table}[t]
  \centering
  \caption{ \small
   Selection criteria for charged particles used for the reconstruction 
  of charm hadrons.
  } \label{tab:basic}
  \vspace*{3mm}
  \setlength{\extrarowheight}{1.5pt}
  \begin{tabular*}{0.9\textwidth}{@{\hspace{15mm}}c@{\extracolsep{\fill}}c@{\hspace{15mm}}}
    \multicolumn{2}{c}{Track selection} \\
    \hline        
$\Pmu^{\pm},\mathrm{h}^{\pm}$ & $\chi^2_{\mathrm{tr}}/\mathrm{ndf}<5~$ \\ 
$\Pmu^{\pm}$                 & $p^{\mathrm{T}}>650~\mathrm{MeV}/c$  \\ 
$\mathrm{h}^{\pm}$           & $p^{\mathrm{T}}>250~\mathrm{MeV}/c~\&~2<\eta<5  ~\&~ \chi^2_{\mathrm{IP}}>9$  \\ 
$\pipm,\mathrm{K}^{\pm} $    & $3.2 < p < 100~\mathrm{GeV}/c$ \\ 
$\mathrm{p}^{\pm}$           & $10  < p < 100~\mathrm{GeV}/c$ \\ 
    \hline 
   \multicolumn{2}{c}{Particle identification} \\
   \hline 
$\Pmu^{\pm}$    & $\Delta  \ln \mathcal{L}_{\Pmu/\mathrm{h}} > 0 $ \\ 
$\Ppi^{\pm}$    & $\Delta  \ln \mathcal{L}_{\Ppi/\mathrm{K}} > 2 $ \\ 
$\Kpm$         & $\Delta  \ln \mathcal{L}_{\mathrm{K}/\Ppi} > 2 $ \\ 
$\proton,\bar{\proton}$ & $\Delta  \ln \mathcal{L}_{\proton/\mathrm{K}} > 10~\&~ 
                  \Delta  \ln \mathcal{L}_{\proton/\Ppi} > 10 $  \\ 
 \end{tabular*}   
\end{table}

The selected charged particles are combined to form 
$\jpsi\to\mumu$,
$\Dz\to\mathrm{K}^-\Ppi^+$, 
$\Dp\to\mathrm{K}^-\Ppi^+\Ppi^+$, 
$\Ds\to\mathrm{K}^-\mathrm{K}^+\Ppi^+$ and 
$\Lc\to\mathrm{p}\mathrm{K}^-\Ppi^+$~candidates. 
A vertex fit is made to all combinations and 
a selection criterion on 
the corresponding $\chi^2_{\mathrm{VX}}$ applied. 
The transverse momentum,
$p^{\mathrm{T}}$, 
for open charm hadron candidates is required to be larger than $3~\mathrm{GeV}/c$.  
To ensure that the long-lived charm
particle originates from a primary
vertex, the minimal value of the charm particle's $\chi^2_{\mathrm{IP}}$ with respect 
to any of the reconstructed proton-proton collision 
vertices is required to be $<9$. 
In addition, the decay time $c\tau$ of long-lived charm mesons is required to be 
in excess of $100\,\mum$, and in the range 
$100< c\tau < 500\,\mum$~for \Lc~candidates.
To suppress the higher combinatorial background for \Lc~candidates, only 
pions, kaons and protons with a transverse momentum in excess of 
$0.5~\mathrm{GeV}/c$ are used in this case.

A global decay chain fit of the selected candidates 
is performed~\cite{Hulsbergen:2005pu}.
For channels containing a \jpsi~meson it is required 
that the muons be consistent with originating from a common 
vertex and that this be compatible with one of the reconstructed 
$\mathrm{pp}$~collision vertices. In the case of long-lived charm hadrons,
 the momentum direction is required to be consistent 
with the flight direction calculated from the locations 
of the primary and secondary vertices. 
To remove background from $\mathrm{b}$-hadron decays 
the reduced $\chi^2$ of this fit, $\chi^2_{\mathrm{fit}}/\mathrm{ndf}$,  
is required to be $<5$. To further reduce the combinatorial background as well as cross-feed due to 
particle misidentification, for the decay mode 
$\Dz\to\Km\pip$ a selection criterion 
on the cosine of the angle between 
the kaon momentum  in the \Dz~centre-of-mass frame and 
the \Dz~flight direction in the laboratory frame, $\theta^*$ is
applied. For $\Ds\to\Kp\Km\pip$ candidates, the invariant 
mass of  the $\Kp\Km$~system is 
required to be consistent with 
the $\Pphi$~meson mass. These selection criteria are 
summarized in Table~\ref{tab:composed}.

\begin{table}[t]
  \centering
  \caption{ \small
    Criteria used for the selection of charm hadrons. 
  } \label{tab:composed}
  \vspace*{3mm}
 \begin{small}
  \setlength{\extrarowheight}{1.5pt}
  \begin{tabular*}{0.98\textwidth}{@{\hspace{1mm}}lc@{\extracolsep{\fill}}ccccc@{\hspace{1mm}}}
    & 
    & \jpsi  & \Dz & \Dp & \Ds  &  \Lc  
    \\ 
    & 
    &  \mumu  
    & $\Km\pip$ 
    & $\Km\pip\pip$ 
    & $\left(\Kp\Km\right)_{\Pphi}\pip$ 
    & $\proton\Km\pip$ 
    \\
    \hline 
    $y$
    &
    & $2< y < 4$ 
    & $2< y < 4$ 
    & $2< y < 4$ 
    & $2< y < 4$ 
    & $2< y < 4$ 
    \\
    $p^{\mathrm{T}}$
    & $\left[\mathrm{GeV}/c\right]$ 
    & $< 12$
    & $ 3 < p^{\mathrm{T}} < 12$ 
    & $ 3 < p^{\mathrm{T}} < 12$ 
    & $ 3 < p^{\mathrm{T}} < 12$ 
    & $ 3 < p^{\mathrm{T}} < 12$ 
    \\ 
    $\chi^2_{\mathrm{VX}}$ 
    & 
    & $ < 20$
    & $ <  9$ 
    & $ < 25$ 
    & $ < 25$ 
    & $ < 25$  
    \\
    $\chi^2_{\mathrm{IP}}$ 
    & 
    & --- 
    & $<9$  
    & $<9$  
    & $<9$  
    & $<9$  
    \\
    $\chi^2_{\mathrm{fit}}/\mathrm{ndf}$ 
    &
    & $<5$
    & $<5$
    & $<5$
    & $<5$
    & $<5$
    \\
    $c\tau$
    & $\left[{}\mum\right]$ 
    & --- 
    & $c\tau > 100$ 
    & $c\tau > 100$ 
    & $c\tau > 100$ 
    & $\begin{array}{c}
      c\tau > 100 \\ 
      c\tau < 500   
    \end{array}$ 
    \\ 
    $\left| \cos\theta^{*}\right|$ 
    & 
    & --- 
    & $< 0.9$
    & --- 
    & --- 
    & --- 
    \\
    $\mathrm{m}_{\Kp\Km}$
    & $\left[ \mathrm{GeV}/c^2\right]$ 
    & --- 
    & --- 
    & --- 
    & $< 1.04$ 
    & --- 
    \\
    $\min p^{\mathrm{T}}_{\mathrm{h}^\pm}$
    & $\left[ \mathrm{GeV}/c\right]$ 
    & --- 
    & --- 
    & --- 
    & --- 
    & $> 0.5$ \\ 
  \end{tabular*}
 \end{small}   
\end{table}

\begin{figure}[t]
  \setlength{\unitlength}{1mm}
  \centering
  \begin{picture}(100,80)
    \put(0,0){
      \includegraphics*[width=100mm,height=80mm,%
      ]{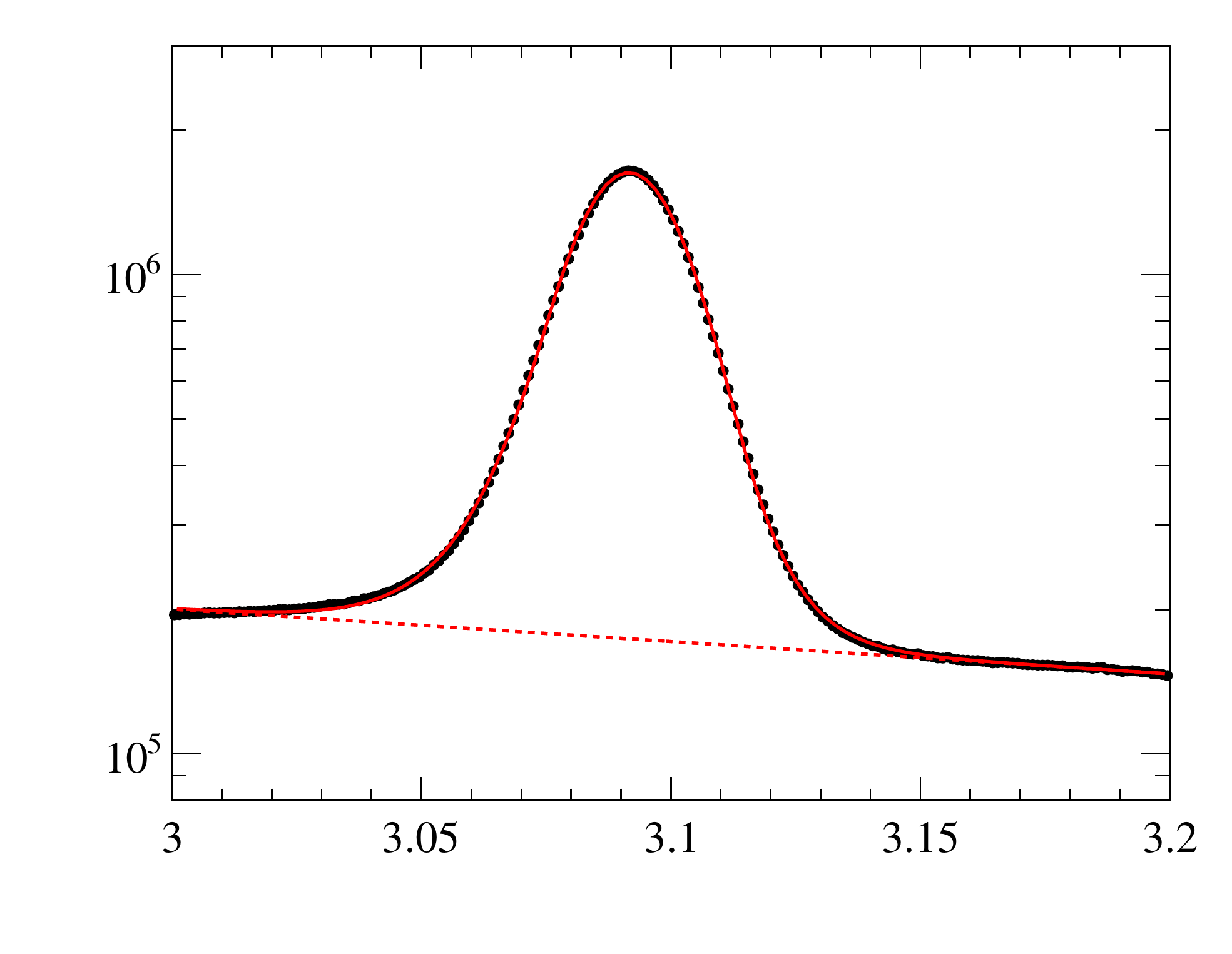}
    }
    \put(20,65)  { \small $\jpsi \to \mumu $  } 
    \put(50,3)   { $m_{\mumu}$  }
    \put(80,3)   { $\left[ \mathrm{GeV}/c^2\right]$}
    \put(-3 ,38)  { \small 
      \begin{sideways}%
        $\dfrac{\mathrm{dN}}{\mathrm{d}m_{\mumu}}~~\left[\dfrac{1}{1~\mathrm{MeV}/c^2}\right]$
      \end{sideways}%
    }
    \put(75,65){  \small  
      $\begin{array}{r}
        \mathrm{LHCb}~
      \end{array}$
    }
  \end{picture}
  \caption {  \small 
    Invariant mass distribution for selected \jpsi~candidates.
    The results of a fit to the model described in the
    text is superimposed on a logarithmic scale. The solid line
    corresponds to the total fitted PDF whilst the dotted line
    corresponds to the background component.
  }
  \label{fig:psi}
\end{figure}
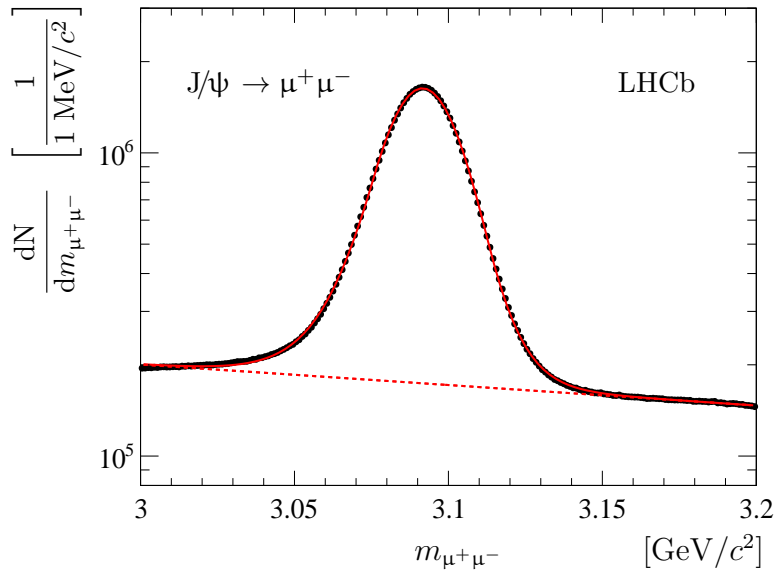

The invariant mass distributions for selected 
\jpsi, \Dz, \Dp, \Ds~and \Lc~candidates 
are presented in Figs.~\ref{fig:psi} and~\ref{fig:charm} for \jpsi~and 
open charm mesons, respectively. 
The distributions are modelled by 
a double-sided Crystal~Ball
function~\cite{Aaij:2011yc,Skwarnicki:1986xj}
for the $\jpsi\to\mumu$, 
and a modified Novosibirsk
function~\cite{Lees:2011gw} for the $\Dz\to\Km\pip$, $\Dp\to\Km\pip\pip$~and
$\Ds\to\Kp\Km \pip$~and 
$\Lc\to\proton\Km\pip$~signals. 
In each case the combinatorial background
component is modelled with an exponential function. The signal yields
are summarized in Table~\ref{tab:charm} together with an estimate of
the contamination from the decays of $\mathrm{b}$~hadrons,
 $f_{\mathrm{b}}^{\mathrm{MC}}$.  The latter has been
estimated using simulated events, normalized to the corresponding
measured cross-sections.

\begin{figure}[hbt]
  \setlength{\unitlength}{1mm}
  \centering
  \begin{picture}(150,120)
    \put(0,60){
      \includegraphics*[width=75mm,height=60mm,%
      ]{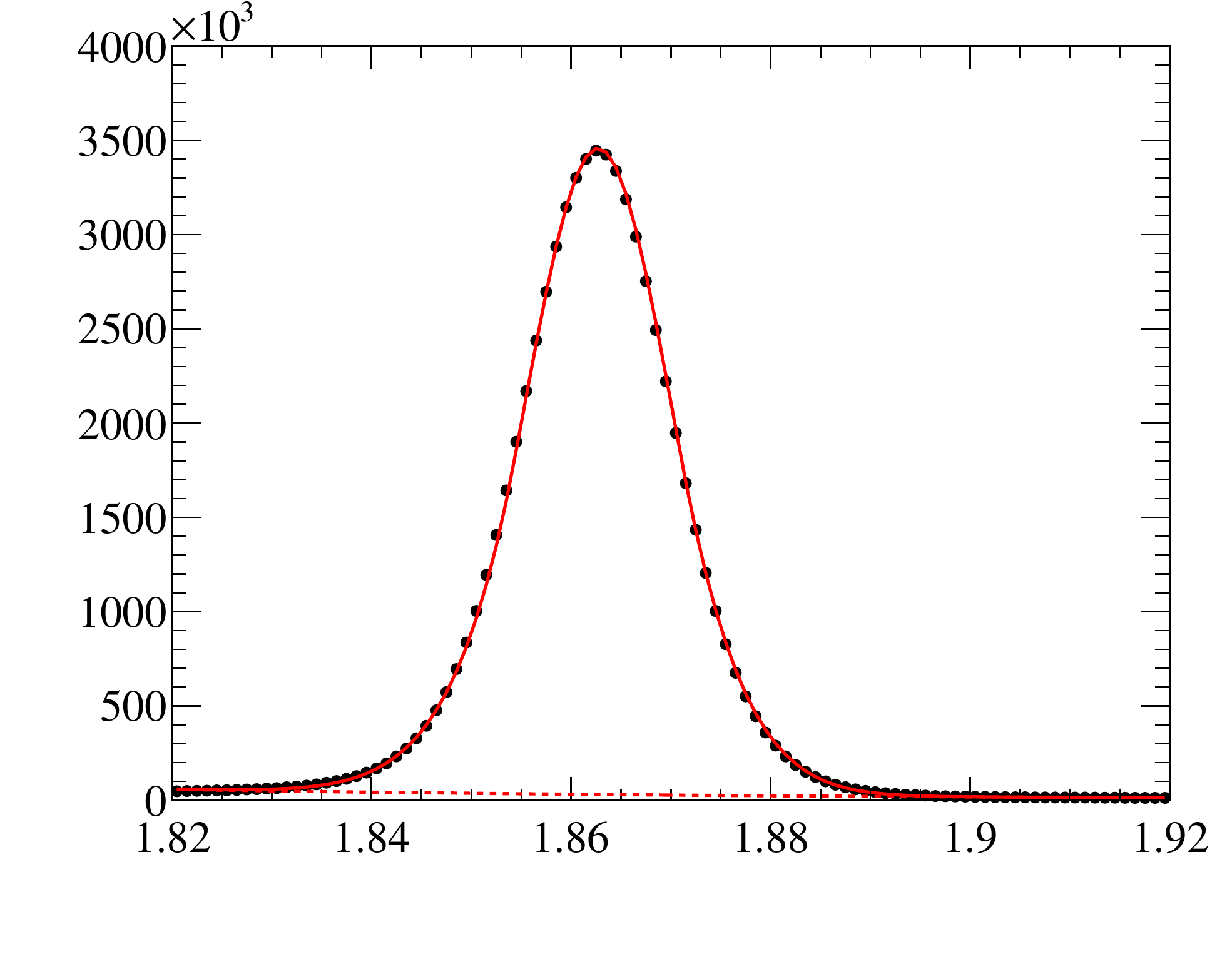}
    }
    \put(77,60){
      \includegraphics*[width=75mm,height=60mm,%
      ]{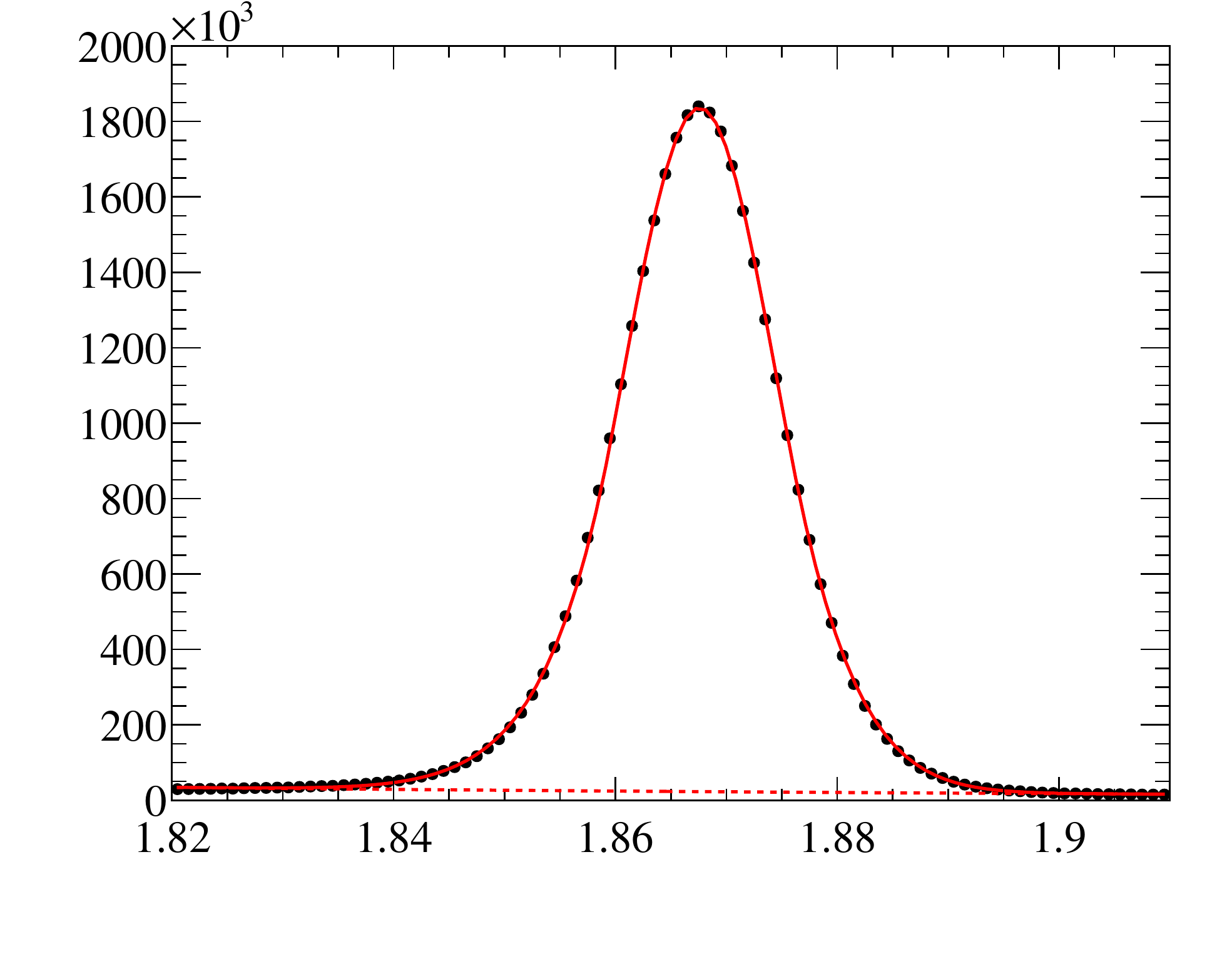}
    }
    \put(0,0){
      \includegraphics*[width=75mm,height=60mm,%
      ]{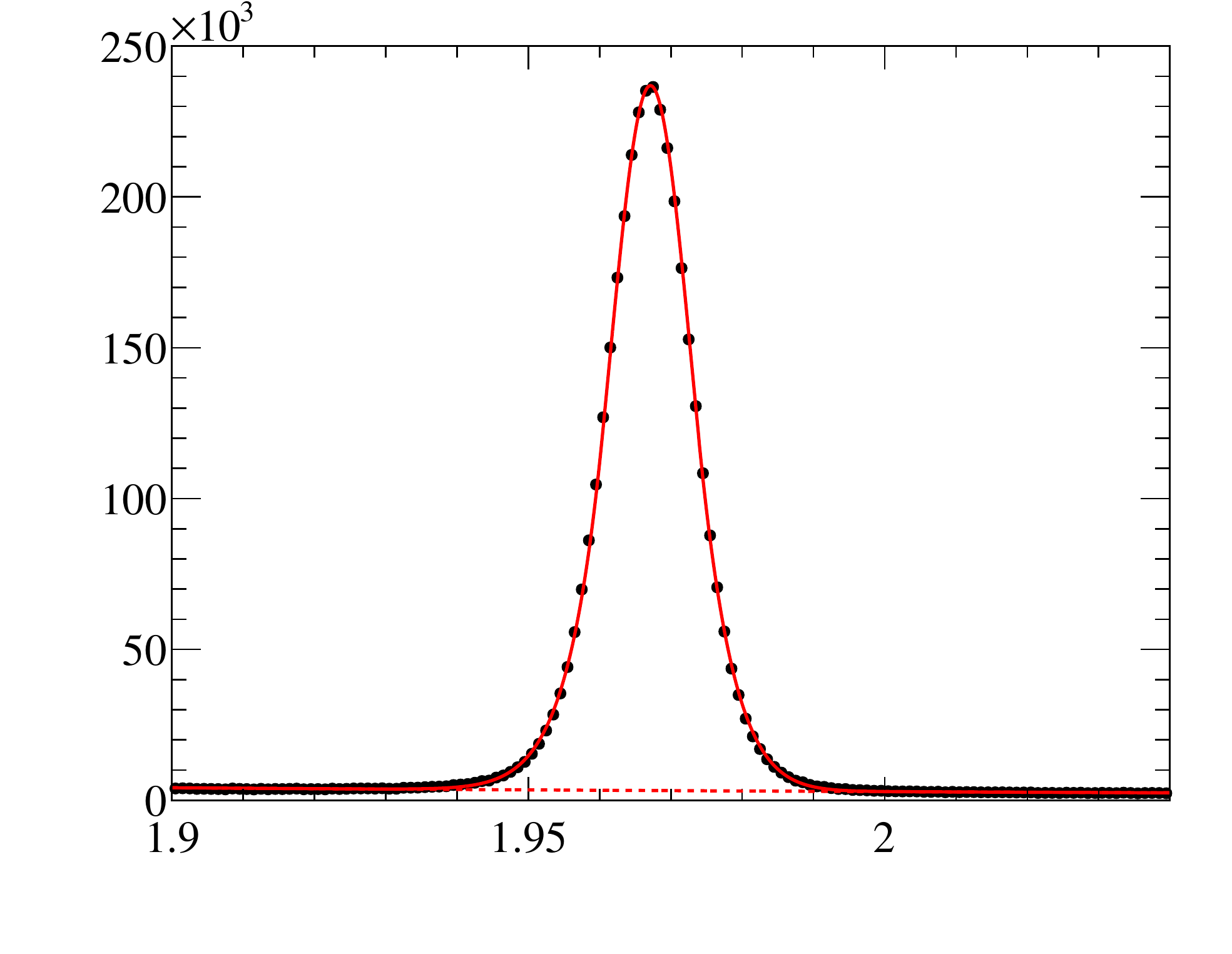}
    }
    \put(77,0){
      \includegraphics*[width=75mm,height=60mm,%
      ]{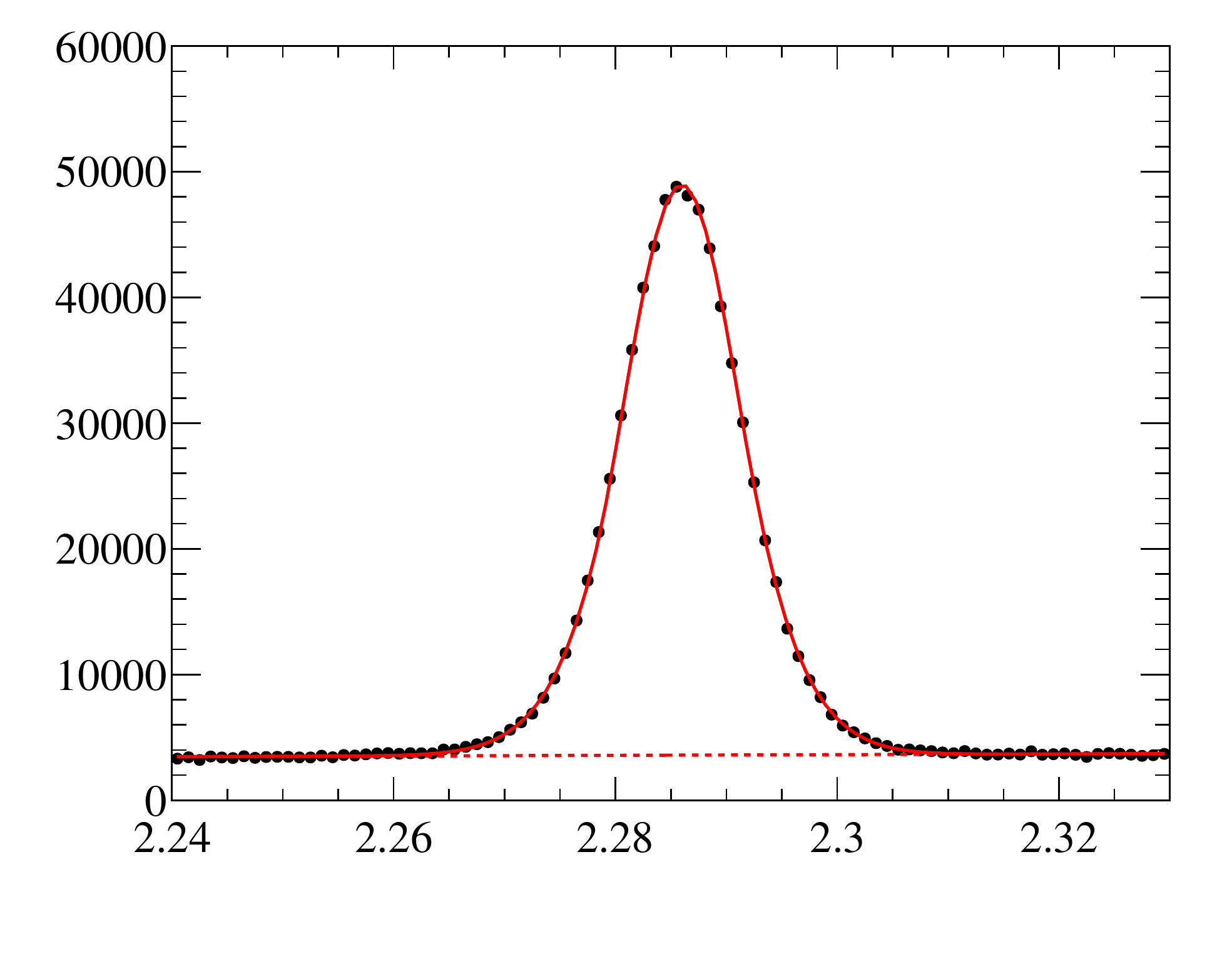}
    }
    \put(42,111) { \small  $\Dz \to \mathrm{K}^- \pip$       } 
    \put(90,112) { \small  $\Dp \to \mathrm{K}^- \pip{}\pip$ }
    \put(42,52)  { \small  $\Ds \to \mathrm{K}^+ \mathrm{K}^- \pip$}
    \put(90,52)  { \small  $\Lc \to \mathrm{p} \mathrm{K}^- \pip$ } 
    \put(30,1)   { $m_{\left(\mathrm{K^+K^-}\right)_{\Pphi}\pip}$}
    \put(105,1)  { $m_{\mathrm{p}\mathrm{K}^{-}\pip}$}
    \put(30,61)  { $m_{\mathrm{K}^{-}\pip}$}
    \put(105,61) { $m_{\mathrm{K}^{-}\pip\pip}$}
    \put(55,1)   { $\left[ \mathrm{GeV}/c^2\right]$}
    \put(132,1)  { $\left[ \mathrm{GeV}/c^2\right]$}
    \put(55,61)  { $\left[ \mathrm{GeV}/c^2\right]$}
    \put(132,61) { $\left[ \mathrm{GeV}/c^2\right]$}
   \put(-3, 27){ \small 
     \begin{sideways}%
       $\tfrac{\mathrm{dN}}{\mathrm{d}m_{\mathrm{KK}\Ppi}}~~\left[\tfrac{1}{1~\mathrm{MeV}/c^2}\right]$
     \end{sideways}%
   }
   \put(-4, 89){ \small
     \begin{sideways}%
       $\tfrac{\mathrm{dN}}{\mathrm{d}m_{\mathrm{K}\Ppi}}~~\left[\tfrac{1}{1~\mathrm{MeV}/c^2}\right]$
     \end{sideways}%
   }
   \put(73, 28){ \small 
     \begin{sideways}%
       $\tfrac{\mathrm{dN}}{\mathrm{d}m_{\mathrm{pK}\Ppi}}~~\left[\tfrac{1}{1~\mathrm{MeV}/c^2}\right]$
     \end{sideways}%
   }
   \put(73 , 87){ \small 
     \begin{sideways}%
       $\tfrac{\mathrm{dN}}{\mathrm{d}m_{\mathrm{K}\Ppi\Ppi}}~~\left[\tfrac{1}{1~\mathrm{MeV}/c^2}\right]$
     \end{sideways}%
   }
    \put(130,47){ \small 
      $\begin{array}{r}
        \mathrm{LHCb}~
      \end{array}$
    }
    \put(130,107){ \small 
      $\begin{array}{r}
        \mathrm{LHCb}~ 
      \end{array}$
    }
    \put( 11,47){ \small 
      $\begin{array}{l}
        \mathrm{LHCb}~
      \end{array}$
    }
    \put( 11,107){ \small 
      $\begin{array}{l}
        \mathrm{LHCb}~
      \end{array}$
    }
   \put( 12,112)   { a) }
   \put(140,112)   { b) }
   \put( 12, 52)   { c) }
   \put(140, 52)   { d) }
  \end{picture}
  \caption { \small 
    Invariant mass distributions 
    for selected a)~\Dz, 
   b)~\Dp,  
   c)~\Ds~and 
   d)~\Lc~candidates. The solid line corresponds to the
   total fitted PDF whilst the dotted line shows the background component.
  }
  \label{fig:charm}
\end{figure}
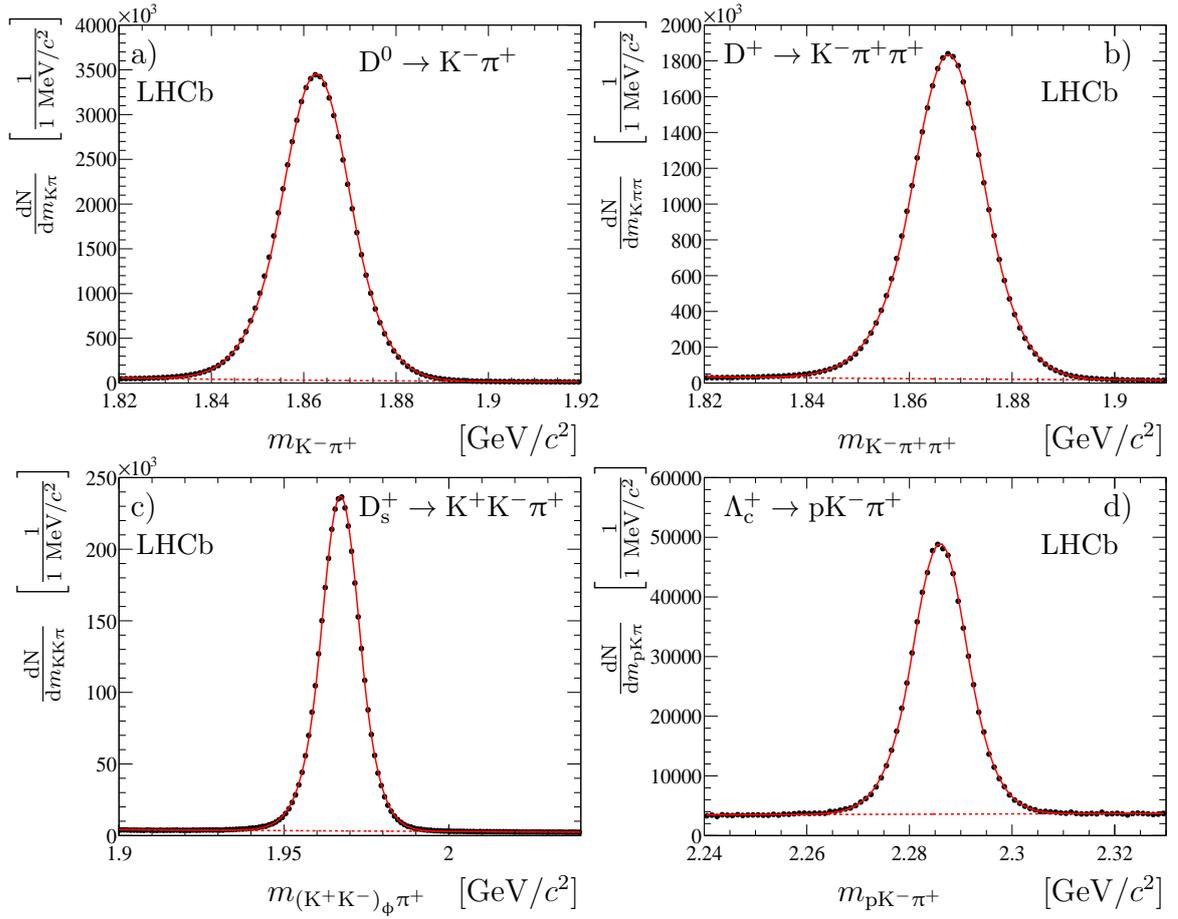
\begin{table}[t]
  \centering
  \caption{ \small
   Yields, $S$, 
   and contamination from $b$-hadron decays,
   $f_{\mathrm{b}}^{\mathrm{MC}}$,
   for the prompt charm signal.
  } \label{tab:charm}
  \vspace*{3mm}
  \setlength{\extrarowheight}{1.5pt}
  \begin{tabular*}{0.95\textwidth}{@{\hspace{3mm}}lc@{\extracolsep{\fill}}ccccc@{\hspace{3mm}}}
    & 
    & \jpsi  & \Dz & \Dp & \Ds  &  \Lc  
    \\ 
    &
    &  \mumu  
    & $\Km\pip$ 
    & $\Km\pip\pip$ 
    & $\left(\Kp\Km\right)_{\Pphi}\pip$ 
    & $\proton\Km\pip$ 
    \\
    \hline
    $S$
    & $\left[10^6\right]$  
    & 49.57  
    & 65.77  
    & 33.25  
    &  3.59  
    & 0.637  
    \\
    $f_{\mathrm{b}}^{\mathrm{MC}}$
    & $\left[ \% \right]$  
    & 1.6 
    & 1.7 
    & 1.3  
    & 2.6  
    & 4.5  \\
  \end{tabular*}   
\end{table}

\begin{figure}[htb]
  \setlength{\unitlength}{1mm}
  \centering
  \begin{picture}(150,60)
    \put(0,0){
      \includegraphics*[width=75mm,height=60mm,%
      ]{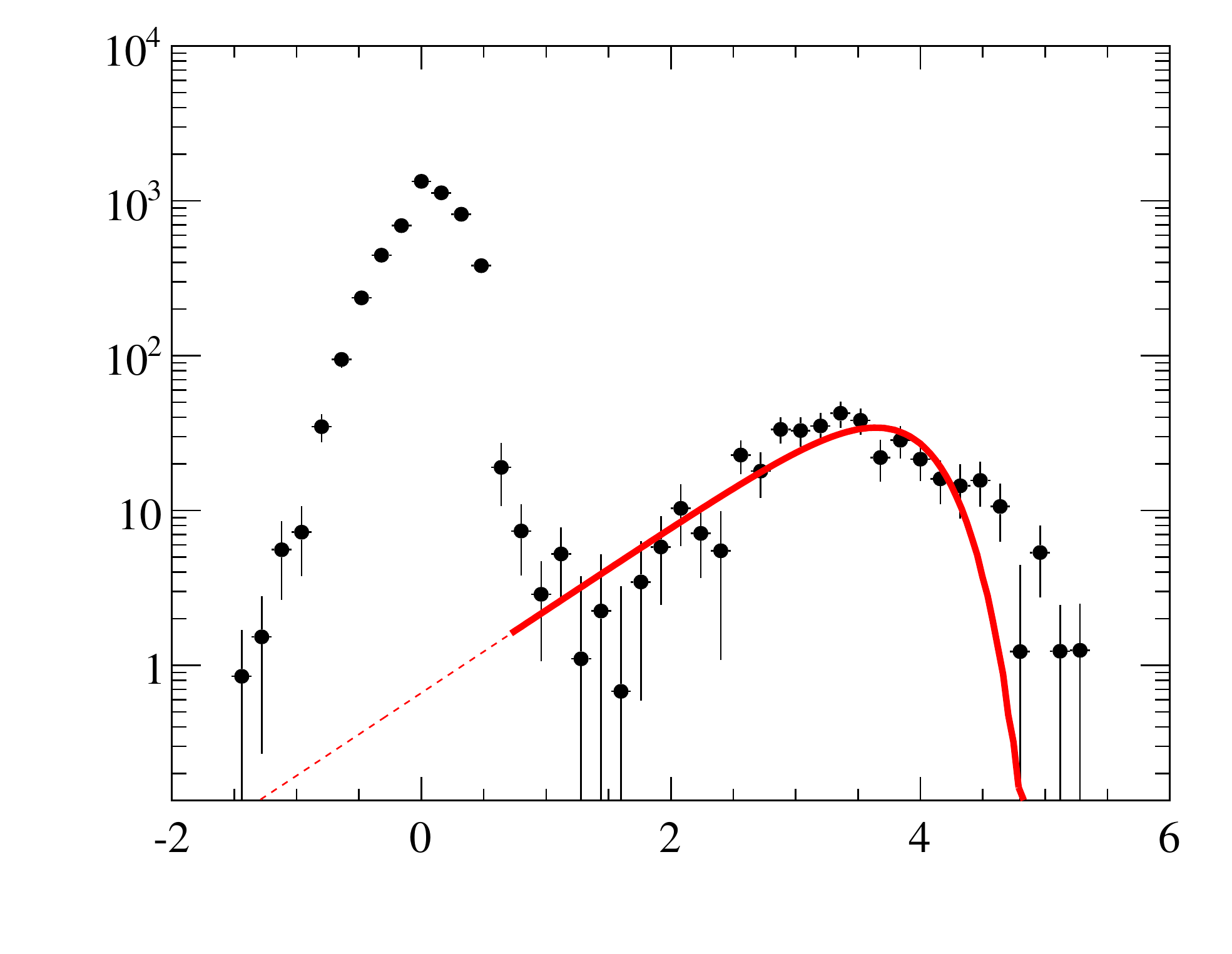}
    }
    \put(75,0){
      \includegraphics*[width=75mm,height=60mm,%
      ]{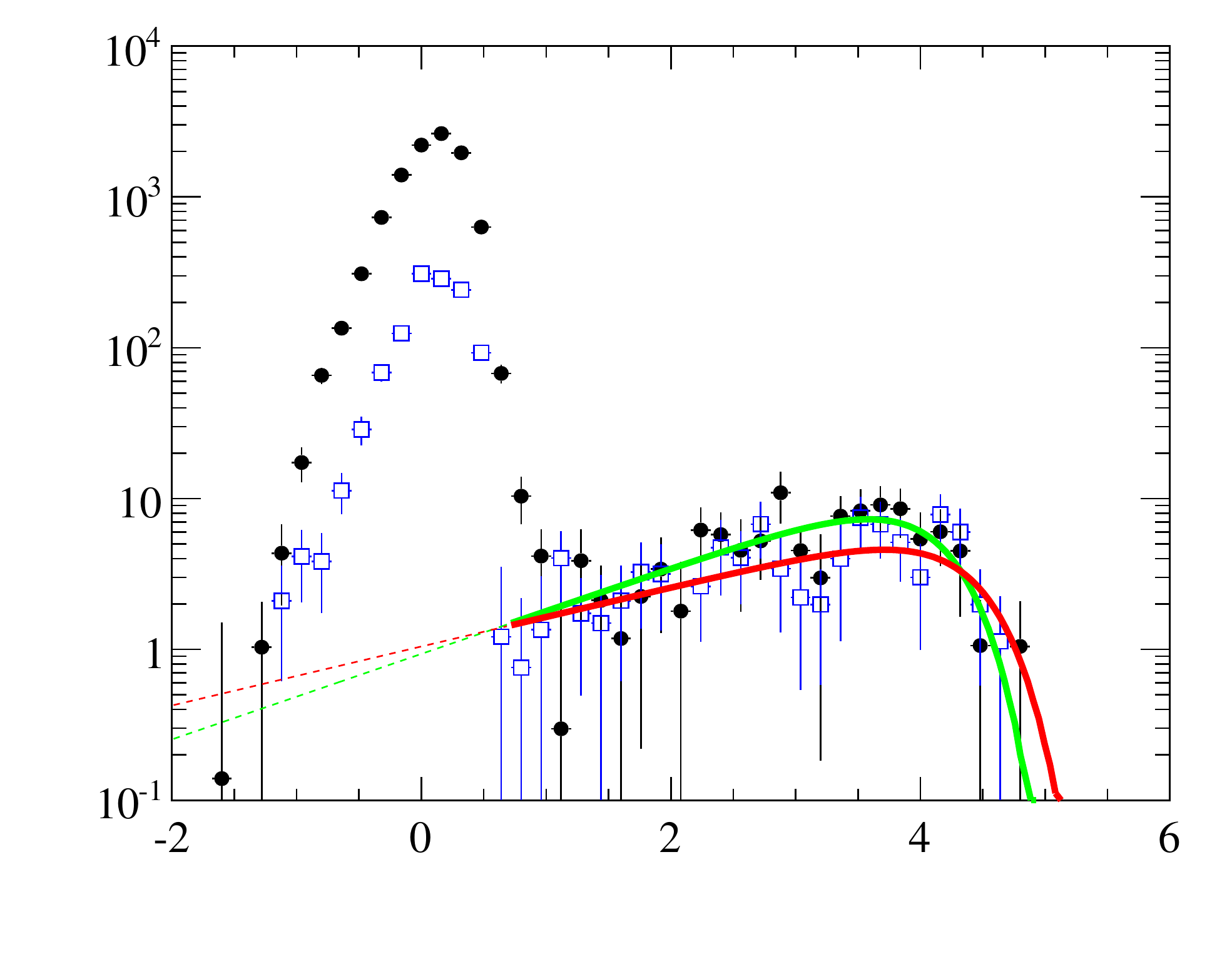}
    }
    \put( 30,1)   { $\log_{10}\chi^2_{\mathrm{global}}/\mathrm{ndf}$}
    \put(105,1)   { $\log_{10}\chi^2_{\mathrm{global}}/\mathrm{ndf}$}
    \put(2.5 , 38){ \small 
      \begin{sideways}%
        Candidates 
      \end{sideways}%
    }
    \put(77.5 , 38){ \small 
      \begin{sideways}%
        Candidates 
      \end{sideways}%
    }
    \put( 55,45){ \small 
      $\begin{array}{r}
        \mathrm{LHCb}~
      \end{array}$
    }
    \put(130,45){ \small 
      $\begin{array}{r}
        \mathrm{LHCb}~ 
      \end{array}$
    }
    \put( 65,52)   { a) }
    \put(140,52)   { b) }
    \put(130,35){ \tiny
      $\begin{array}{cl}
        {                \text{\ding{108}} } & \DzDzb  \\
        {\color{blue}    \square } & \DzDz  
       \end{array}$
    } 
  \end{picture}
  \caption { \small 
    a)~Background subtracted distribution of 
    $\log_{10}\chi^2_{\mathrm{global}}/\mathrm{ndf}$ 
for \psiDz events.
    The solid line corresponds to the fit result in the region 
    $\chi^2_{\mathrm{global}}/\mathrm{ndf}>5$ by the function described 
    in the text, 
    the dashed line corresponds to the extrapolation of the fit results to
    the $\chi^2_{\mathrm{global}}/\mathrm{ndf}<5$ region.
    b)~Likewise for~\DzDz~(blue squares and red line) 
       and \DzDzb~(black circles and green line). 
  }
  \label{fig:pileup}
\end{figure}

The selected charm candidates are paired to form di-charm candidates:
\psiC,  \CC~and \CCbar. 
A global fit of the di-charm candidates 
is performed~\cite{Hulsbergen:2005pu},
similar to that described above for single charm hadrons,
which requires both
hadrons to be consistent with originating from a 
common vertex. The reduced $\chi^2$~of this fit, 
$\chi^2_{\mathrm{global}}/\mathrm{ndf}$, is required to be less than 5.
This reduces the background from the pile-up of two
interactions each producing a charm hadron to 
a negligible level. The remaining 
contamination from 
the pile-up and decays from beauty hadrons
is extracted directly from the data as follows. 
The distributions of $\chi^2_{\mathrm{global}}/\mathrm{ndf}$ for 
\psiDz,
\DzDz and \DzDzb events are shown in Fig.~\ref{fig:pileup}.
For the region $\chi^2_{\mathrm{global}}/\mathrm{ndf}>5$ the distributions 
are well described by functions of the form\footnote{The functional
  form is inspired by the $\chi^2$ distribution.}
\begin{equation}
f \left( x \right) \propto   (\alpha x)^{\frac{n}{2}-1}\mathrm{e}^{-\frac{\alpha x}{2}},
\end{equation} 
where $\alpha$ and $n$ are free parameters.  Fits with this functional
form are used to extrapolate the yield 
in the region 
$\chi^2_{\mathrm{global}}/\mathrm{ndf}>5$ 
to the region 
$\chi^2_{\mathrm{global}}/\mathrm{ndf}<5$. 
Based on these studies we
conclude that background from pile-up is negligible.

\begin{figure}[htb]
  \setlength{\unitlength}{1mm}
  \centering
  \begin{picture}(150,120)
    \put(0,60){
      \includegraphics*[width=75mm,height=60mm,%
      ]{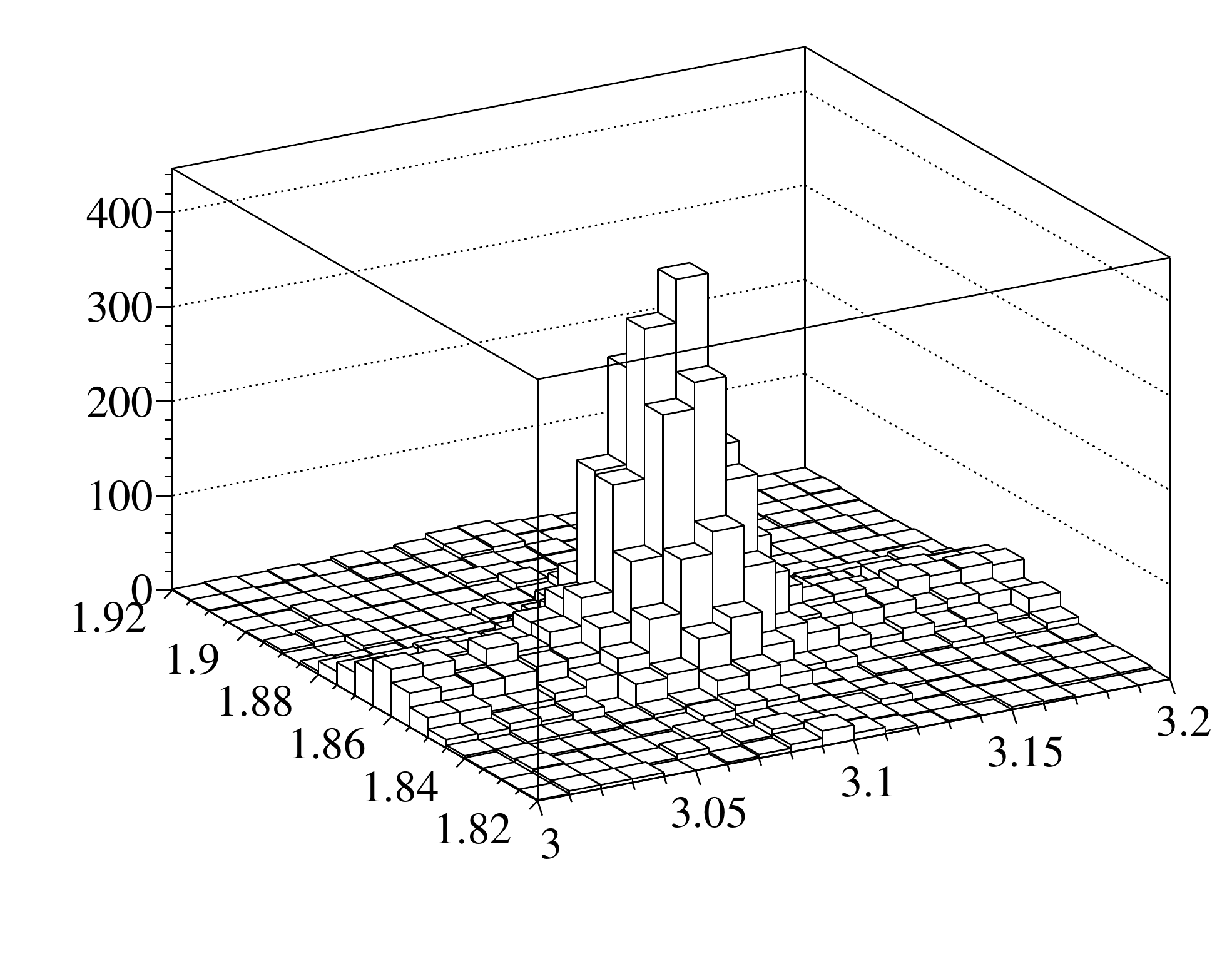}
    }
    \put(75,60){
      \includegraphics*[width=75mm,height=60mm,%
      ]{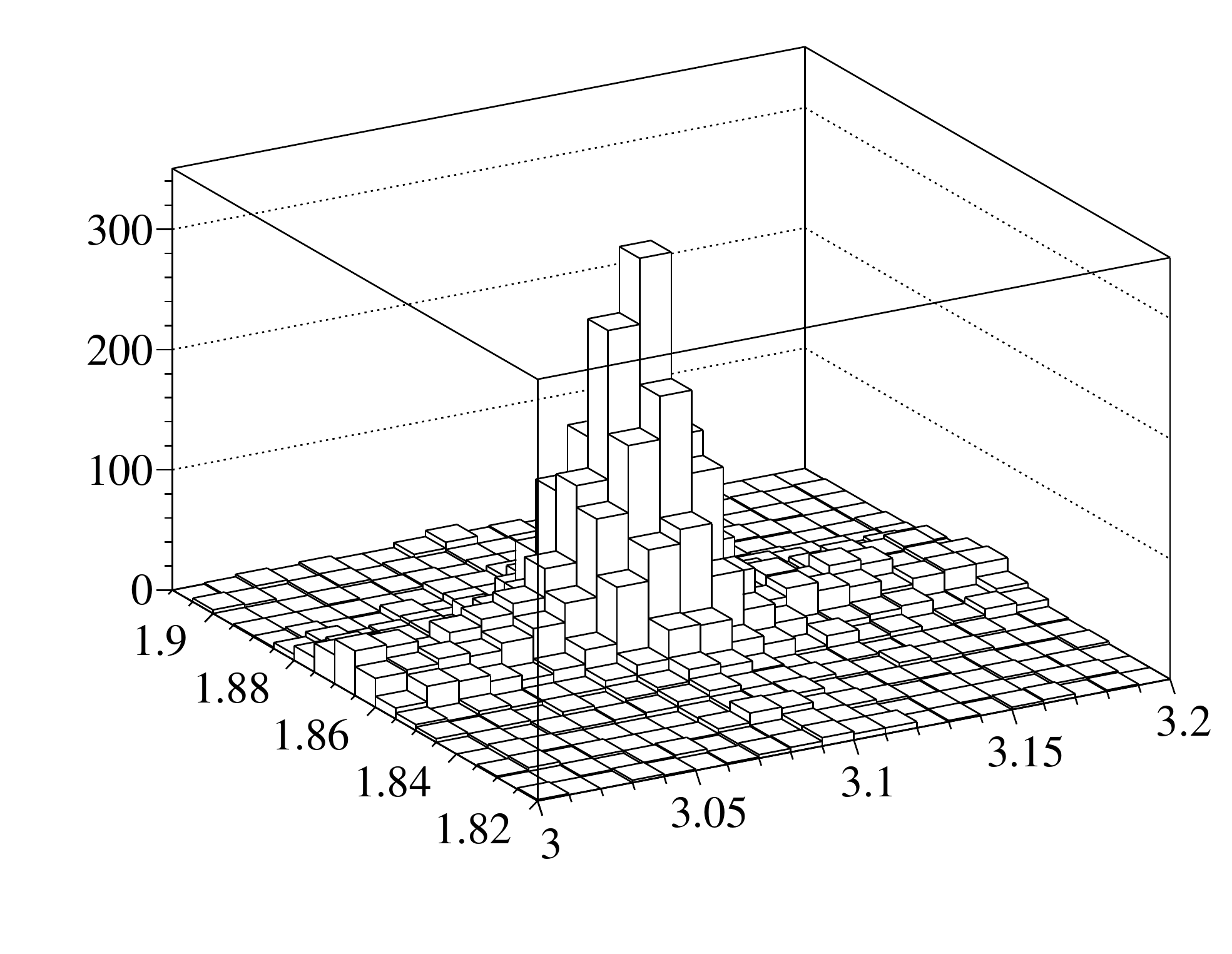}
    }
    \put(0,0){
      \includegraphics*[width=75mm,height=60mm,%
      ]{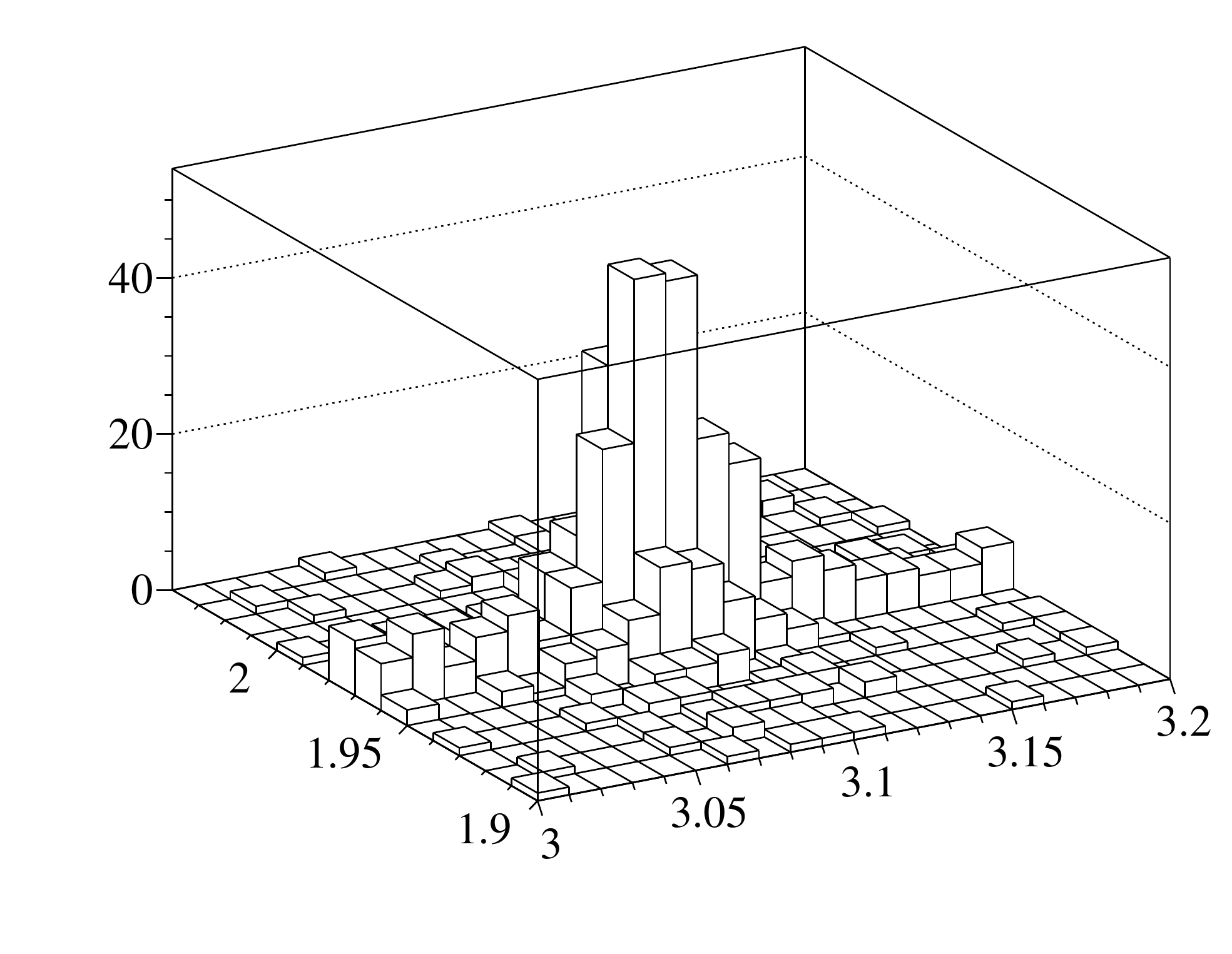}
    }
    \put(75,0){
      \includegraphics*[width=75mm,height=60mm,%
      ]{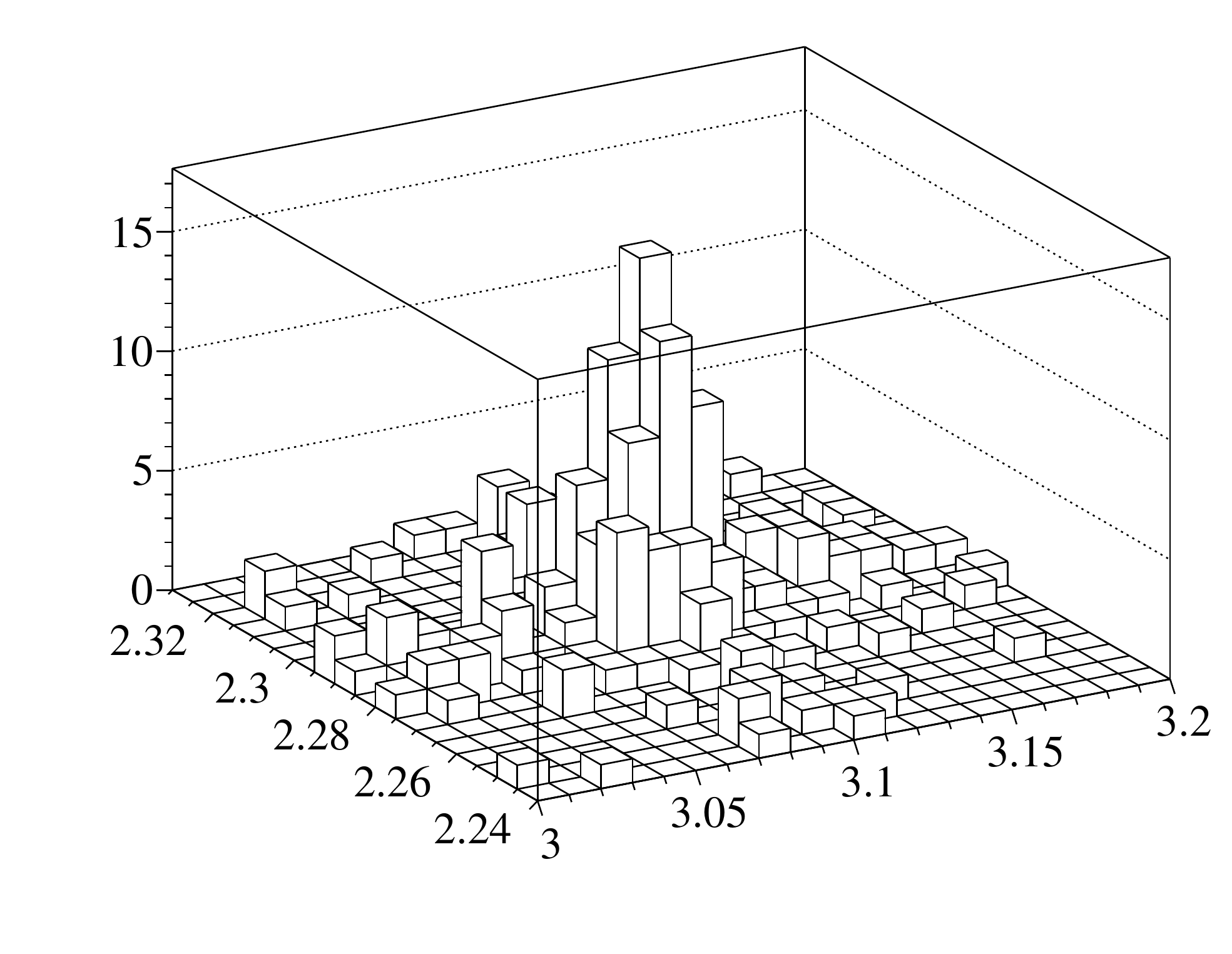}
    }
    \put(7.5,74.5)  { 
      \begin{rotate}{-32} \small 
        $m_{\mathrm{K}\Ppi}~\left[\mathrm{GeV}/c^2\right]$
      \end{rotate}
    }
    \put(82.4,74.5)  { 
      \begin{rotate}{-32} \small 
        $m_{\mathrm{K}\Ppi{}\Ppi}~\left[\mathrm{GeV}/c^2\right]$
      \end{rotate}
    }
    \put(7.5,14.5)  { 
      \begin{rotate}{-32} \small 
        $m_{\mathrm{KK}\Ppi}~\left[\mathrm{GeV}/c^2\right]$
      \end{rotate}
    }
    \put(82.5,14.5)  { 
      \begin{rotate}{-32} \small 
        $m_{\mathrm{pK}\Ppi}~\left[\mathrm{GeV}/c^2\right]$
      \end{rotate}
    }
    \put(46,63.5)  { 
      \begin{rotate}{12} \small 
        $m_{\mumu}~~~\left[\mathrm{GeV}/c^2\right]$
      \end{rotate}
    }
    \put(121,63.5)  { 
      \begin{rotate}{12} \small 
        $m_{\mumu}~~~\left[\mathrm{GeV}/c^2\right]$
      \end{rotate}
    }
    \put(46,3.5)  { 
      \begin{rotate}{12} \small 
        $m_{\mumu}~~~\left[\mathrm{GeV}/c^2\right]$
      \end{rotate}
    }
    \put(121,3.5)  { 
      \begin{rotate}{12} \small 
        $m_{\mumu}~~~\left[\mathrm{GeV}/c^2\right]$
      \end{rotate}
    }
    \put(1 ,  88 )  { \small 
      \begin{sideways}%
        Candidates 
      \end{sideways}%
    }
    \put(76 , 88 )  { \small 
      \begin{sideways}%
        Candidates 
      \end{sideways}%
    }
    \put(1 ,  28 )  { \small 
      \begin{sideways}%
        Candidates 
      \end{sideways}%
    }
    \put(76 , 28 )  { \small 
      \begin{sideways}%
        Candidates 
      \end{sideways}%
    }
    \put(65  , 114){a)}
    \put(140 , 114){b)}
    \put(65  ,  54){c)}
    \put(140 ,  54){d)}
    \put(10,115){ \small 
      $\begin{array}{l}
        \mathrm{LHCb}~ 
      \end{array}$
    }
    \put(10,55){ \small 
      $\begin{array}{l}
        \mathrm{LHCb}~ 
      \end{array}$
    }
    \put(85,115){ \small
      $\begin{array}{l}
        \mathrm{LHCb}~ 
      \end{array}$
    }
    \put(85,55){ \small 
      $\begin{array}{l}
        \mathrm{LHCb}~ 
      \end{array}$
    }
  \end{picture}
  \caption { \small
    Invariant 
    mass distributions for 
    a)~\psiDz,
    b)~\psiDp,
    c)~\psiDs and 
    d)~\psiLc candidates. 
  }
  \label{fig:psid_trg_pt3_2d}
\end{figure}

\begin{figure}[htb]
  \setlength{\unitlength}{1mm}
  \centering
  \begin{picture}(150,120)
    \put(0,60){
      \includegraphics*[width=75mm,height=60mm,%
      ]{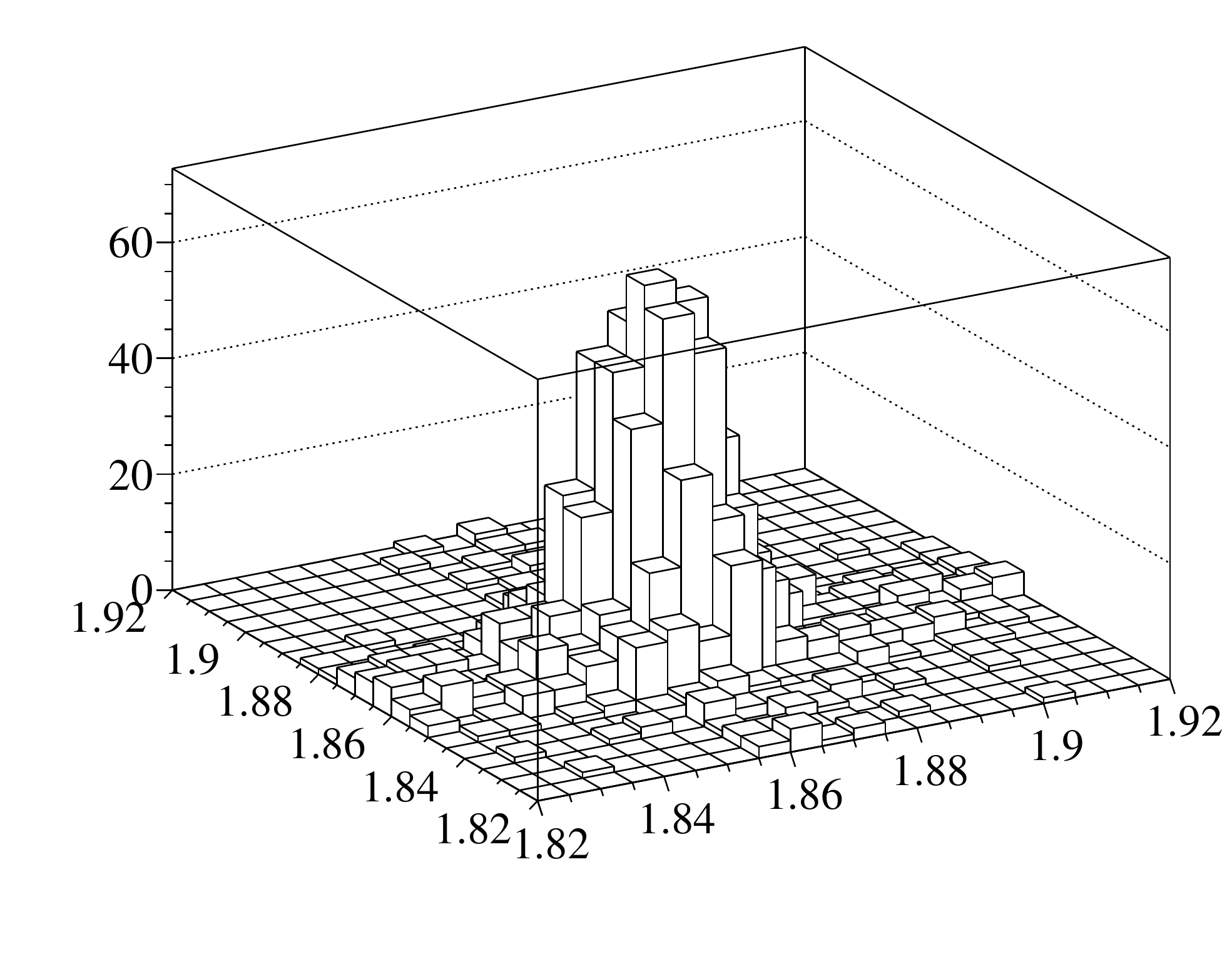}
    }
    \put(75,60){
      \includegraphics*[width=75mm,height=60mm,%
      ]{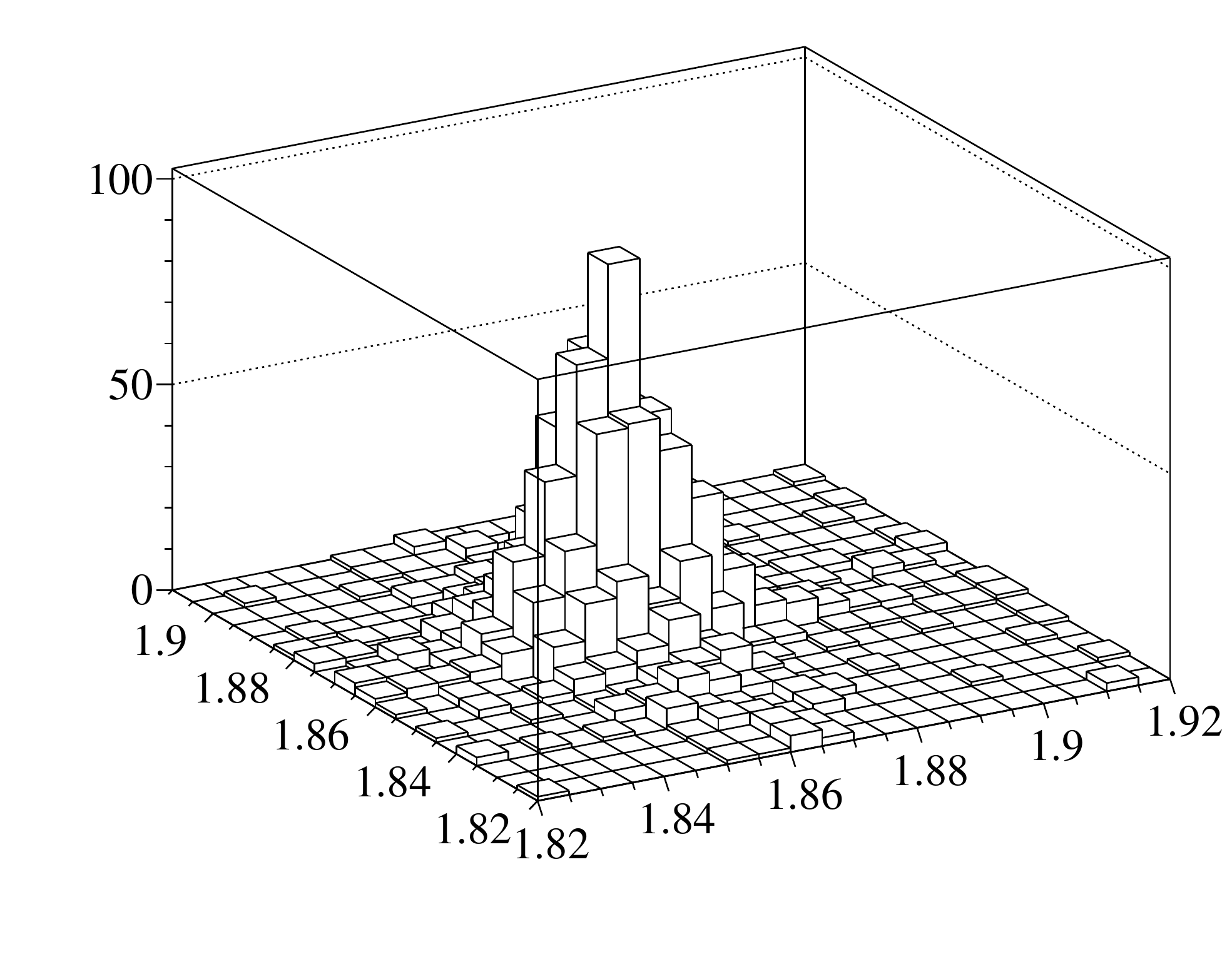}
    }
    \put(0,0){
      \includegraphics*[width=75mm,height=60mm,%
      ]{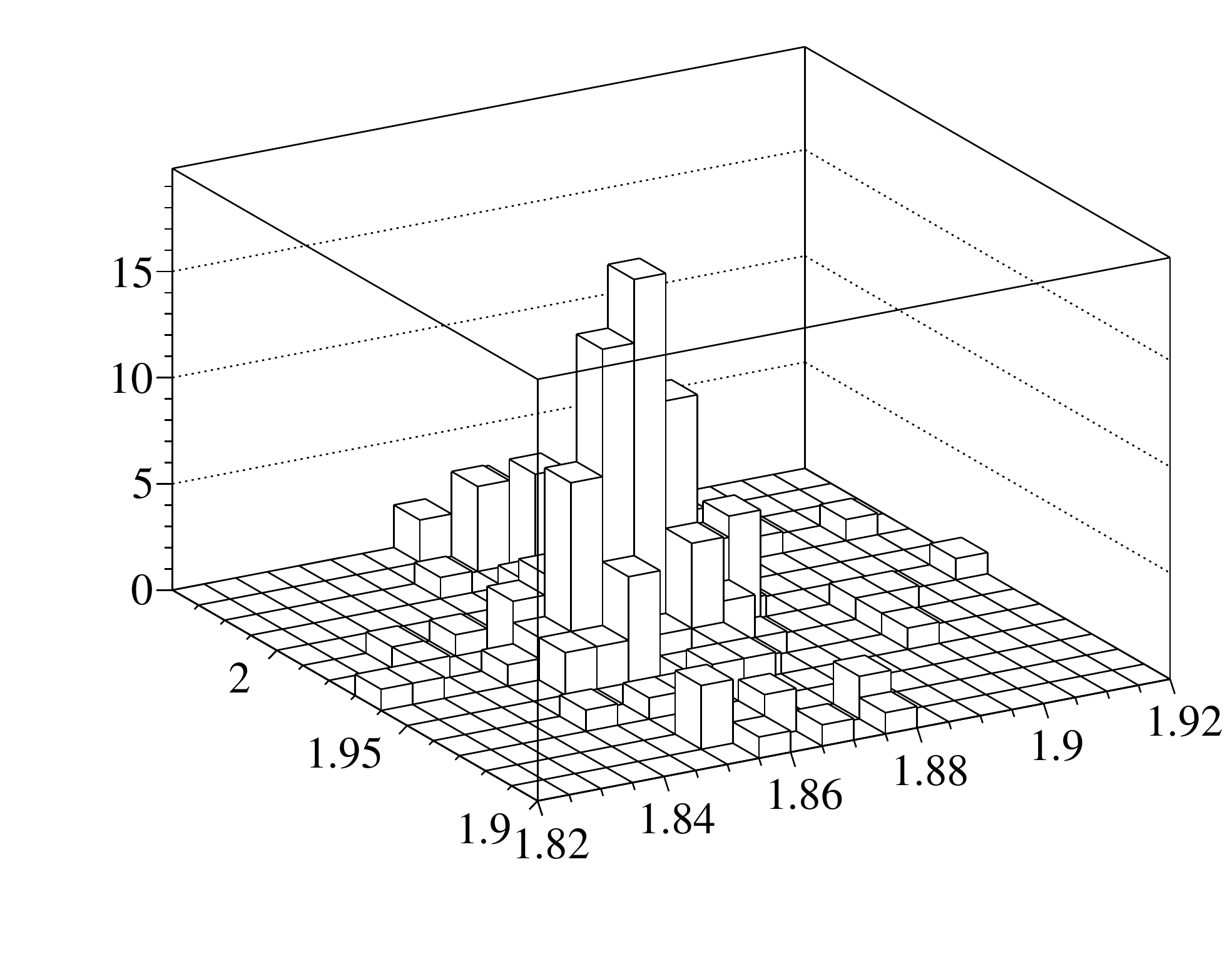}
   }
    \put(75,0){
      \includegraphics*[width=75mm,height=60mm,%
      ]{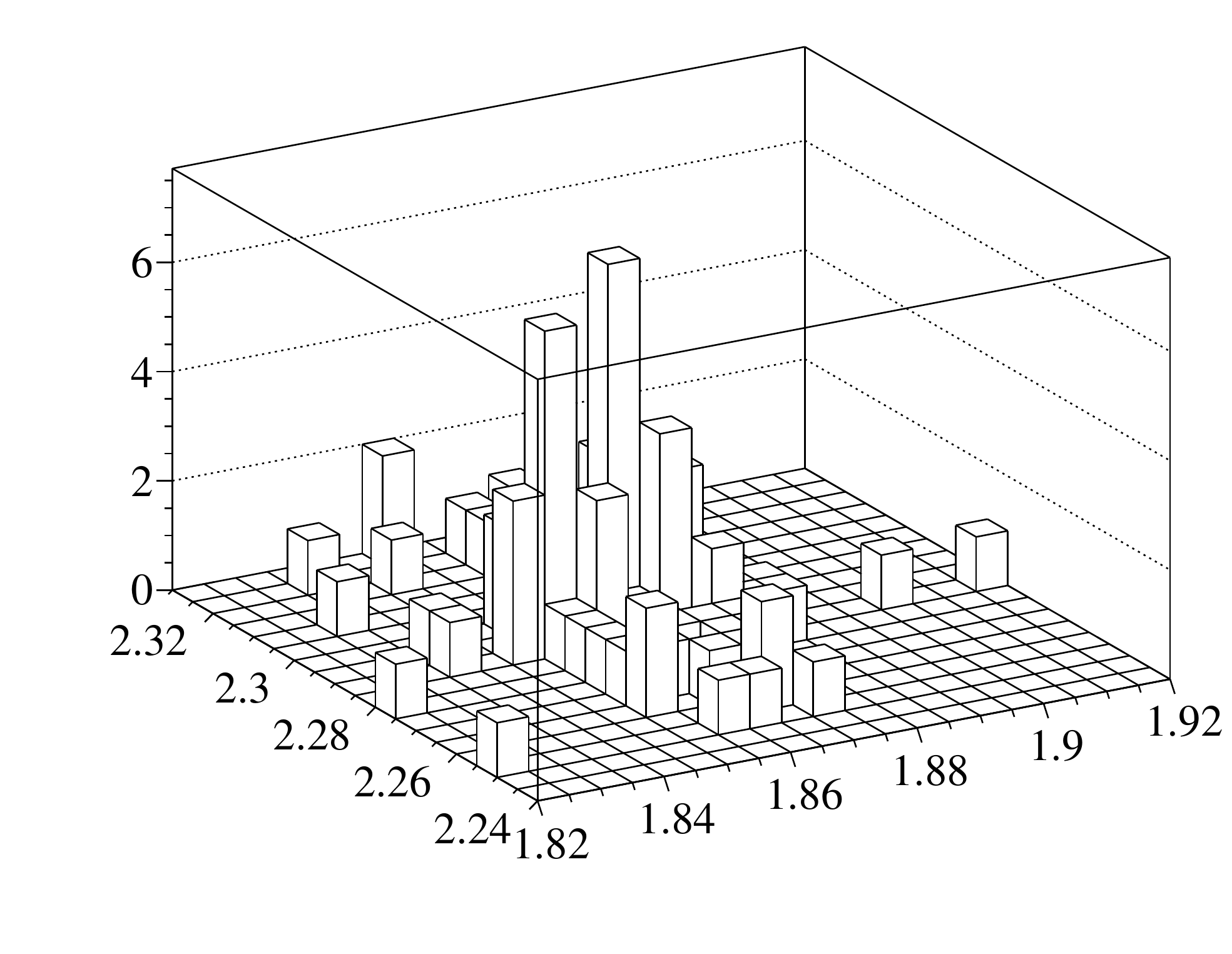}
    }
    \put(7.5,74.5)  { 
      \begin{rotate}{-32} \small 
        $m_{\mathrm{K}\Ppi}~\left[\mathrm{GeV}/c^2\right]$
      \end{rotate}
    }
    \put(82.5,74.5)  { 
      \begin{rotate}{-32} \small 
        $m_{\mathrm{K}\Ppi{}\Ppi}~\left[\mathrm{GeV}/c^2\right]$
      \end{rotate}
    }
    \put(7.5,14.5)  { 
      \begin{rotate}{-32} \small 
        $m_{\mathrm{KK}\Ppi}~\left[\mathrm{GeV}/c^2\right]$
      \end{rotate}
    }
    \put(82.5,14.5) { 
      \begin{rotate}{-32} \small 
        $m_{\mathrm{pK}\Ppi}~\left[\mathrm{GeV}/c^2\right]$
      \end{rotate}
    }
    \put(46,64)  { 
      \begin{rotate}{12} \small 
        $m_{\mathrm{K}\Ppi}~~~\left[\mathrm{GeV}/c^2\right]$
      \end{rotate}
    }
    \put(121,64)  { 
      \begin{rotate}{12} \small 
        $m_{\mathrm{K}\Ppi}~~~\left[\mathrm{GeV}/c^2\right]$
      \end{rotate}
    }
    \put(46,4)  { 
      \begin{rotate}{12} \small 
        $m_{\mathrm{K}\Ppi}~~~\left[\mathrm{GeV}/c^2\right]$
      \end{rotate}
    }
    \put(121,4)  { 
      \begin{rotate}{12} \small 
        $m_{\mathrm{K}\Ppi}~~~\left[\mathrm{GeV}/c^2\right]$
      \end{rotate}
    }
    \put(1 , 88 )  { \small 
      \begin{sideways}%
        Candidates 
      \end{sideways}%
    }
    \put(76 , 88 )  { \small 
      \begin{sideways}%
        Candidates 
      \end{sideways}%
    }
    \put(1 , 28 )  { \small 
      \begin{sideways}%
        Candidates 
      \end{sideways}%
    }
    \put(76 , 28 )  { \small 
      \begin{sideways}%
        Candidates 
      \end{sideways}%
    }
    \put(65  , 114){a)}
    \put(140 , 114){b)}
    \put(65  ,  54){c)}
    \put(140 ,  54){d)}
    \put(10,115){ \small 
      LHCb
    }
    \put(10,55){ \small 
      LHCb
    }
    \put(85,115){ \small 
      LHCb
    }
    \put(85,55){ \small 
      LHCb
    }
  \end{picture}
  \caption { \small
    Invariant 
    mass distributions for $\Dz{}\mathrm{C}$ candidates:
    a)~\DzDz,
    b)~\DzDp,
    c)~\DzDs and 
    d)~\DzLc.
  }
  \label{fig:cc_trg_pt3_2d1}
\end{figure}

\begin{figure}[htb]
  \setlength{\unitlength}{1mm}
  \centering
  \begin{picture}(150,120)
    \put(0,60){
      \includegraphics*[width=75mm,height=60mm,%
      ]{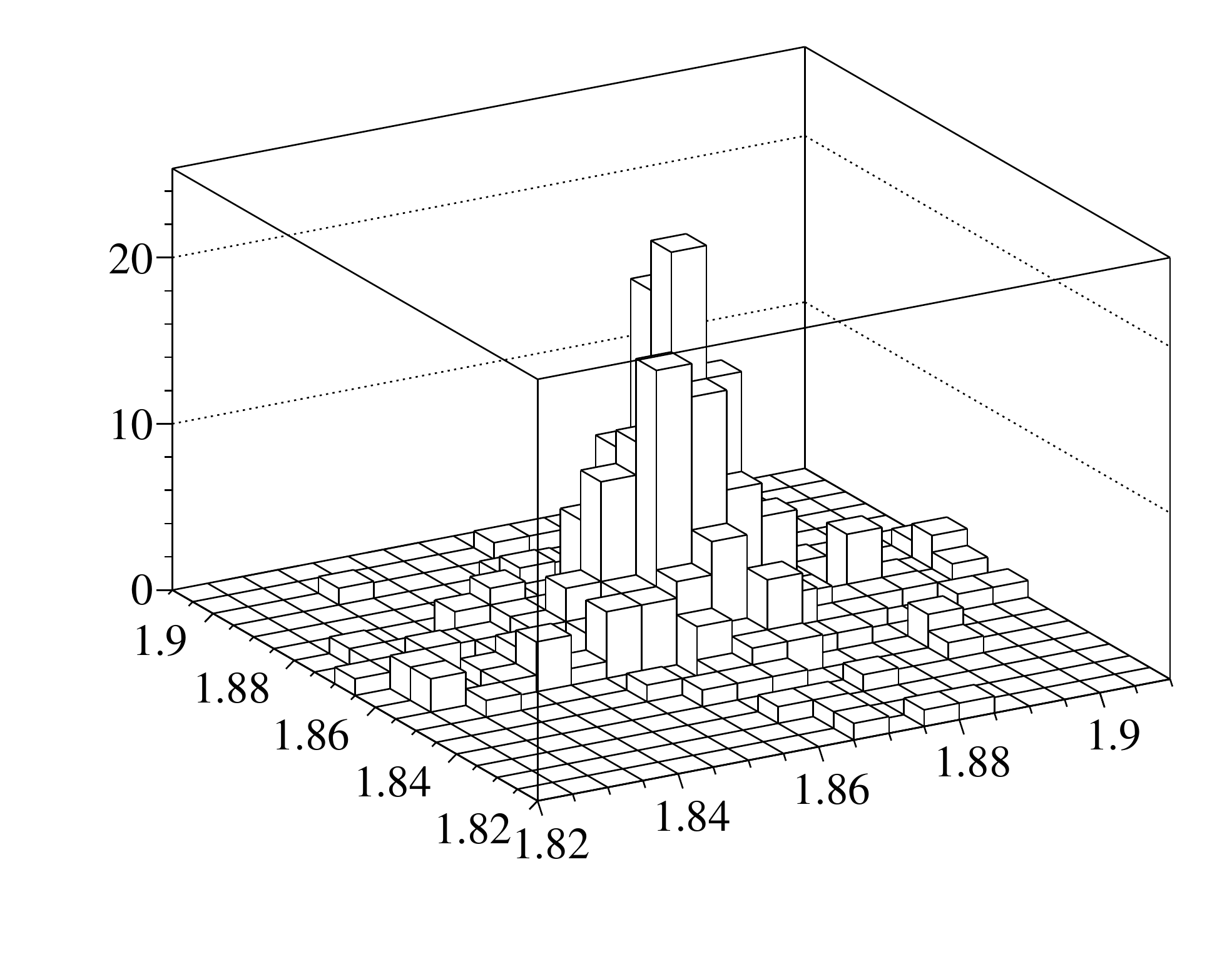}
    }
    \put(75,60){
      \includegraphics*[width=75mm,height=60mm,%
      ]{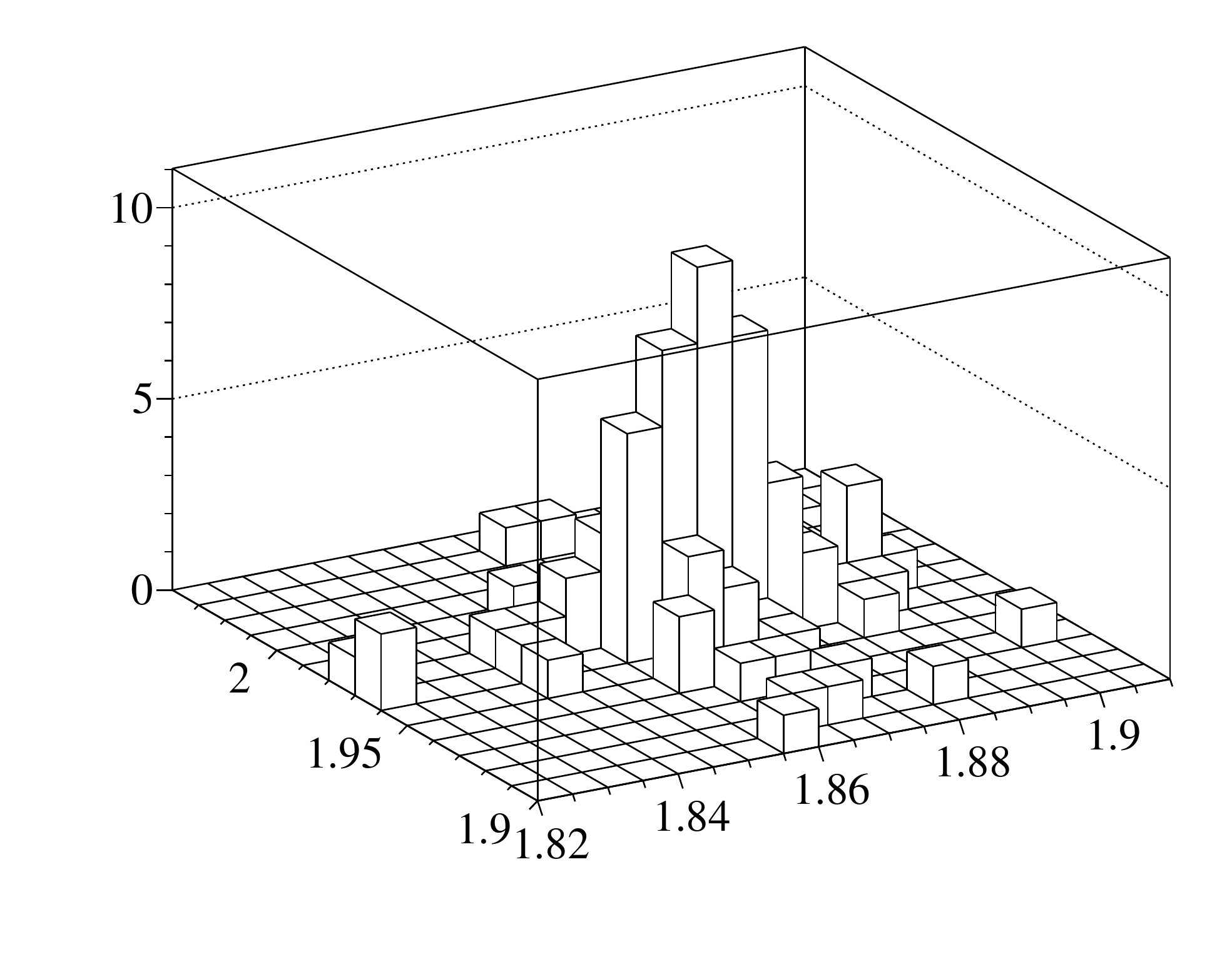}
    }
    \put(0, 0){
      \includegraphics*[width=75mm,height=60mm,%
      ]{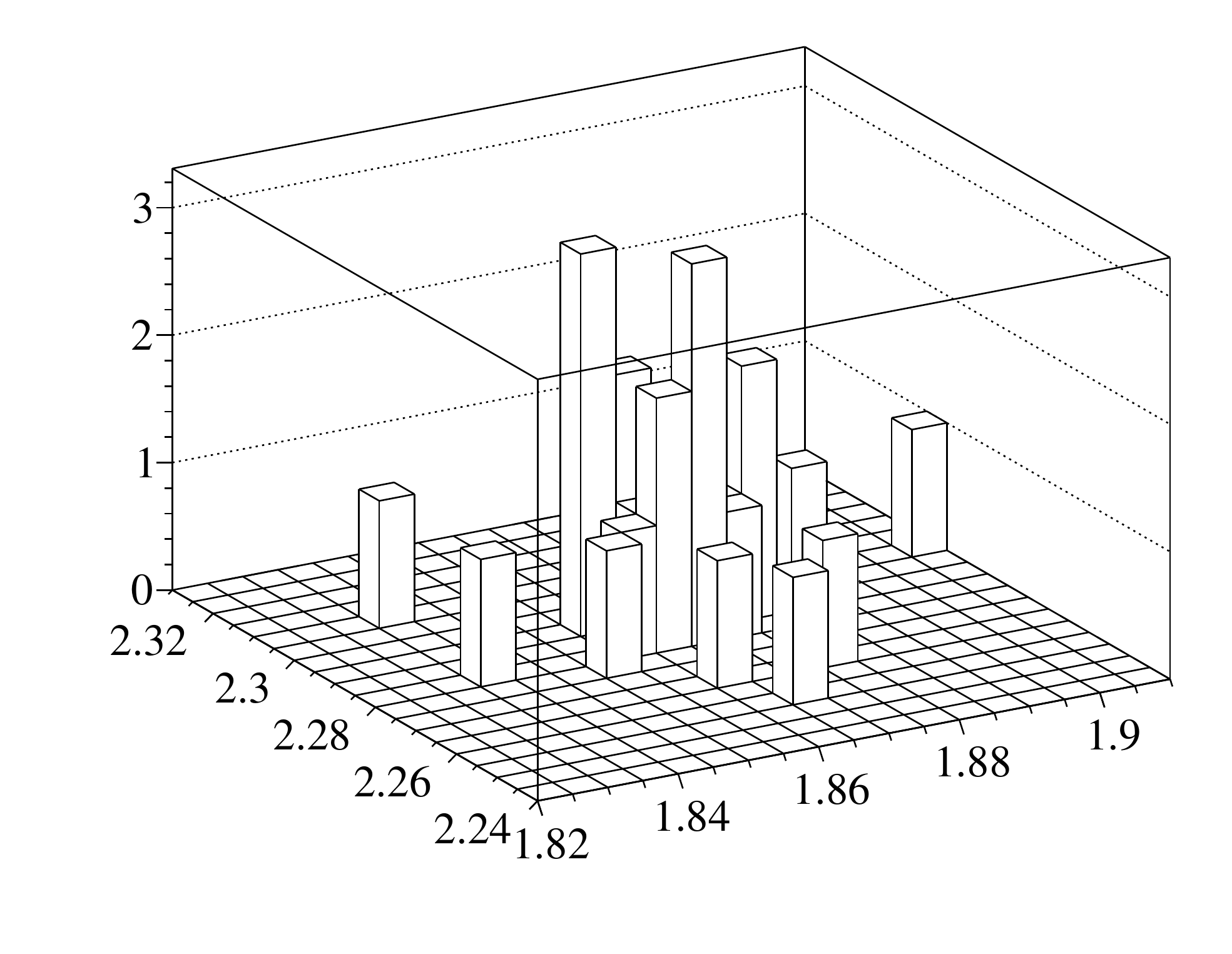}
    }
    \put(7.5,74.55)  { 
      \begin{rotate}{-32} \small 
        $m_{\mathrm{K}\Ppi{}\Ppi}~\left[\mathrm{GeV}/c^2\right]$
      \end{rotate}
    }
    \put(83,74.5)  { 
      \begin{rotate}{-32} \small 
        $m_{\mathrm{KK}\Ppi}~\left[\mathrm{GeV}/c^2\right]$
      \end{rotate}
    }
    \put(7.5,14.5)  { 
      \begin{rotate}{-32} \small 
        $m_{\mathrm{pK}\Ppi}~\left[\mathrm{GeV}/c^2\right]$
      \end{rotate}
    }
    \put(46,64)  { 
      \begin{rotate}{12} \small 
        $m_{\mathrm{K}\Ppi{}\Ppi}~\left[\mathrm{GeV}/c^2\right]$
      \end{rotate}
    }
    \put(121,64)  { 
      \begin{rotate}{12} \small 
        $m_{\mathrm{K}\Ppi{}\Ppi}~\left[\mathrm{GeV}/c^2\right]$
      \end{rotate}
    }
    \put(46,4)  { 
      \begin{rotate}{12} \small 
        $m_{\mathrm{K}\Ppi{}\Ppi}~\left[\mathrm{GeV}/c^2\right]$
      \end{rotate}
    }
    \put(1 ,88 )  { \small 
      \begin{sideways}%
        Candidates 
      \end{sideways}%
    }
    \put(76,88 )  { \small 
      \begin{sideways}%
        Candidates 
      \end{sideways}%
    }
    \put(1,28 )  { \small 
      \begin{sideways}%
        Candidates 
      \end{sideways}%
    }
    \put(65  , 114){a)}
    \put(140 , 114){b)}
    \put(65  ,  54){c)}
    \put(10,115){ \small 
      LHCb 
    }
    \put(10,55){ \small 
      LHCb
    }
    \put(85,115){ \small 
      LHCb
    }
  \end{picture}
  \caption { \small 
    Invariant 
    mass distributions for $\Dp{}\mathrm{C}$~candidates:
    a)~\DpDp,
    b)~\DpDs, and 
    c)~\DpLc.
  }
  \label{fig:cc_trg_pt3_2d2}
\end{figure}

\begin{figure}[htb]
  \setlength{\unitlength}{1mm}
  \centering
  \begin{picture}(150,120)
    \put(0,60){
      \includegraphics*[width=75mm,height=60mm,%
      ]{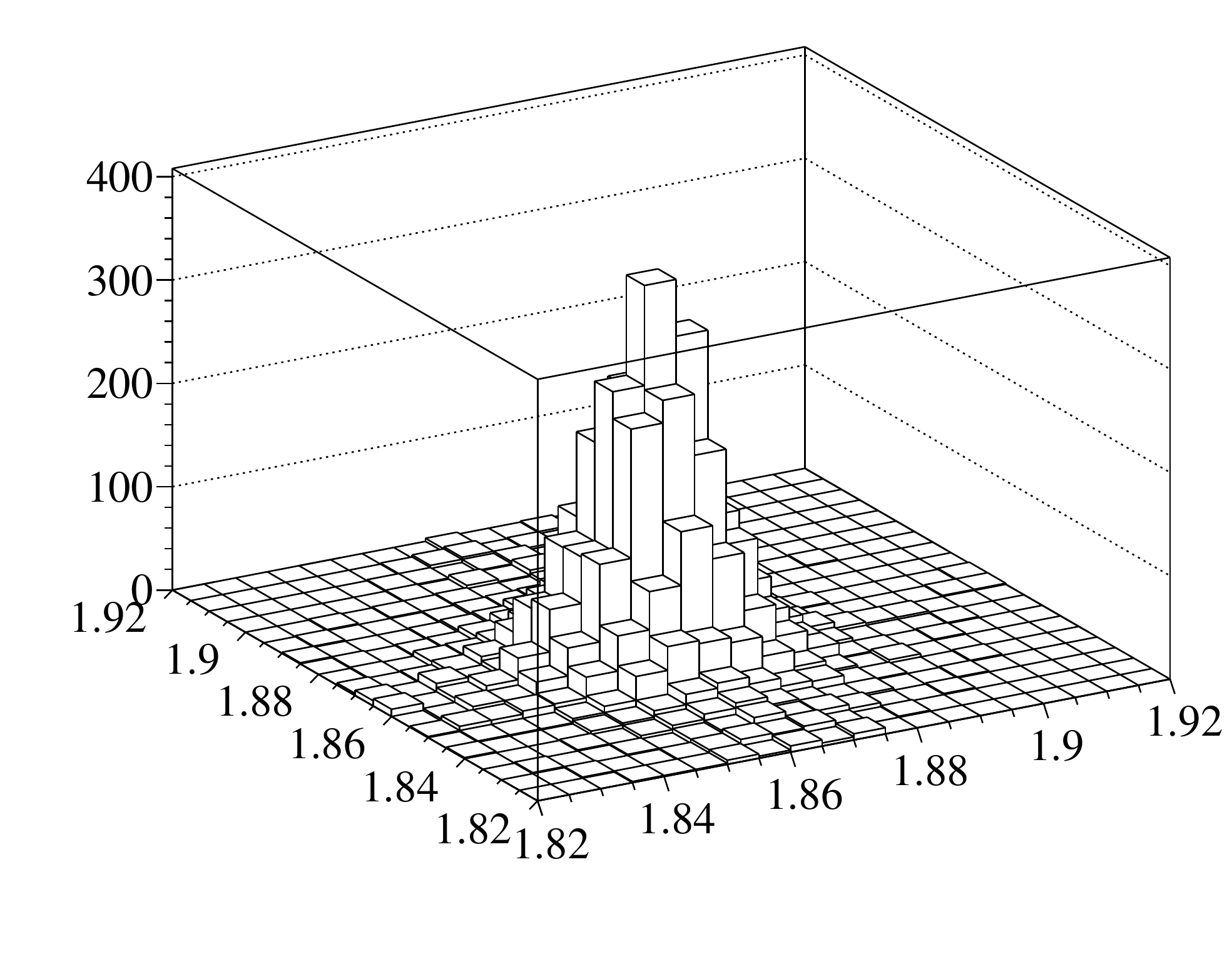}
    }
    \put(75,60){
      \includegraphics*[width=75mm,height=60mm,%
      ]{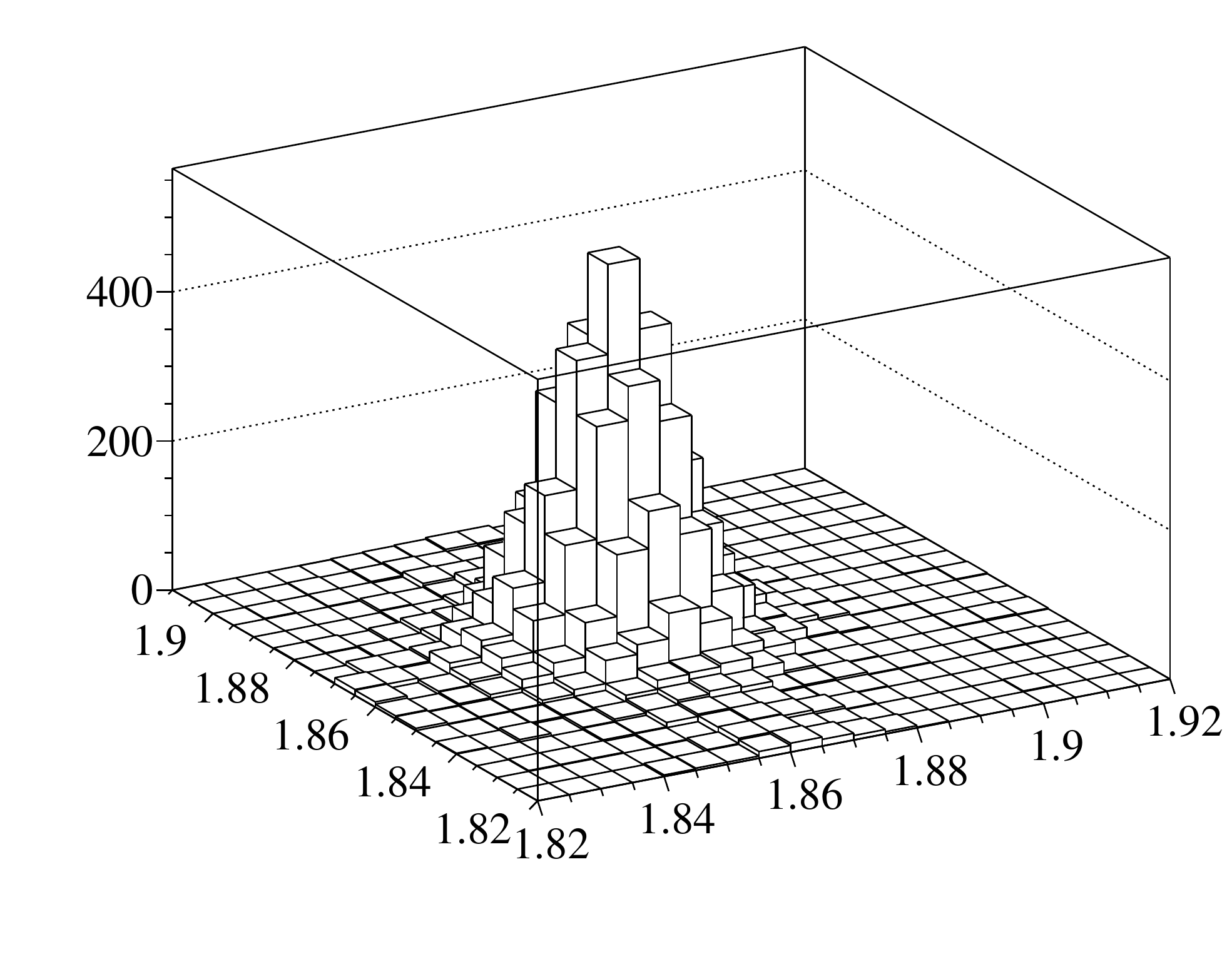}
    }
    \put(0,0){
      \includegraphics*[width=75mm,height=60mm,%
      ]{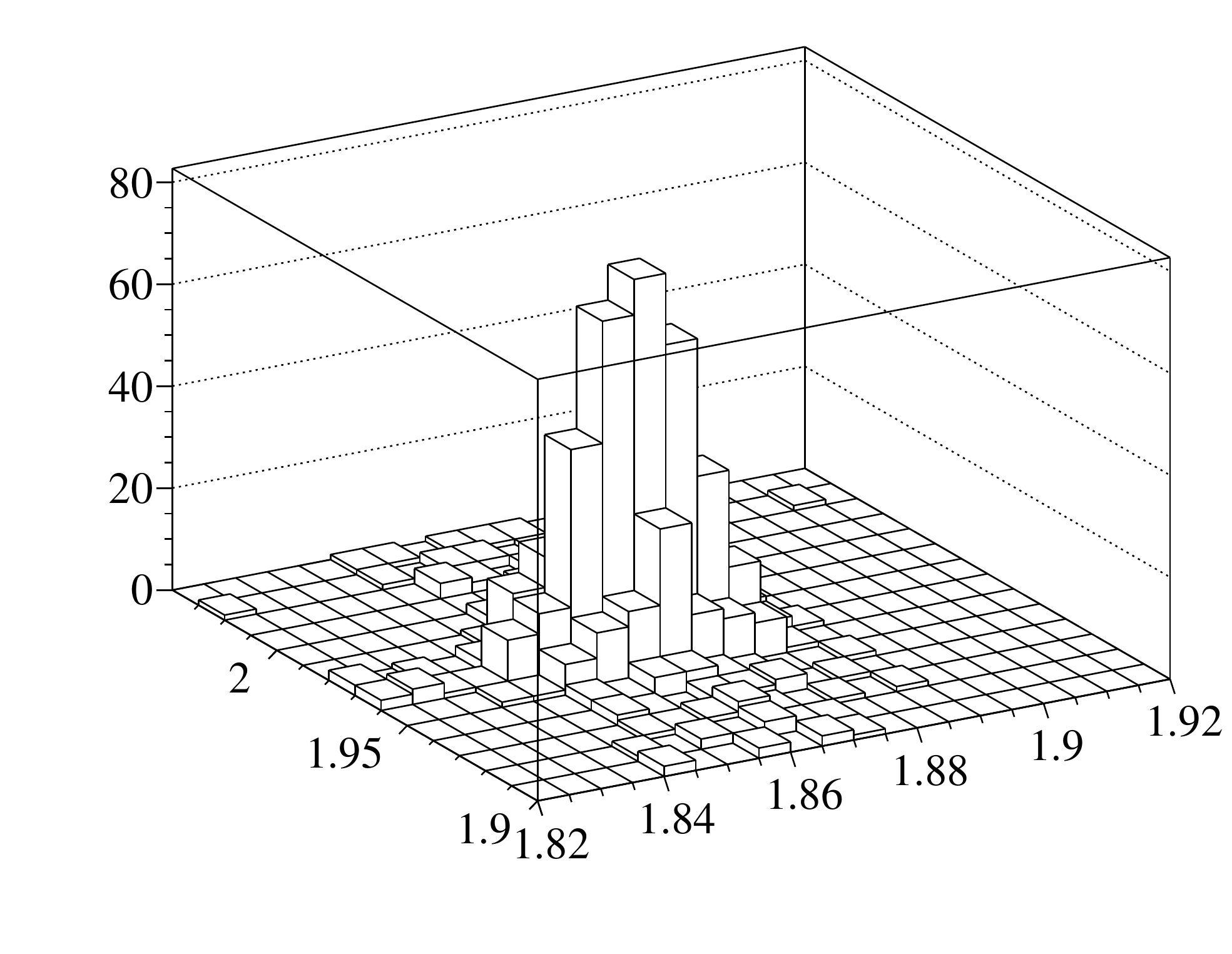}
    }
    \put(75,0){
      \includegraphics*[width=75mm,height=60mm,%
      ]{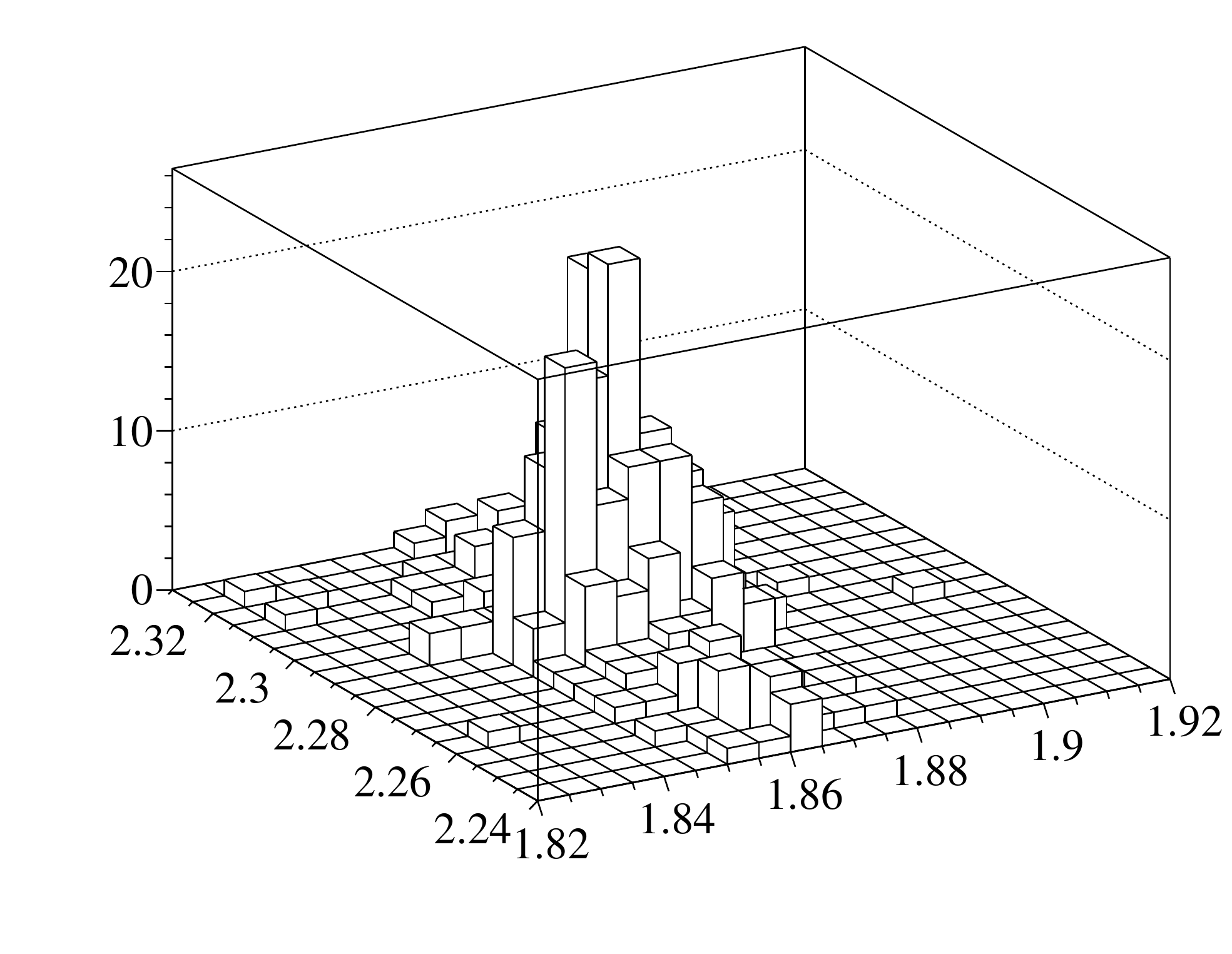}
    }
    \put(7.5,74.5)  { 
      \begin{rotate}{-32} \small 
        $m_{\mathrm{K}\Ppi}~\left[\mathrm{GeV}/c^2\right]$
      \end{rotate}
    }
    \put(82.5,74.5)  { 
      \begin{rotate}{-32} \small 
        $m_{\mathrm{K}\Ppi\Ppi}~\left[\mathrm{GeV}/c^2\right]$
      \end{rotate}
    }
    \put(7.5,14.5)  { 
      \begin{rotate}{-32} \small 
        $m_{\mathrm{KK}\Ppi}~\left[\mathrm{GeV}/c^2\right]$
      \end{rotate}
    }
    \put(82.5,14.5)  { 
      \begin{rotate}{-32} \small 
        $m_{\mathrm{pK}\Ppi}~\left[\mathrm{GeV}/c^2\right]$
      \end{rotate}
    }
    \put(46,64)  { 
      \begin{rotate}{12} \small 
        $m_{\mathrm{K}\Ppi}~~~\left[\mathrm{GeV}/c^2\right]$
      \end{rotate}
    }
    \put(121,64)  { 
      \begin{rotate}{12} \small 
        $m_{\mathrm{K}\Ppi}~~~\left[\mathrm{GeV}/c^2\right]$
      \end{rotate}
    }
    \put(46,4)  { 
      \begin{rotate}{12} \small 
        $m_{\mathrm{K}\Ppi}~~~\left[\mathrm{GeV}/c^2\right]$
      \end{rotate}
    }
    \put(121,4)  { 
      \begin{rotate}{12} \small 
        $m_{\mathrm{K}\Ppi}~~~\left[\mathrm{GeV}/c^2\right]$
      \end{rotate}
    }
    \put( 1 , 88 )  { \small 
      \begin{sideways}%
        Candidates 
      \end{sideways}%
    }
    \put(76 , 88 )  { \small 
      \begin{sideways}%
        Candidates 
      \end{sideways}%
    }
    \put( 1 , 28 )  { \small 
      \begin{sideways}%
        Candidates 
      \end{sideways}%
    }
    \put(76 , 28 )  { \small 
      \begin{sideways}%
        Candidates 
      \end{sideways}%
    }
    \put(65  , 114){a)}
    \put(140 , 114){b)}
    \put(65  ,  54){c)}
    \put(140 ,  54){d)}
    \put(10,115){ \small 
      LHCb
    }
    \put(10,55){ \small 
      LHCb
    }
    \put(85,115){ \small 
      LHCb
    }
    \put(85,55){ \small 
      LHCb
    }
  \end{picture}
  \caption { \small 
    Invariant 
    mass distributions for 
    $\Dz{}\bar{\mathrm{C}}$ candidates:
    a)~\DzDzb,
    b)~\DzDpb,
    c)~\DzDsb and 
    d)~\DzLcb.
  }
  \label{fig:ccbar_trg_pt3_2d1}
\end{figure}

\begin{figure}[htb]
  \setlength{\unitlength}{1mm}
  \centering
  \begin{picture}(150,120)
    \put( 0,60){
      \includegraphics*[width=75mm,height=60mm,%
      ]{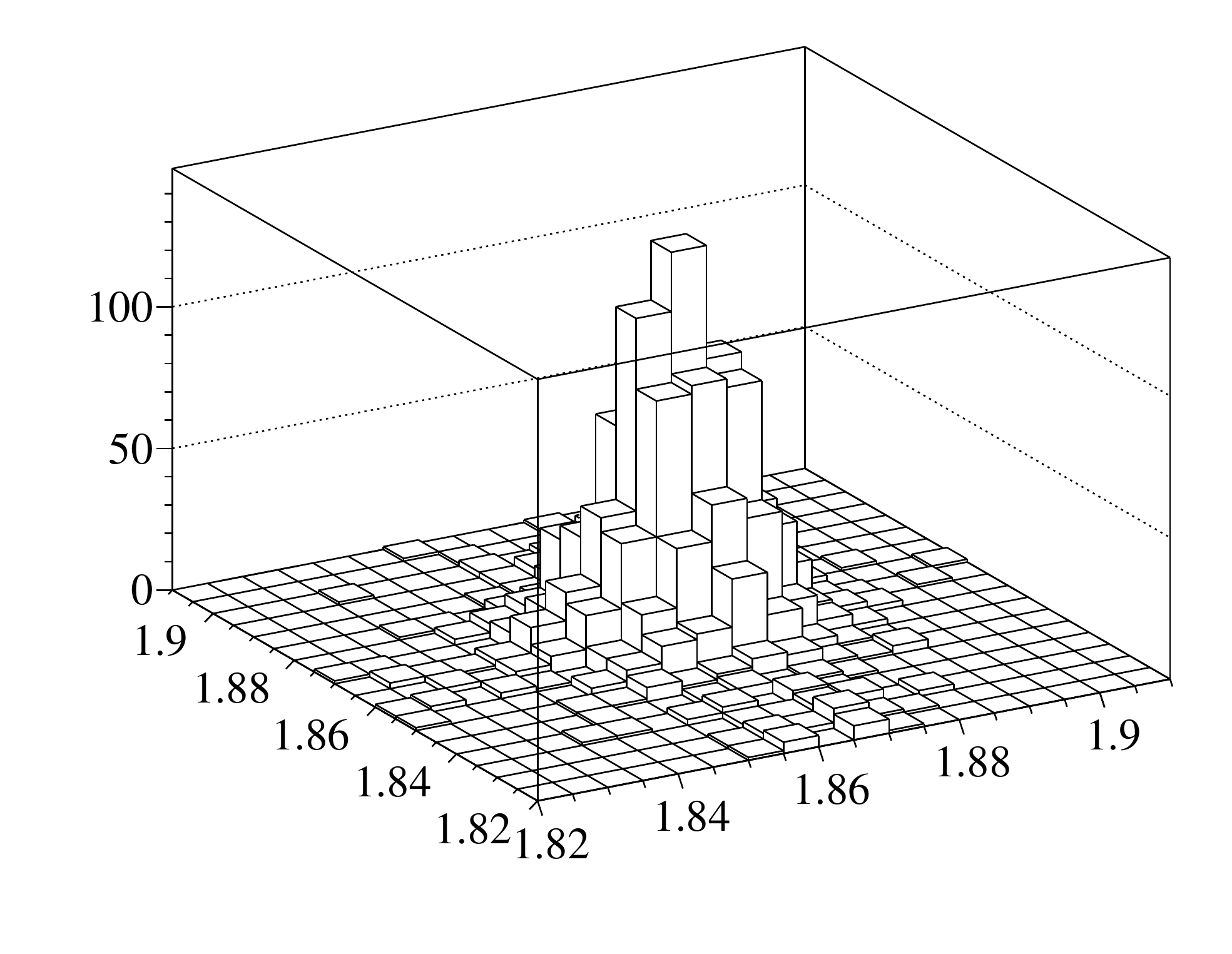}
    }
    \put(75,60){
      \includegraphics*[width=75mm,height=60mm,%
      ]{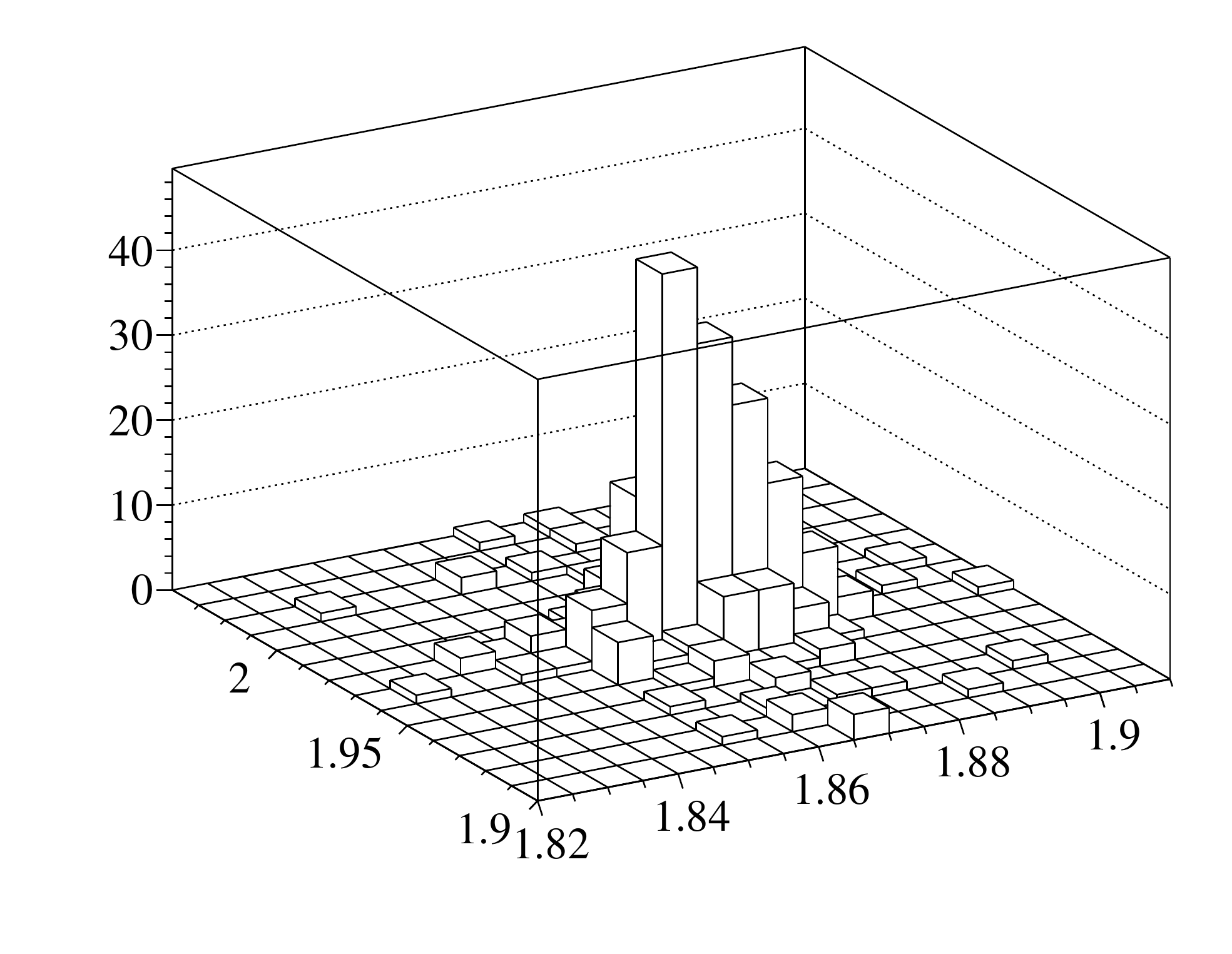}
    }
    \put(0,00){
      \includegraphics*[width=75mm,height=60mm,%
      ]{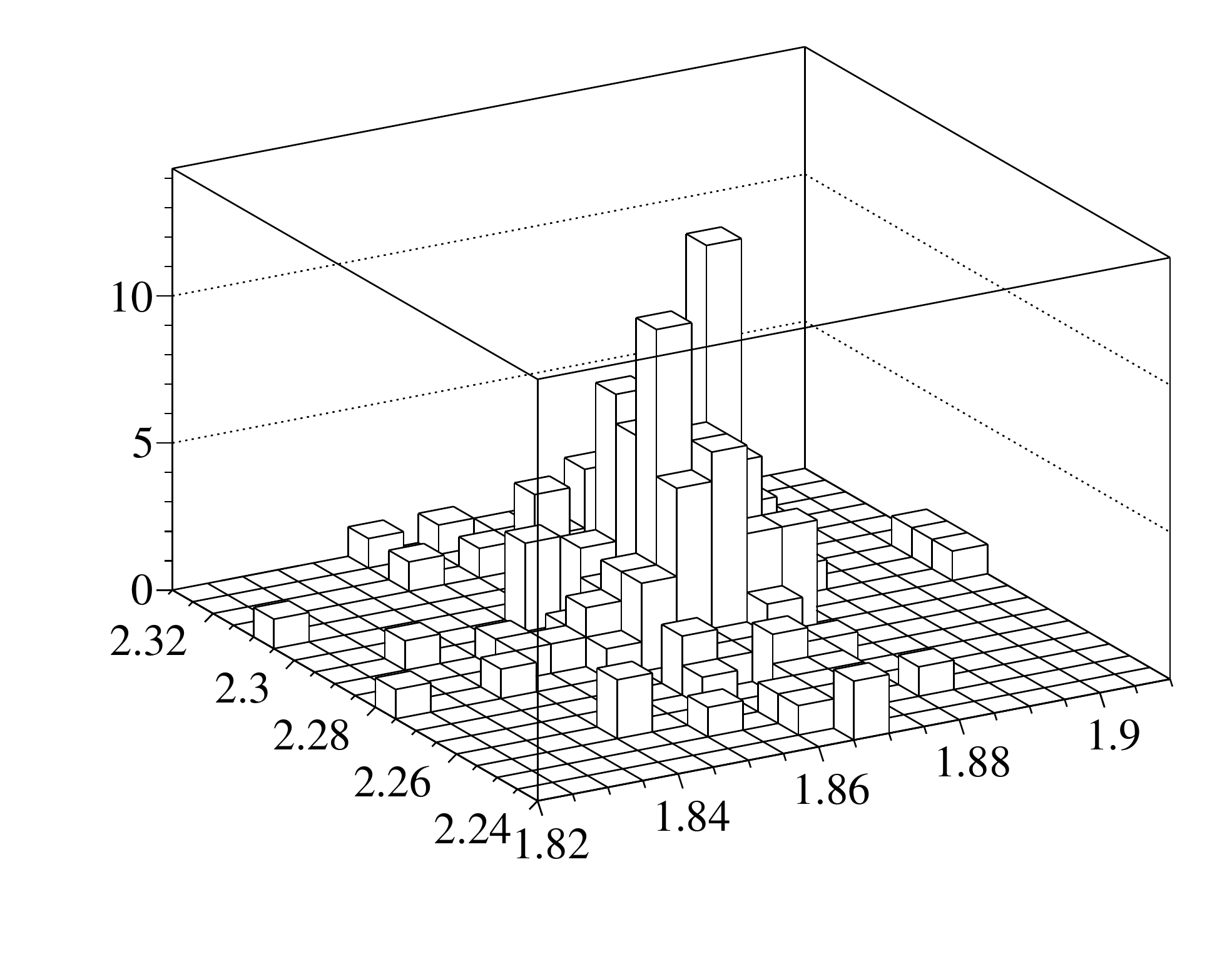}
    }
    \put(7.5,74.5)  { 
      \begin{rotate}{-32} \small 
        $m_{\mathrm{K}\Ppi{}\Ppi}~\left[\mathrm{GeV}/c^2\right]$
      \end{rotate}
    }
    \put(82.5,74.5)  { 
      \begin{rotate}{-32} \small 
        $m_{\mathrm{KK}\Ppi}~\left[\mathrm{GeV}/c^2\right]$
      \end{rotate}
    }
    \put(7.5,14.5)  { 
      \begin{rotate}{-32} \small 
        $m_{\mathrm{pK}\Ppi}~\left[\mathrm{GeV}/c^2\right]$
      \end{rotate}
    }
    \put(46,64)  { 
      \begin{rotate}{12} \small 
        $m_{\mathrm{K}\Ppi{}\Ppi}~\left[\mathrm{GeV}/c^2\right]$
      \end{rotate}
    }
    \put(121,64)  { 
      \begin{rotate}{12} \small 
        $m_{\mathrm{K}\Ppi{}\Ppi}~\left[\mathrm{GeV}/c^2\right]$
      \end{rotate}
    }
    \put(46,4)  { 
      \begin{rotate}{12} \small 
        $m_{\mathrm{K}\Ppi{}\Ppi}~\left[\mathrm{GeV}/c^2\right]$
      \end{rotate}
    }
    \put(1 , 88 )  { \small 
      \begin{sideways}%
        Candidates 
      \end{sideways}%
    }
    \put(76 ,88)  { \small 
      \begin{sideways}%
        Candidates 
      \end{sideways}%
    }
    \put(1 , 28 )  { \small 
      \begin{sideways}%
        Candidates 
      \end{sideways}%
    }
    \put(65  , 114){a)}
    \put(140 , 114){b)}
    \put(65  ,  54){c)}
    \put(10,115){ \small 
      LHCb
    }
    \put(10,55){ \small 
      LHCb
    }
    \put(85,115){ \small 
      LHCb
    }
  \end{picture}
  \caption { \small 
    Invariant 
    mass distributions for \CCbar candidates:
    a)~\DpDpb,
    b)~\DpDsb and 
    c)~\DpLcb.
  }
  \label{fig:ccbar_trg_pt3_2d2}
\end{figure}

The mass distributions for all pairs after these criteria are applied are shown 
in Figs.~\ref{fig:psid_trg_pt3_2d} to~\ref{fig:ccbar_trg_pt3_2d2} for channels
with sufficiently large data samples.

\clearpage 
\section{Signal determination}
\label{sec:extractionl}
The event yields are determined using unbinned extended maximum likelihood
fits to the mass distributions of the di-charm sample. 
The fit model is based on the probability density functions (PDFs) for single open or 
hidden charm production described in Section~\ref{sec:EventSelection}.  
These basic PDFs are used to build the components of 
the two dimensional mass fit. Let $i$ and $j$ denote the two resonance species. 
The reconstructed signal samples consist
of the following components:
\begin{itemize}
\item Di-charm signal. This is modelled by a product
  PDF of the individual signal components for the first and the second particle.
\item Combinatorial background. This is modelled by a product
  PDF of the individual background components $i$ and
  $j$ denoted by $B_i(m_{i})$ and $B_j(m_{j})$.
\item Single production of component $i$ together with combinatorial
  background for component $j$. This is modelled by a product PDF of
  the signal component $i$ denoted $S_i(m_i)$ and the background component $j$ denoted 
  $B_{j}(m_j)$.
\item Single production of component $j$ together with combinatorial
  background for component $i$. This is modelled by a product PDF of
  the signal component $j$ denoted $S_j(m_j)$ and the background component $i$
  denoted $B_{i}(m_i)$.
\end{itemize}
The total PDF is then
\begin{eqnarray}
 F(m_i,m_j) & \propto & 
  N^{S_i \times S_j}  \times S_{i} ( m_i ) S_{j} ( m_j )  +
  N^{S_i \times B_j}  \times S_{i} ( m_i ) B_{j} ( m_j )  \nonumber \\
   & + &  N^{B_i \times S_j}  \times B_{i} ( m_i ) S_{j} ( m_j )  +
        N^{B_i \times B_j}  \times B_{\mathrm{i}} ( m_i ) B_{j} ( m_j ), 
\end{eqnarray}
where 
$N^{S_i \times S_j}$,
$N^{S_i\times B_j}$,
$N^{B_i \times S_j}$ and  
$N^{B_i \times B_j}$ are 
the yields of the four components described above.
The correctness of the fitting procedure is evaluated 
in simulation studies. As discussed in Section~\ref{sec:EventSelection} both the contribution of
pile-up background and $\mathrm{b}$-hadron decays is small and can be neglected. 
The goodness of fit is 
found to be acceptable 
using the distance to the nearest neighbour method
described in Refs.~\cite{Bickel:1983,Williams:2010vh}.

As a cross-check of the results, the signal yields have been determined from 
the single charm hadron mass spectra using the technique described 
in Ref.~\cite{Aaij:2011yc}.
In this approach, for each pair of charm species 
the invariant mass distributions 
of the first charm candidate 
are fitted to obtain the yield in bins of the invariant mass of 
the second candidate and vice versa.
This technique gives signal yields consistent 
within 10\% of the statistical uncertainty 
and also allows the statistical
significance of the result
to be easily evaluated.
This exceeds five standard deviations
for most of the modes considered. 
The signal yields for \psiC, \CC~and \CCbar~events are presented in
Tables~\ref{tab:raw_psid}~and~\ref{tab:raw_d2} together with the estimate of
the goodness of fit. 
\begin{table}[htb]
  \centering
  \caption{ \small
    Yields of \psiC~events, 
    $S$, 
    statistical 
    significance of the signals, 
    $S_{\sigma}$, determined
    from fits based on the technique described in Ref.~\cite{Aaij:2011yc},
    and goodness-of-fit characteristic ($\chi^2$~probability),~$P$.
    When no significance is quoted, it is in excess of $8\sigma$.
   } \label{tab:raw_psid}
  \vspace*{3mm}
  \setlength{\extrarowheight}{1.5pt}
  \begin{tabular*}{0.75\textwidth}{@{\hspace{5mm}}c@{\extracolsep{\fill}}ccc@{\hspace{5mm}}}
    Mode
    & $S$
    & $S_{\sigma}$
    & $P~\left[\%\right]$ 
    \\
    \hline
    \psiDz  
    & $ 4875 \pm  86 $ 
    &  
    & 59            
    \\
    \psiDp
    & $ 3323 \pm  71 $ 
    &  
    & 26            
    \\
    \psiDs
    &  $ \phantom{0}328 \pm  22 $ 
    &  
    & 65            
    \\
    \psiLc
    &  $ \phantom{0}116 \pm  14 $ 
    & $7.3\sigma$   
    & 98            
    \\
  \end{tabular*}   
\end{table}

\begin{table}[htb]
  \centering
  \caption{ \small
    Yields of \CC~and \CCbar~events, 
    $S$, 
    statistical 
    significance of the signals, 
    $S_{\sigma}$, determined
    from fits based on the technique, described in Ref.~\cite{Aaij:2011yc},
    and goodness-of-fit characteristic,~$P$.
    When no significance is quoted, it is in excess of $8\sigma$.
   } \label{tab:raw_d2}
  \vspace*{3mm}
  \begin{tabular*}{0.75\textwidth}{@{\hspace{5mm}}c@{\extracolsep{\fill}}ccc@{\hspace{5mm}}}
    Mode
    & $S$
    & $S_{\sigma}$
    & $P~\left[\%\right]$ 
    \\
    \hline
    \DzDz
    & $  \phantom{0}1087 \pm 37\phantom{0} $  
    &  
    &  4.5            
    \\
    \DzDzb
    & $ 10080 \pm 105 $ 
    &  
    & 33                 
    \\
    \hline 
    \DzDp
    & $  \phantom{0}1177 \pm  39\phantom{0} $ 
    &  
    & 24              
    \\
    \DzDpb
    & $ 11224 \pm 112 $ 
    &  
    & 36              
    \\
    \hline 
    \DzDs
    & $ \phantom{00}111 \pm  12\phantom{0} $  
    & $8\sigma$       
    & 10              
    \\
    \DzDsb
    & $  \phantom{00}859 \pm 31\phantom{0} $ 
    &  
    & 13              
    \\
    \hline 
    \DzLc
    & $  \phantom{000}41 \pm  8\phantom{00} $ 
    & $5\sigma$       
    &  9              
    \\
    \DzLcb
    & $  \phantom{00}308 \pm 19\phantom{0} $ 
    &  
    & 35              
    \\
    \hline 
    \DpDp
    & $  \phantom{00}249 \pm 19\phantom{0} $ 
    &  
    & 15              
    \\
    \DpDpb
    & $  \phantom{0}3236 \pm  61\phantom{0} $ 
    &  
    & 67              
    \\
    \hline 
    \DpDs   
    & $  \phantom{000}52 \pm  9\phantom{00} $ 
    & $5\sigma$       
    & 54              
    \\
    \DpDsb   
    & $  \phantom{00}419 \pm 22\phantom{0} $ 
    & 
    & 59              
    \\
    \hline 
    \DpLc
    & $  \phantom{000}21 \pm  5\phantom{00} $ 
    & $2.5\sigma$      
    &  36              
    \\
    \DpLcb
    & $  \phantom{00}137 \pm 14\phantom{0} $ 
    & $8\sigma$       
    &  7              
    \\
  \end{tabular*}   
\end{table}

\clearpage 
\section{Efficiency correction}
\label{sec:effcorrection}  
The yields are corrected for the detection efficiency to obtain
the measured cross-sections. The efficiency for \psiC,
\CCbar~and \CC~events $\varepsilon^{\mathrm{tot}}$ 
is computed for each signal event and 
is decomposed into three factors
\begin{equation}
  \varepsilon^{\mathrm{tot}}
  =
  \varepsilon^{\mathrm{reco}} \times 
  \varepsilon^{\mathrm{ID}}  \times 
  \varepsilon^{\mathrm{trg}}, 
\label{eq:total_eff}
\end{equation}
where $\varepsilon^{\mathrm{reco}}$ is the efficiency for acceptance, reconstruction and selection, 
$\varepsilon^{\mathrm{ID}}$ is the efficiency for particle identification and 
$\varepsilon^{\mathrm{trg}}$ is the trigger efficiency. 
The first term in Eq.~\eqref{eq:total_eff}, $\varepsilon^{\mathrm{reco}}$  
is factorized into the
product of 
efficiencies 
for the first
and second charm particle and a correction factor
\begin{equation}
  \varepsilon^{\mathrm{reco}}=
   \varepsilon^{\mathrm{reco}}_{1} \times 
   \varepsilon^{\mathrm{reco}}_{2} \times 
   \xi^{\mathrm{trk}},
  \label{eq:eff_reco}
 \end{equation}
where the efficiencies $\varepsilon^{\mathrm{reco}}_{(1,2)}$ are
evaluated using the simulation, and 
the correction factor\footnote{This is the
  product of the individual corrections for each track.}  
$\xi^{\mathrm{trk}}$ 
is determined from the \jpsi~data
using a {\it tag-and-probe} method and 
accounts for relative differences in the track reconstruction 
efficiency between data and simulation.

The efficiency $\varepsilon^{\mathrm{reco}}_{i}$ is determined using
the simulation in bins of rapidity $y$ and transverse momentum
$p^{\mathrm{T}}$ of the charm hadron. In the case of the \jpsi~meson,
the effect of the unknown polarization on the efficiency is
accounted for by binning in $| \cos \theta^*_{\jpsi}|$, 
where $\theta^{*}_{\jpsi}$ is the angle 
between the \mup~momentum in the \jpsi~centre-of-mass frame and 
the \jpsi~flight direction in the laboratory frame.  

The efficiency for hadron identification as a function of momentum and
pseudorapidity is determined from the data using
samples of $\mathrm{D}^{*+} \to \left( \Dz \to \mathrm{K}^- \pi^+\right) \pi^+$, 
and $\Lambda \to \mathrm{p} \pi^-$~\cite{Powell:2011zz,LHCb-PROC-2011-008}.
The efficiency for dimuon identification, 
$\varepsilon^{\mathrm{ID}}_{\jpsi}$ is obtained from the analysis of 
the $\jpsi\to\mumu$~sample as a function of 
transverse momentum and rapidity of the~\jpsi. 

For the \psiC~sample the \jpsi~particle is required to trigger the event
whilst for the \CC~and \CCbar case either of the two charm mesons
could trigger the event. The trigger efficiency for the di-charm system in the two
cases is thus
\begin{subequations}
\label{eq:eff_trg}
\begin{eqnarray}
\varepsilon^{\mathrm{trg}}_{\psiC} & =  & \varepsilon^{\mathrm{trg}}_{\jpsi} \label{eq:eff_trg_psi} \\
\varepsilon^{\mathrm{trg}}_{\CCbar,\CC}   & =  & 
1 - ( 1 - \varepsilon^{\mathrm{trg}}_{\mathrm{C}_1})\times
( 1 - \varepsilon^{\mathrm{trg}}_{\mathrm{C}_2}). \label{eq:eff_trg_cc} 
\end{eqnarray}
\end{subequations}
In both cases the trigger efficiency for a single charm hadron 
$\varepsilon^{\mathrm{trg}}_{\jpsi}$ or 
$\varepsilon^{\mathrm{trg}}_{\mathrm{C}}$ 
is determined  directly from the data using the inclusive prompt charm  sample  as a 
function of $y$ and $p^{\mathrm{T}}$. This is done using a method that 
exploits the fact that events with prompt charm hadrons can be triggered either 
by the decay products of the charm hadron, or by the rest of the
event~\cite{Aaij:2011yc,Aaij:2010nx}. The overlap between the two cases allows 
the trigger efficiency to be estimated.

As discussed in Sect.~\ref{sec:Detector},  global event cuts are
applied in the trigger on the sub-detector hit multiplicites to reject complex events.
The efficiency of these cuts $\varepsilon^{\mathrm{GEC}}$ is studied using the 
distributions of  hit multiplicity  
after background subtraction.
These distributions have been 
extrapolated from the regions unaffected by the cuts into the potentially affected regions and 
compared with the observed distributions in order to determine $\varepsilon^{\mathrm{GEC}}$.

The efficiency-corrected signal yield $N^{\mathrm{corr}}$ is determined using the
\sPlot~\cite{Pivk:2004ty} technique. 
Each candidate is given a weight for 
it  to be signal, $\omega_i$, based on the result of the fit to the
mass distributions described before. 
The weight is then divided by the total event  
efficiency and summed to give the efficiency-corrected yield
\begin{equation}
N^{\mathrm{corr}}= \displaystyle\sum\limits_{i}
\frac{\omega_{i}}{\varepsilon^{\mathrm{tot}}_i }. 
\label{eq:ncorr}
\end{equation}

In the case of the $\Dz{}\mathrm{C}$ and $\Dz{}\overline{\mathrm{C}}$ final
states the corresponding yields  have been corrected to take into account 
the double Cabibbo-suppressed decay (DCS)~mode 
$\Dz{}\to \mathrm{K}^+ \pi^-$, which mixes the $\Dz{}\mathrm{C}$ and 
$\Dz{}\overline{\mathrm{C}}$~reconstructed final states
\begin{equation}
       \begin{pmatrix}
        N^{\prime}_{\Dz{}\mathrm{C}}     \\ 
        N^{\prime}_{\Dz{}\bar{\mathrm{C}}} 
       \end{pmatrix}
        = 
       \frac{1}{\sqrt{1 - r^2}}
       \begin{pmatrix}
          1   & - r \\ 
         -r   &   1 
       \end{pmatrix} \times 
        \begin{pmatrix}
        N^{\mathrm{corr}}_{\Dz{}\mathrm{C}}     \\ 
        N^{\mathrm{corr}}_{\Dz{}\bar{\mathrm{C}}} 
        \end{pmatrix},  
\end{equation}
where $r$ is
$r^{\mathrm{DCS}} = \dfrac { \Gamma \left( \Dz{}\to \mathrm{K}^+ \pi^- \right) }   
        { \Gamma \left( \Dz{}\to \mathrm{K}^- \pi^+ \right) } 
         = \left( 3.80 \pm 0.18\right)\times 10^{-3}$~\cite{Nakamura:2010zzi}.
This value of $r^{\mathrm{DCS}}$~accounts also for the effect 
of 
\Dz-\Dzb~mixing.
For the \DzDz and \DzDzb cases the value of $r = 2 r^{\mathrm{DCS}}$ is used.

%
\section{Systematic uncertainties}
\label{sec:syst}
The sources of systematic uncertainty that enter 
into the cross-section determination in addition to those related to
the knowledge of branching ratios and luminosity are discussed
below. The dominant source of systematic uncertainty arises from possible 
differences in the
track reconstruction efficiency between data and simulation which are
not accounted for in the per-event efficiency. This includes the
knowledge of the hadronic interaction length of the detector which 
results in an uncertainty
of 2\%~per final state hadron~\cite{Aaij:2010nx}. 
An additional
uncertainty is due to the statistical uncertainty on the determination 
of the per-event efficiency 
due to the finite size of the simulation and calibration
samples. This is estimated by varying the obtained efficiencies within
their corresponding uncertainties.
The unknown polarization of \jpsi~mesons affects the 
acceptance, reconstruction and selection efficiency
$\varepsilon^{\mathrm{reco}}_{\jpsi}$~\cite{Aaij:2011jh}. 
In this analysis the effect is reduced by explicitly taking into
account the dependence of $\varepsilon^{\mathrm{reco}}_{\jpsi}$ 
on $|\cos\theta^*_{\jpsi}|$ in the efficiency determination. 
The remaining dependence results in a systematic uncertainty 
of 3\% for channels containing a \jpsi.

Additional uncertainties are due to differences 
between data and simulation, 
uncertainty on the global event cuts, 
knowledge of the branching fractions of charm hadrons, $\BR_i$. 
Uncertainties due to the parameterization of
the signal and background components are found to be negligible. 

The absolute luminosity scale was measured at specific periods during 
the data taking, using both van~der~Meer scans~\cite{vandermeer}
where colliding  beams are moved transversely  across each other 
to determine the beam profile, 
and a beam-gas imaging method~\cite{Ferroluzzi:2005em,LHCb-PAPER-2011-015}.
For the latter, reconstructed  
beam-gas interaction vertices near the beam crossing point determine 
the beam profile. 
The knowledge of the  absolute luminosity scale is used to calibrate the number of 
tracks in the silicon-strip vertex detector, which is found to be stable 
throughout the data-taking period  and can therefore  be used to monitor 
the instantaneous luminosity 
of the entire data sample.
The dataset for this analysis corresponds to an integrated 
luminosity of $355\pm13~\mathrm{pb}^{-1}$.

The sources of systematic uncertainty on the \psiC~production
cross-section measurements are summarized in Table~\ref{tab:syst_psiC}
and those for open charm in Tables~\ref{tab:syst_d2} 
and~\ref{tab:syst_d3}.
The total systematic uncertainties have been evaluated 
taking correlations into account where appropriate.
\begin{table}[htb]
  \centering
  \caption{ \small
    Relative systematic uncertainties  ($\%$)
    for the \psiC~cross-sections.
  } \label{tab:syst_psiC}
  \vspace*{3mm}
  \setlength{\extrarowheight}{1.5pt}
  \begin{tabular*}{0.90\textwidth}{@{\hspace{5mm}}lc@{\extracolsep{\fill}}ccccc@{\hspace{5mm}}}
    \multicolumn{2}{c}{Source}
    & \psiDz 
    & \psiDp
    & \psiDs
    & \psiLc
    \\
    \hline 
    \jpsi~reconstruction 
    & $\varepsilon^{\mathrm{reco}}_1$
    & $1.3$ & $1.3$ & $1.3$ & $1.3$
    \\
    $\mathrm{C}$~reconstruction 
    & $\varepsilon^{\mathrm{reco}}_2$ 
    & $0.7$
    & $0.8$
    & $1.7$
    & $3.3$
    \\
    Muon ID 
    & $\varepsilon^{\mathrm{ID}}_{\jpsi}$
    & $1.1$ & $1.1$ & $1.1$ & $1.1$ 
    \\
    Hadron ID 
    & $\varepsilon^{\mathrm{ID}}_{\mathrm{had}}$
    & $1.1$
    & $1.9$
    & $1.1$
    & $1.5$
    \\
    Tracking 
    & $\xi^{\mathrm{trk}}$
    & $4.9$  
    & $7.0$  
    & $7.0$  
    & $7.0$  
    \\
    Trigger 
    & $\varepsilon^{\mathrm{trg}}_{\psiC}$
    & $3.0$ & $3.0$ & $3.0$ & $3.0$
    \\
    \jpsi~polarization 
    & $\varepsilon^{\mathrm{reco}}_{\jpsi}$
    & $3.0$ & $3.0$ & $3.0$ & $3.0$
    \\
    Global event cuts
    & $\varepsilon^{\mathrm{GEC}}$
    & $0.7$ & $0.7$ & $0.7$ & $0.7$
    \\
    Luminosity
    & $\mathcal{L}$
    & $3.7$ & $3.7$ & $3.7$ & $3.7$
    \\
    $\BR (\jpsi\to\mumu)$
    & $\BR_1$
    & $1.0$ & $1.0$ & $1.0$ & $1.0$
    \\
    $\mathrm{C}$ branching fractions
    & $\BR_2$
    & 1.3 
    & 4.3 
    & 6.0 
    & 26
    \\
    \hline 
    Total 
    & 
    &  8
    & 10 
    & 11
    & 28
    \\
  \end{tabular*}   
\end{table}

\begin{table}[htb]
  \centering
  \caption{ \small
    Relative systematic uncertainties ($\%$) for the $\Dz\mathrm{C}$~cross-sections.
    The uncertainties for \CC~and \CCbar~are equal. 
  } \label{tab:syst_d2}
  \vspace*{3mm}
  \setlength{\extrarowheight}{1.5pt}
  \begin{tabular*}{0.9\textwidth}{@{\hspace{5mm}}lc@{\extracolsep{\fill}}ccccc@{\hspace{5mm}}}
    \multicolumn{2}{c}{Source}
    & \DzDz
    & \DzDp
    & \DzDs
    & \DzLc
    \\
    \hline 
    $\Dz\mathrm{C}$~reconstruction 
    & $\varepsilon^{\mathrm{reco}}_{1}\times\varepsilon^{\mathrm{reco}}_{2}$ 
    & $1.4$
    & $1.4$
    & $2.3$
    & $3.6$
    \\
    Hadron ID 
    & $\varepsilon^{\mathrm{ID}}_{\mathrm{had}}$
    & $1.2$
    & $1.8$
    & $1.6$
    & $2.4$
    \\
    Tracking
    & $\xi^{\mathrm{trk}}$
    & $8.5$ 
    & $10.7\phantom{0}$  
    & $10.6\phantom{0}$  
    & $10.6\phantom{0}$  
    \\
    Trigger 
    & $\varepsilon^{\mathrm{trg}}_{\CC,\CCbar}$
    & 1.8
    & 2.5 
    & 3.9
    & 5.2 
    \\
    Global event cuts
    & $\varepsilon^{\mathrm{GEC}}$
    & $1.0$ & $1.0$ & $1.0$ & $1.0$
    \\
    Luminosity
    & $\mathcal{L}$
    & $3.7$ & $3.7$ & $3.7$ & $3.7$
    \\
    $\BR( \Dz \to \mathrm{K}^- \pi^+ )$
    & $\BR_1$
    & $1.3$ & $1.3$ & $1.3$ & $1.3$
    \\
    $\mathrm{C}$ branching fractions
    & $\BR_2$
    & 1.3 
    & 4.3 
    & 6.0 
    & 26
    \\
    \hline 
    Total 
    & 
    & 10
    & 12 
    & 14
    & 30
    \\
  \end{tabular*}   
\end{table}

\begin{table}[htb]
  \centering
  \caption{ \small
    Relative systematic uncertainties ($\%$) for the $\Dp\mathrm{C}$~cross-sections.
    The  uncertainties for the \CC~and \CCbar~are equal.
  } \label{tab:syst_d3}
  \vspace*{3mm}
  \setlength{\extrarowheight}{1.5pt}
  \begin{tabular*}{0.9\textwidth}{@{\hspace{5mm}}lc@{\extracolsep{\fill}}cccc@{\hspace{5mm}}}
    \multicolumn{2}{c}{Source}
    & \DpDp
    & \DpDs
    & \DpLc
    \\
    \hline 
    $\Dp\mathrm{C}$~reconstruction 
    & $\varepsilon^{\mathrm{reco}}_1\times\varepsilon^{\mathrm{reco}}_2$ 
    & $1.4$
    & $2.2$
    & $4.0$
    \\
    Hadron ID 
    & $\varepsilon^{\mathrm{ID}}_{\mathrm{had}}$
    & $2.3$
    & $2.4$
    & $3.0$
    \\
    Tracking 
    & $\xi^{\mathrm{trk}}$
    & $12.8\phantom{0}$ 
    & $12.8\phantom{0}$ 
    & $12.8\phantom{0}$ 
    \\
    Trigger 
    & $\varepsilon^{\mathrm{trg}}_{\CC,\CCbar}$
    & 3.7
    & 5.8 
    & 5.0
    \\
    Global event cuts
    & $\varepsilon^{\mathrm{GEC}}$
    & $1.0$ & $1.0$ & $1.0$ 
    \\
    Luminosity
    & $\mathcal{L}$
    & $3.7$ & $3.7$ & $3.7$ 
    \\
    $\BR( \Dp \to \mathrm{K}^- \pi^+ \pi^+)$
    & $\BR_1$
    & $4.3$ & $4.3$ & $4.3$ 
    \\
    $\mathrm{C}$ branching fractions
    & $\BR_2$
    & 4.3 
    & 6.0 
    & 26
    \\
    \hline 
    Total 
    & 
    & 17 
    & 17
    & 31
    \\
  \end{tabular*}   
\end{table}

\clearpage 
%

\cleardoublepage 
\section{Results}
The model-independent cross-section for double charm production 
in the fiducial range is computed as
\begin{equation}
  \sigma = 
  \dfrac{ N^{\mathrm{corr}}}
        {\mathcal{L}\times
	  \BR_1 \times 
	  \BR_2 \times 
          \varepsilon^{\mathrm{GEC}}
        }, 
        \label{eq:xsect}
\end{equation}
where 
$\mathcal{L}$ 
is the integrated luminosity obtained as described in Sect.~\ref{sec:syst},
$\BR_{(1,2)}$ stand for the corresponding branching ratios,
$\varepsilon^{\mathrm{GEC}}$ is the efficiency of the 
global event cuts, and  $N^{\mathrm{corr}}$~is 
the efficiency-corrected event yield, calculated according 
to~Eq.~\eqref{eq:ncorr}. 
The branching ratios 
used for these calculations are taken 
from Ref.~\cite{Nakamura:2010zzi}. 
We reiterate that the inclusion of charge conjugate 
processes is implied, so that e.g., $\sigma_{\psiC}$ is the sum of 
production cross-sections for $\jpsi\mathrm{C}$ and 
$\jpsi\overline{\mathrm{C}}$. 

\begin{figure}[htb]
  \setlength{\unitlength}{1mm}
  \centering
  \begin{picture}(120,160)
    \put(0,0){
      \includegraphics*[height=160mm,width=130mm,%
      ]{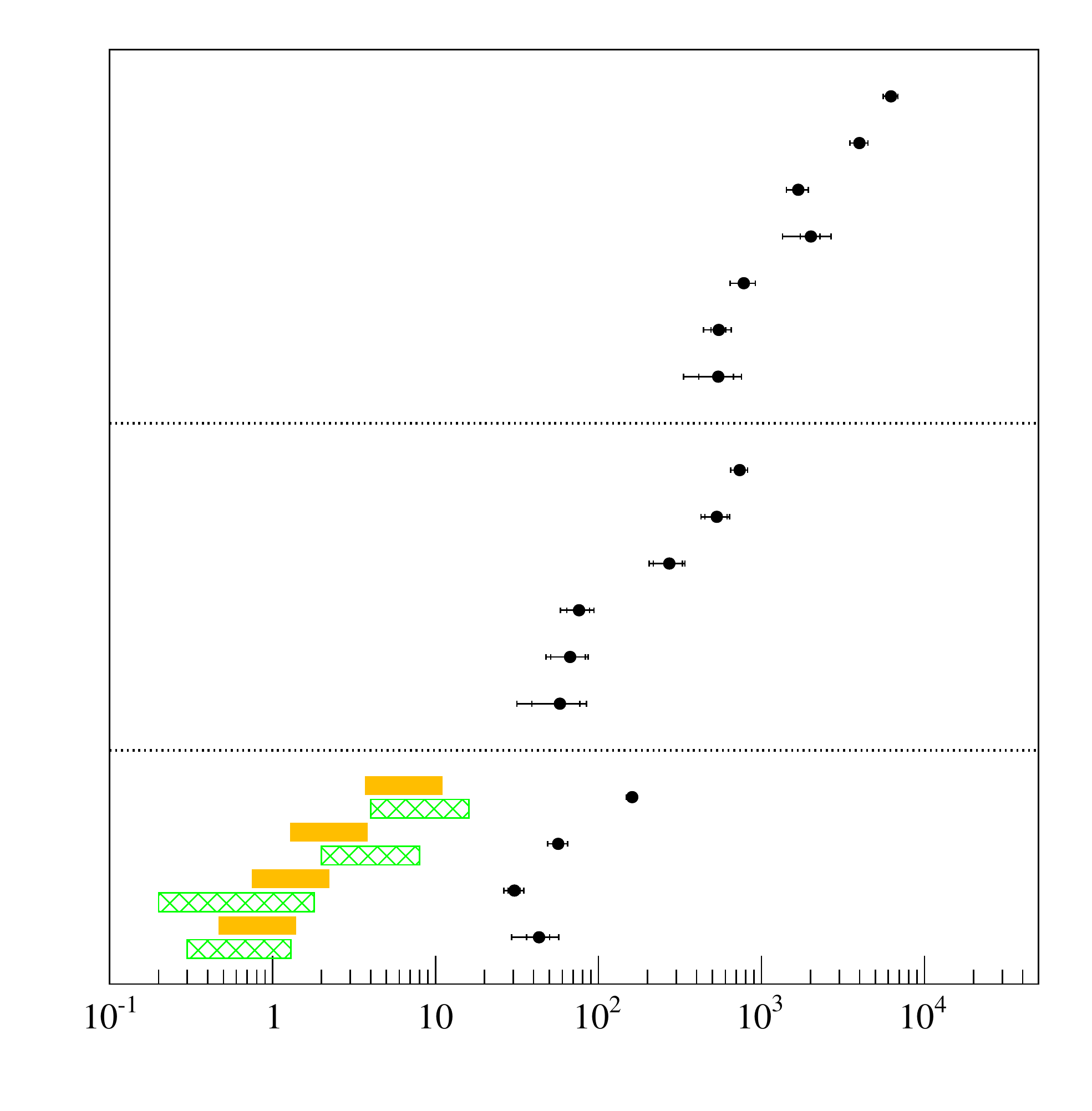} 
    }
    \put(53,143.5) {  \DzDzb  } 
    \put(53,137.0) {  \DzDpb  } 
    \put(53,130.5) {  \DzDsb  } 
    \put(53,124.0) {  \DzLcb  } 
    \put(53,117.5) {  \DpDpb  } 
    \put(53,111.0) {  \DpDsb  } 
    \put(53,104.5) {  \DpLcb  } 
    \put(18, 89.5) {  \DzDz  } 
    \put(18, 83.0) {  \DzDp  } 
    \put(18, 76.5) {  \DzDs  } 
    \put(18, 70.0) {  \DpDp  } 
    \put(18, 63.5) {  \DpDs  } 
    \put(18, 57.0) {  \DpLc  } 
    \put(90, 44.0) {  \psiDz  } 
    \put(90, 37.5) {  \psiDp  } 
    \put(90, 31.0) {  \psiDs  } 
    \put(90, 24.5) {  \psiLc  } 
    \put( 60 ,2)  { \large $\sigma$}
    \put(110 ,2)  { \large $\left[ \mathrm{nb} \right]$}
    \put(16,140){ \small   
      $\begin{array}{l}
        \mathrm{LHCb}~
      \end{array}$
    }
  \end{picture}
  \caption { \small
    Measured cross-sections
    $\sigma_{\psiC}$,
    $\sigma_{\CC}$~and 
    $\sigma_{\CCbar}$~(points with error bars)
    compared, in \psiC~channels, to the calculations in Refs.~\cite{Berezhnoy:1998aa,Baranov:2006dh}
    (hatched areas) and Ref.~\cite{Lansberg:2008gk} 
    (shaded areas).
    The inner error bars indicate 
    the statistical uncertainty whilst the outer error bars indicate
    the sum of the statistical and systematic uncertainties in quadrature.
    Charge-conjugate modes are included. 
  }
  \label{fig:sigma_cmp}
\end{figure}
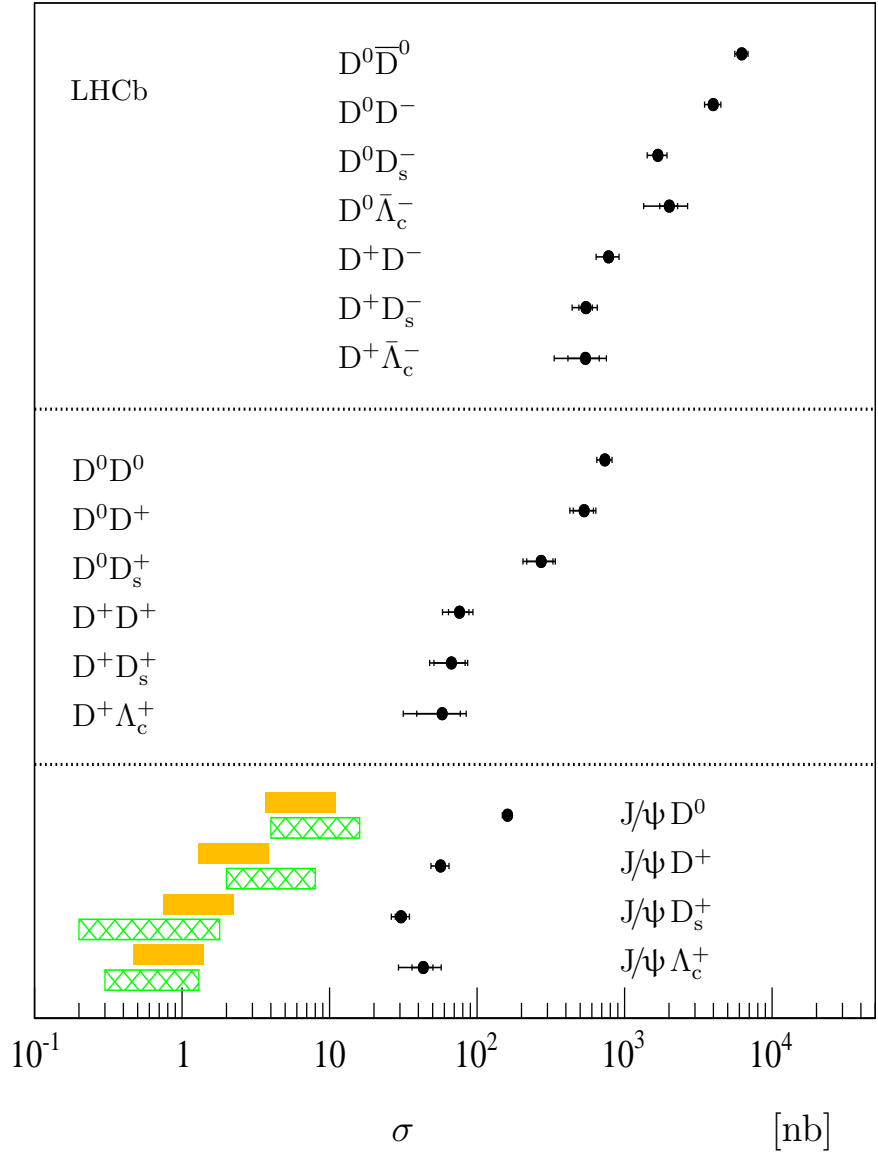

The cross-sections for the production of \jpsi~and associated open charm,
$\sigma_{\psiC}$,  
are measured in the fiducial volume 
$2<y_{\jpsi}, y_{\mathrm{C}}<4$,
$p^{\mathrm{T}}_{\jpsi}<12~\mathrm{GeV}/c$,
$3<p^{\mathrm{T}}_{\mathrm{C}}<12~\mathrm{GeV}/c$. The results are summarized in 
Table~\ref{tab:results_1} and Fig.~\ref{fig:sigma_cmp}.

The systematic uncertainties related to the reconstruction and trigger are
reduced if ratios to the cross-sections for prompt \jpsi,
$\sigma_{\jpsi}$, 
and 
prompt open charm production,
$\sigma_{\mathrm{C}}$, 
with the same fiducial requirements are considered (taking into 
account correlated uncertainties)~\cite{LHCb-CONF-2010-013,Aaij:2011jh}.
These ratios are presented in Table~\ref{tab:results_2}.

The cross-sections for \CC~and \CCbar~events  
in the fiducial volume
$2<y_{\mathrm{C}}<4$,
$3<p^{\mathrm{T}}_{\mathrm{C}}<12~\mathrm{GeV}/c$
are measured and listed in Table~\ref{tab:results_3}
and Fig.~\ref{fig:sigma_cmp}.
The Table also includes the ratio of 
\CC~and \CCbar~production cross-sections,
$\sigma_{\CC}/\sigma_{\CCbar}$, and 
the ratios of the product of the prompt
open charm cross-sections to the \CC(\CCbar)~cross-sections, 
$\sigma_{\mathrm{C}_1}\sigma_{\mathrm{C}_2}/\sigma_{\mathrm{C}_1\mathrm{C}_2}$.

Several of the estimations given in Table~\ref{tab:intro_theo} are also shown 
in Fig.~\ref{fig:sigma_cmp} to compare with our measurements. 
The expectations from gluon-gluon fusion 
processes~\cite{Berezhnoy:1998aa,Baranov:2006dh,Lansberg:2008gk}
are significantly below the measured cross-sections while the DPS 
estimates qualitatively agree with them.  
The observed ratio of \CC/\CCbar events is relatively large, 
e.g. compared with 
$\sigma_{\jpsi\jpsi}/\sigma_{\jpsi} = (5.1\pm1.0\pm1.1)\times10^{-4}$~\cite{Aaij:2011yc}.

\begin{figure}[htb]
  \setlength{\unitlength}{1mm}
  \centering
  \begin{picture}(120,160)
    \put(0,0){
      \includegraphics*[height=160mm,width=130mm,%
      ]{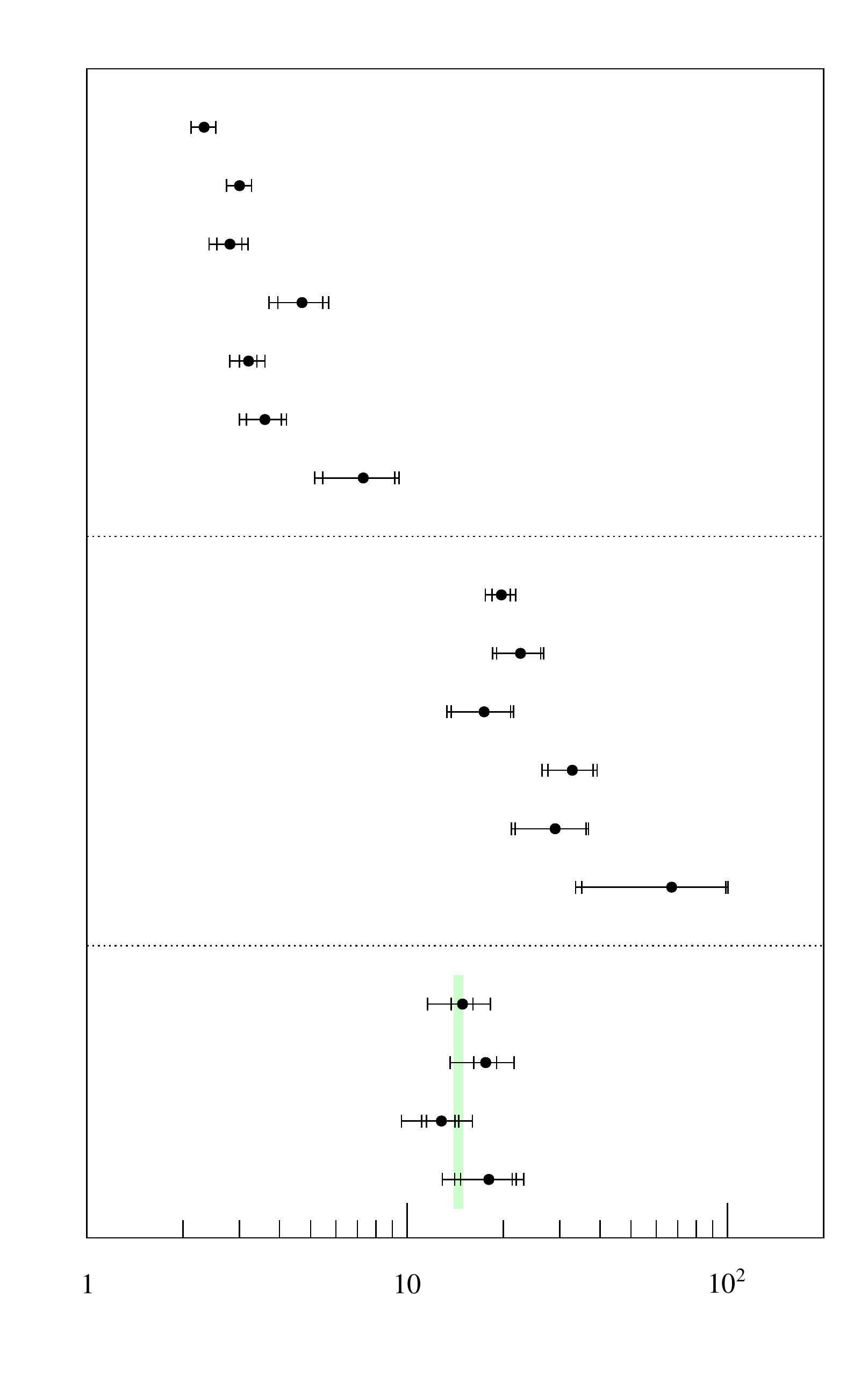} 
    }
    %
    \put(73,143.5) {  \DzDzb  } 
    \put(73,137.0) {  \DzDpb  } 
    \put(73,130.5) {  \DzDsb  } 
    \put(73,124.0) {  \DzLcb  } 
    %
    \put(73,117.5) {  \DpDpb  } 
    \put(73,111.0) {  \DpDsb  } 
    \put(73,104.5) {  \DpLcb  } 
    %
    \put(18, 89.5) {  \DzDz  } 
    \put(18, 83.0) {  \DzDp  } 
    \put(18, 76.5) {  \DzDs  } 
    %
    \put(18, 70.0) {  \DpDp  } 
    \put(18, 63.5) {  \DpDs  } 
    \put(18, 57.0) {  \DpLc  } 
    \put(18, 44.0) {  \psiDz  } 
    \put(18, 37.5) {  \psiDp  } 
    \put(18, 31.0) {  \psiDs  } 
    \put(18, 24.5) {  \psiLc  } 
    \put( 60 ,2)  { \large $\mathcal{R}_{\mathrm{C}_1\mathrm{C}_2}$}
    \put(115 ,2)  { \large $\left[ \mathrm{mb} \right]$}
    \put(100,140){ \small   
     $\begin{array}{r}
        \mathrm{LHCb}~
      \end{array}$
    }
  \end{picture}
  \caption { \small 
    Measured ratios $\mathcal{R}_{\mathrm{C}_1\mathrm{C}_2}$
    (points with error bars)
    in comparison with the expectations from DPS using the cross-section 
    measured at Tevatron for multi-jet events (light green shaded area).
    The inner error bars indicate 
    the statistical uncertainty whilst the outer error bars indicate
    the sum of the statistical and systematic uncertainties in
    quadrature. For the \psiC~case the outermost error bars 
    correspond to the total uncertainties
    including the uncertainties due to  
    the unknown polarization of the prompt \jpsi~mesons. 
  }
  \label{fig:sigma_dps_cmp_scaled}
\end{figure}
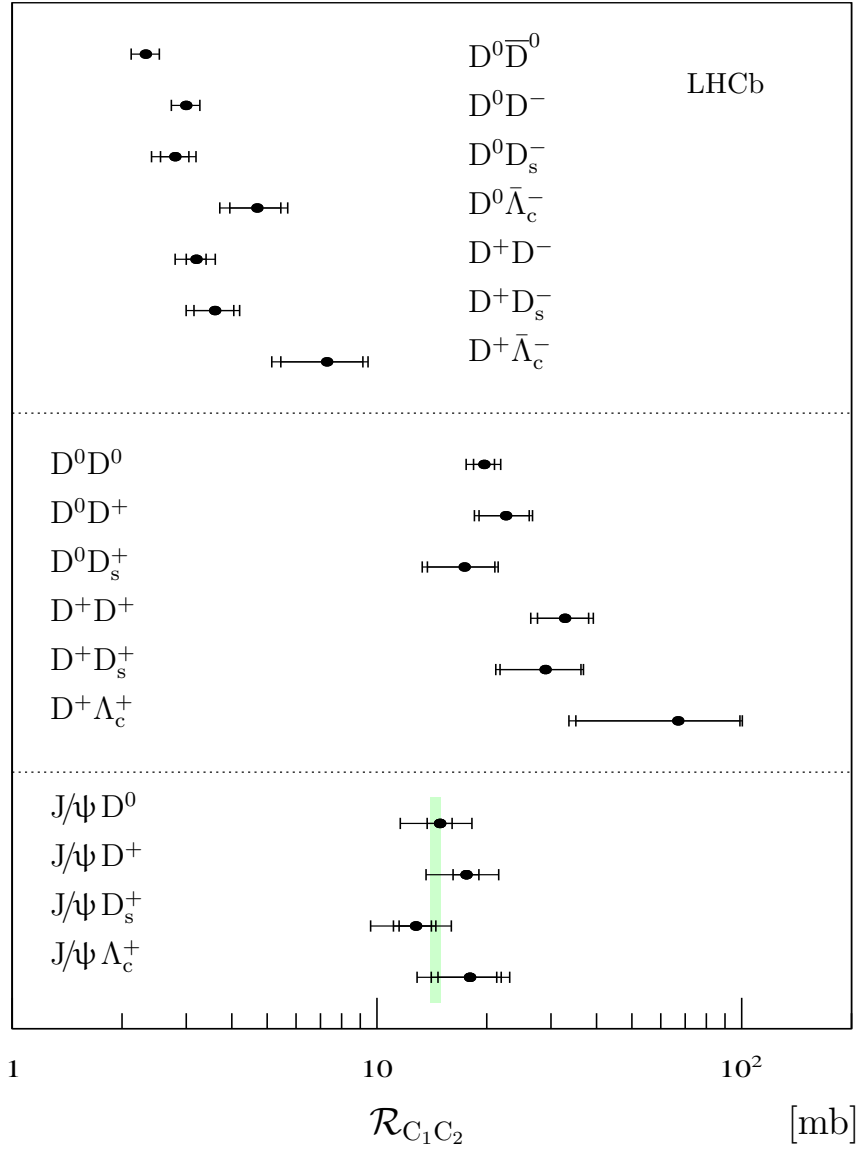

For the ratios  $\sigma_{\jpsi}\sigma_{\mathrm{C}}/\sigma_{\psiC}$~and 
$\sigma_{\mathrm{C}_1}\sigma_{\mathrm{C}_2}/\sigma_{\mathrm{C}_1\mathrm{C}_2}$
listed in Tables~\ref{tab:results_2} and~\ref{tab:results_3}, 
the systematic uncertainties largely cancel. In addition, theoretical 
inputs such as the choice of the strong coupling constant and the charm quark 
fragmentation fractions should cancel allowing 
a more reliable comparison between theory and data. 
%
Figure~\ref{fig:sigma_dps_cmp_scaled} shows the 
ratios $\mathcal{R}_{\mathrm{C}_1\mathrm{C}_2}$ defined as 
\begin{equation*}
\mathcal{R}_{\mathrm{C}_1\mathrm{C}_2} \equiv 
    \upalpha^{\prime} 
    \dfrac { \sigma_{\mathrm{C}_1} \times \sigma_{\mathrm{C}_2} }
                       {\sigma_{\mathrm{C}_1\mathrm{C}_2} },
\label{eq:r}
\end{equation*}
where $\upalpha^{\prime}$ is defined similarly to $\upalpha$~in~Eq.~\eqref{eq:dps}
for the \psiC~and \CC~cases.  When considering \CCbar~production,
$\upalpha^{\prime}=\tfrac{1}{4}$ is used for the~\DzDzb and \DpDm~cases
and $\upalpha^{\prime}=\tfrac{1}{2}$~for the other \CCbar~modes.
%
%
For the \psiC~and \CC~cases these ratios have a clear interpretation in 
the DPS approach~\cite{Kom:2011bd,Baranov:2011ch,Novoselov:2011ff} 
as the effective cross-section of Eq.~\eqref{eq:dps} which should be the 
same for all modes. 
For the \CCbar~case, neglecting the contribution from
$\mathrm{c}\bar{\mathrm{c}}\mathrm{c}\bar{\mathrm{c}}$~production,
this ratio is related by a model-dependent kinematical factor to the total
charm production cross-section and should be 
independent of the final state under consideration.
The values for the effective DPS cross-section
calculated from the \psiC~cross-section 
are in good agreement
with the value measured in multi-jet production at the Tevatron 
$\sigma^{\mathrm{DPS}}_{\mathrm{eff}}=
14.5\pm1.7^{+1.7}_{-2.3}~\mathrm{mb}$~\cite{Abe:1997xk}. 

\begin{table}[t]
  \centering
  \caption{ \small
    Production cross-sections for \psiC.
    The first uncertainty is statistical, 
    and the second is systematic.           
  } \label{tab:results_1}
  \vspace*{3mm}
  \setlength{\extrarowheight}{1.5pt}
  \begin{tabular*}{0.75\textwidth}{@{\hspace{10mm}}c@{\extracolsep{\fill}}c@{\hspace{10mm}}}
    Mode 
    &  $\sigma~~\left[\mathrm{nb}\right]$ \\
    \hline         
    \psiDz 
    & $161.0 \pm 3.7 \pm 12.2$ 
    \\
    \psiDp  
    &  $\phantom{0}56.6 \pm 1.7 \pm   \phantom{0}5.9 $ 
    \\
    \psiDs  
    &  $\phantom{0}30.5 \pm 2.6 \pm   \phantom{0}3.4$ 
    \\
    \psiLc  
    &  $\phantom{0}43.2 \pm 7.0 \pm  12.0$ 
    \\
  \end{tabular*}   
\end{table}

\begin{table}[t]
  \centering
  \caption{ \small
   Ratios of \psiC~production cross-section to prompt \jpsi~cross-section and 
   prompt open charm cross-section, and ratios of the product 
   of prompt \jpsi~and open charm cross-sections to the \psiC~cross-section.
   The first uncertainty is statistical, 
   the second is systematic, 
   and the third is due to the unknown polarization of the
   prompt \jpsi~\cite{Aaij:2011jh}.
  } \label{tab:results_2}
  \vspace*{3mm}
  \setlength{\extrarowheight}{3.5pt}
  \begin{tabular*}{0.95\textwidth}{@{\hspace{3mm}}c@{\extracolsep{\fill}}ccc@{\hspace{3mm}}}
    Mode 
    & $\sigma_{\psiC}/\sigma_{\jpsi}     ~\left[10^{-3} \right]$ 
    & $\sigma_{\psiC}/\sigma_{\mathrm{C}} ~\left[ 10^{-4}  \right]$
    & $\sigma_{\jpsi}\sigma_{\mathrm{C}}/\sigma_{\psiC}~\left[ \mathrm{mb}\right]$    
    \\
    \hline         
    \psiDz 
     & $16.2 \pm 0.4 \pm 1.3^{+3.4}_{-2.5}$   
     & $6.7 \pm 0.2 \pm 0.5$ 
     & $14.9 \pm 0.4 \pm 1.1^{+2.3}_{-3.1} $ 
     \\   
    \psiDp  
     & $\phantom{0}5.7 \pm 0.2 \pm 0.6^{+1.2}_{-0.9}$   
     & $5.7 \pm 0.2 \pm 0.4$ 
     & $17.6 \pm 0.6 \pm 1.3^{+2.8}_{-3.7}$ 
     \\  
    \psiDs  
     & $\phantom{0}3.1 \pm 0.3 \pm 0.4^{+0.6}_{-0.5}$     
     & $7.8 \pm 0.8 \pm 0.6$ 
     & $12.8 \pm 1.3 \pm 1.1 ^{+2.0}_{-2.7}$ 
    \\
    \psiLc  
     & $\phantom{0}4.3 \pm 0.7 \pm 1.2^{+0.9}_{-0.7}$     
     & $5.5 \pm 1.0 \pm 0.6$ 
     & $18.0 \pm 3.3 \pm 2.1 ^{+2.8}_{-3.8}$ 
     \\
  \end{tabular*}   
\end{table}

\begin{table}[t]
  \centering
  \caption{ \small
   Production cross-sections for \CC~and \CCbar, 
    ratios of the \CC~and \CCbar~cross-sections and
    ratios of the product of prompt open charm cross-sections to the \CC(\CCbar)~cross-sections. 
    The first uncertainty is statistical 
    and the second is systematic.
    The symmetry factor 2 is explicitly indicated for 
    the \DzDz, \DzDzb, \DpDp and \DpDpb~ratios.
  } \label{tab:results_3}
  \vspace*{3mm}
  \setlength{\extrarowheight}{1.5pt}
   \begin{tabular*}{0.95\textwidth}{@{\hspace{5mm}}l@{\extracolsep{\fill}}r@{\extracolsep{0mm}}@{$\,\pm\,$}c@{$\,\pm\,$}l@{\extracolsep{\fill}}r@{\extracolsep{0pt}}c@{\extracolsep{0pt}}l@{\extracolsep{\fill}}r@{\extracolsep{0mm}}@{$\,\pm\,$}c@{$\,\pm\,$}l@{\hspace{5mm}}}
    Mode  
    & \multicolumn{3}{c}{$\sigma~~\left[\mathrm{nb}\right]$} 
    & \multicolumn{3}{c}{$\sigma_{\CC}/\sigma_{\CCbar}~~\left[\%\right]$}
    & \multicolumn{3}{c}{$\sigma_{\mathrm{C}_1}\sigma_{\mathrm{C}_2}/\sigma_{\mathrm{C}_1\mathrm{C}_2}~\left[ \mathrm{mb}\right]$}    
    \\
    \hline 
    \DzDz
    &  690 & 40 & 70  
    & \multirow{2}{*}{~~~~10.9}          
    & \multirow{2}{*}{$\,\pm\,$}     
    & \multirow{2}{*}{0.8}          
    & $2 \times(42$ & 3 & 4)  
    \\
    \DzDzb
    & 6230 & 120 & 630 
    &
    &
    &
    & $2 \times(4.7$ & 0.1 & 0.4)
    \\
    \hline 
    \DzDp
    & 520 & 80 & 70
    & \multirow{2}{*}{12.8}     
    & \multirow{2}{*}{$\,\pm\,$}     
    & \multirow{2}{*}{2.1}     
    &  47 & 7 & 4 
    \\
    \DzDpb
    &  3990 & 90 & 500
    &
    &
    &
    &   6.0 & 0.2 & 0.5
    \\
    \hline 
    \DzDs
    & 270  & 50 & 40 
    & \multirow{2}{*}{15.7}     
    & \multirow{2}{*}{$\,\pm\,$}     
    & \multirow{2}{*}{3.4}     
    & 36 & 8 & 4   
    \\
    \DzDsb
    &  1680 & 110 & 240 
    &
    &
    &
    & 5.6 & 0.5 & 0.6 
    \\
    \hline 
    \DzLcb   
    & 2010  & 280 & 600
    & \multicolumn{3}{c}{---}
    &  9 & 2 & 1 
    \\
    \hline 
    \DpDp   
    & 80 & 10 & 10
    & \multirow{2}{*}{9.6}
    & \multirow{2}{*}{$\,\pm\,$}     
    & \multirow{2}{*}{1.6}
    & $2 \times(66$ & 11 & 7)
    \\
    \DpDpb   
    & 780 & 40 & 130 
    &
    &
    &
    & $2 \times(6.4$ &  0.4 & 0.7)
    \\
    \hline 
    \DpDs   
    & 70 & 15 & 10
    & \multirow{2}{*}{12.1}
    & \multirow{2}{*}{$\,\pm\,$}     
    & \multirow{2}{*}{3.3}
    & 59 & 15 & 6 
    \\
    \DpDsb   
    & 550 & 60 & 90 
    &
    &
    &
    & 7 & 1 &1 
    \\
    \hline 
    \DpLc   
    & 60 & 30 & 20
    & \multirow{2}{*}{10.7}        
    & \multirow{2}{*}{$\,\pm\,$}     
    & \multirow{2}{*}{5.9}        
    & 140 & 70 & 20 
    \\
    \DpLcb   
    & 530 & 130 & 170 
    &
    &
    &
    &  15 & 4 & 2 
    \\
  \end{tabular*}   
\end{table}

\clearpage 
%
\section{Properties of \boldmath\psiC, \boldmath\CC, and \boldmath\CCbar~events}
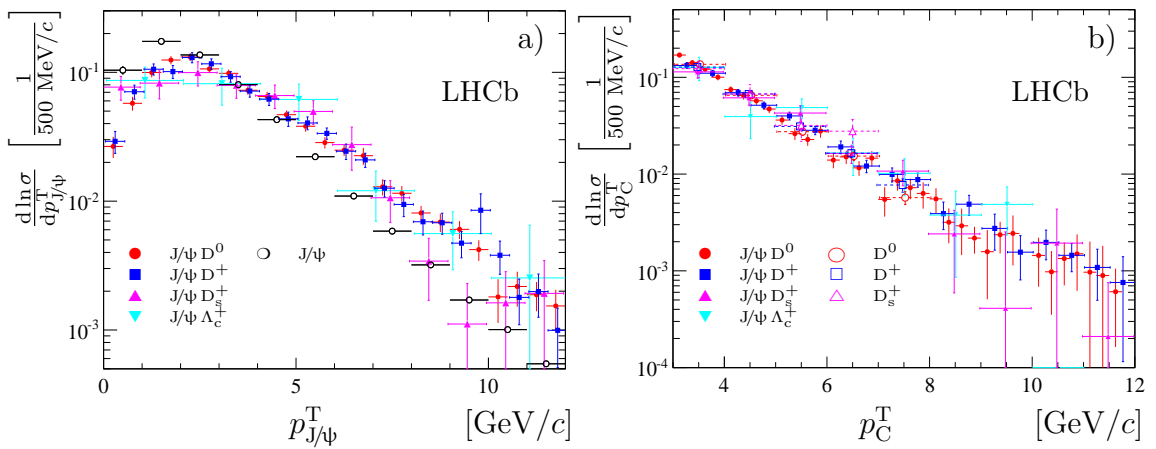
\begin{figure}[b]
  \setlength{\unitlength}{1mm}
  \centering
  \begin{picture}(150,60)
    \put(0,0){
      \includegraphics*[width=75mm,height=60mm,%
      ]{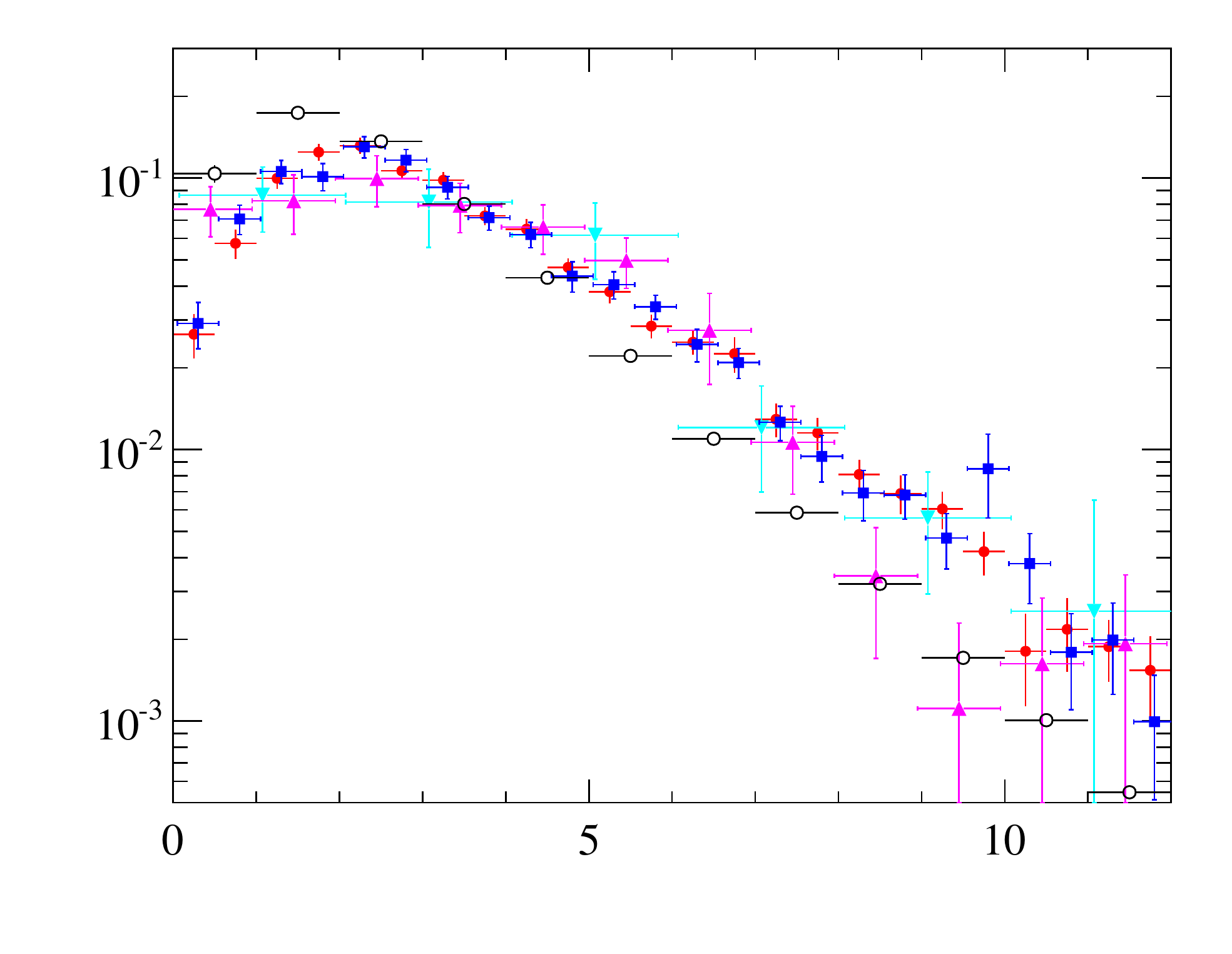}
    }
    \put(75,0){
      \includegraphics*[width=75mm,height=60mm,%
      ]{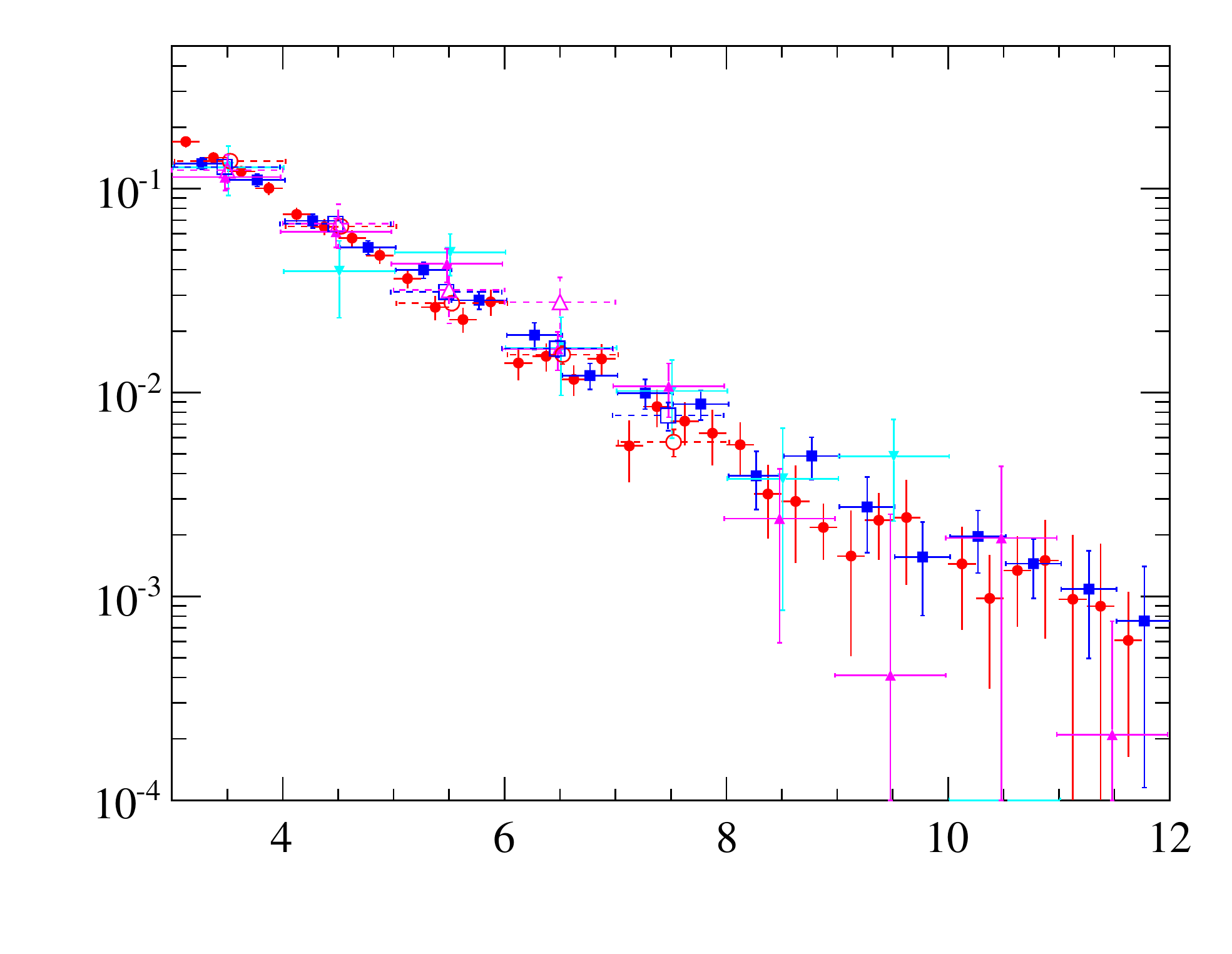}
    }
    \put( 35,1)   { $p^{\mathrm{T}}_{\jpsi}$}
    \put(110,1)   { $p^{\mathrm{T}}_{\mathrm{C}}$}
    \put( 58,1)   { $\left[ \mathrm{GeV}/c\right]$}
    \put(133,1)   { $\left[ \mathrm{GeV}/c\right]$}
    \put(-2 , 28 )  { \small 
      \begin{sideways}%
        $\tfrac{\mathrm{d}\ln\sigma}{{\mathrm{d}}p^{\mathrm{T}}_{\jpsi}}%
        ~\left[ \tfrac{1}{500~\mathrm{MeV}/c}\right]$
      \end{sideways}%
    }
    \put(73,  28 )  { \small 
      \begin{sideways}%
        $\tfrac{\mathrm{d}\ln\sigma}{{\mathrm{d}}p^{\mathrm{T}}_{\mathrm{C}}}%
        ~\left[ \tfrac{1}{500~\mathrm{MeV}/c}\right]$
      \end{sideways}%
    }
    \put( 55,45){ \small 
      LHCb 
    }
    \put(130,45){ \small 
      LHCb
    }
    \put( 65,52)   { a) }
    \put(140,52)   { b) }
    \put(12,20){  \tiny
      $\begin{array}{clcl}
        {\color{red}      \text{\ding{108}} } & \psiDz & { \text{\ding{109}} } & \jpsi \\ 
        {\color{blue}     \text{\ding{110}} } & \psiDp & &  \\
        {\color{RootSix}  \text{\ding{115}} } & \psiDs & &  \\
        {\color{RootSeven}\text{\ding{116}} } & \psiLc & &  \\
       \end{array}$
    } 
    \put(87,20){ \tiny
      $\begin{array}{clcl}
        {\color{red}     \text{\ding{108}} } & \psiDz & {\color{red}    \bigcirc}  & \Dz \\
        {\color{blue}    \text{\ding{110}} } & \psiDp & {\color{blue}   \square}   & \Dp \\
        {\color{RootSix} \text{\ding{115}} } & \psiDs & {\color{RootSix}\triangle} & \Ds \\
        {\color{RootSeven}    \text{\ding{116}} } & \psiLc & &  
       \end{array}$
    } 
  \end{picture}
  \caption { \small
    a)~Transverse momentum 
    spectra of \jpsi for  
    \psiC~and prompt \jpsi~events.
    b)~Transverse momentum 
    spectra for open charm hadrons 
    for  
    \psiC~and prompt 
    \Dz,
    \Dp and 
    \Ds~events.
  }
  \label{fig:pt_psi_pt3}
\end{figure}

 The data samples available also allow the properties of the multiple charm
events to be studied. The transverse momentum spectra for \jpsi~and open 
charm mesons in \psiC~events 
are presented in Fig.~\ref{fig:pt_psi_pt3}.
 
\begin{figure}[htb]
  \setlength{\unitlength}{1mm}
  \centering
  \begin{picture}(150,60)
    \put(0,0){
      \includegraphics*[width=75mm,height=60mm,%
      ]{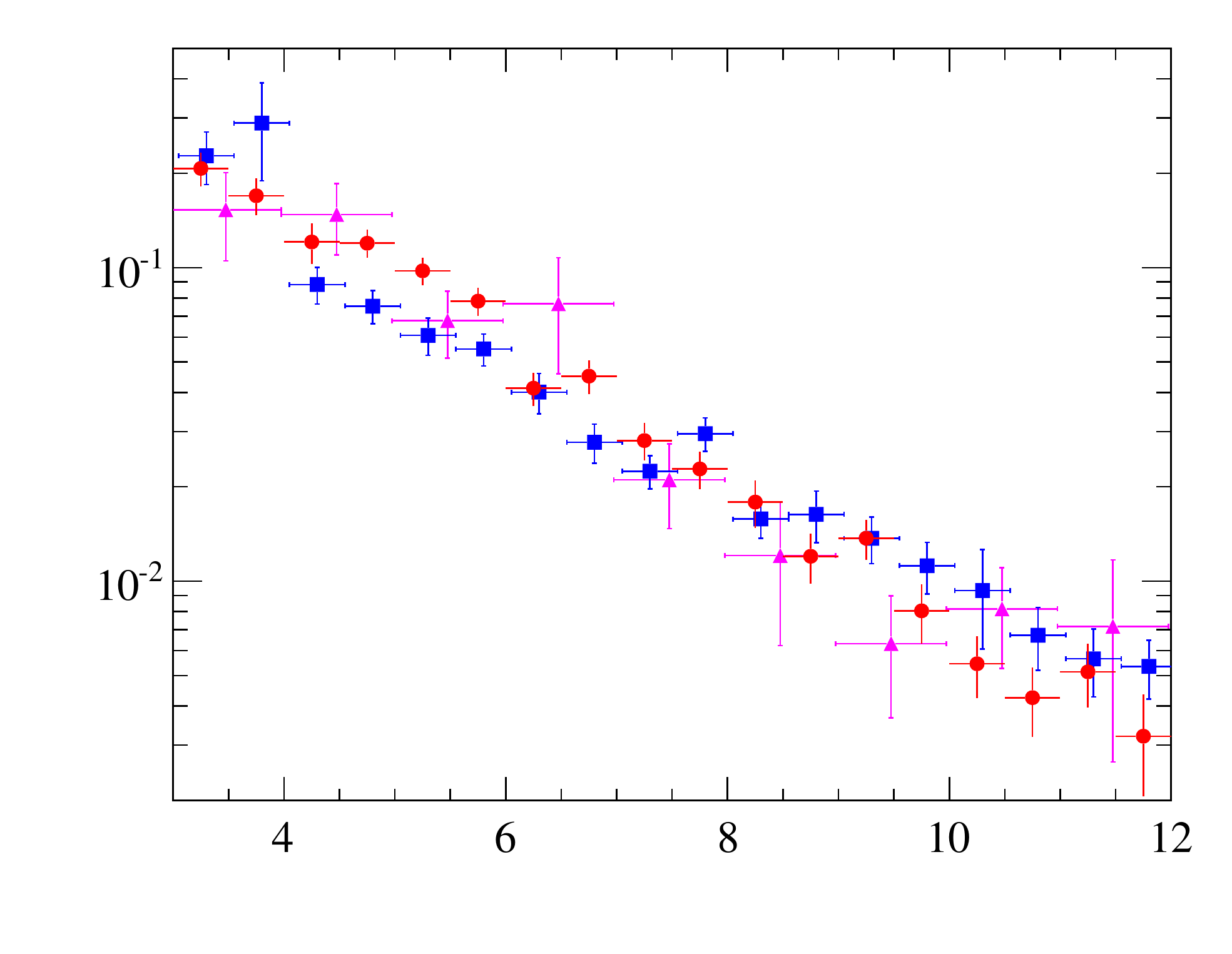}     
    }
    \put(75,0){
      \includegraphics*[width=75mm,height=60mm,%
      ]{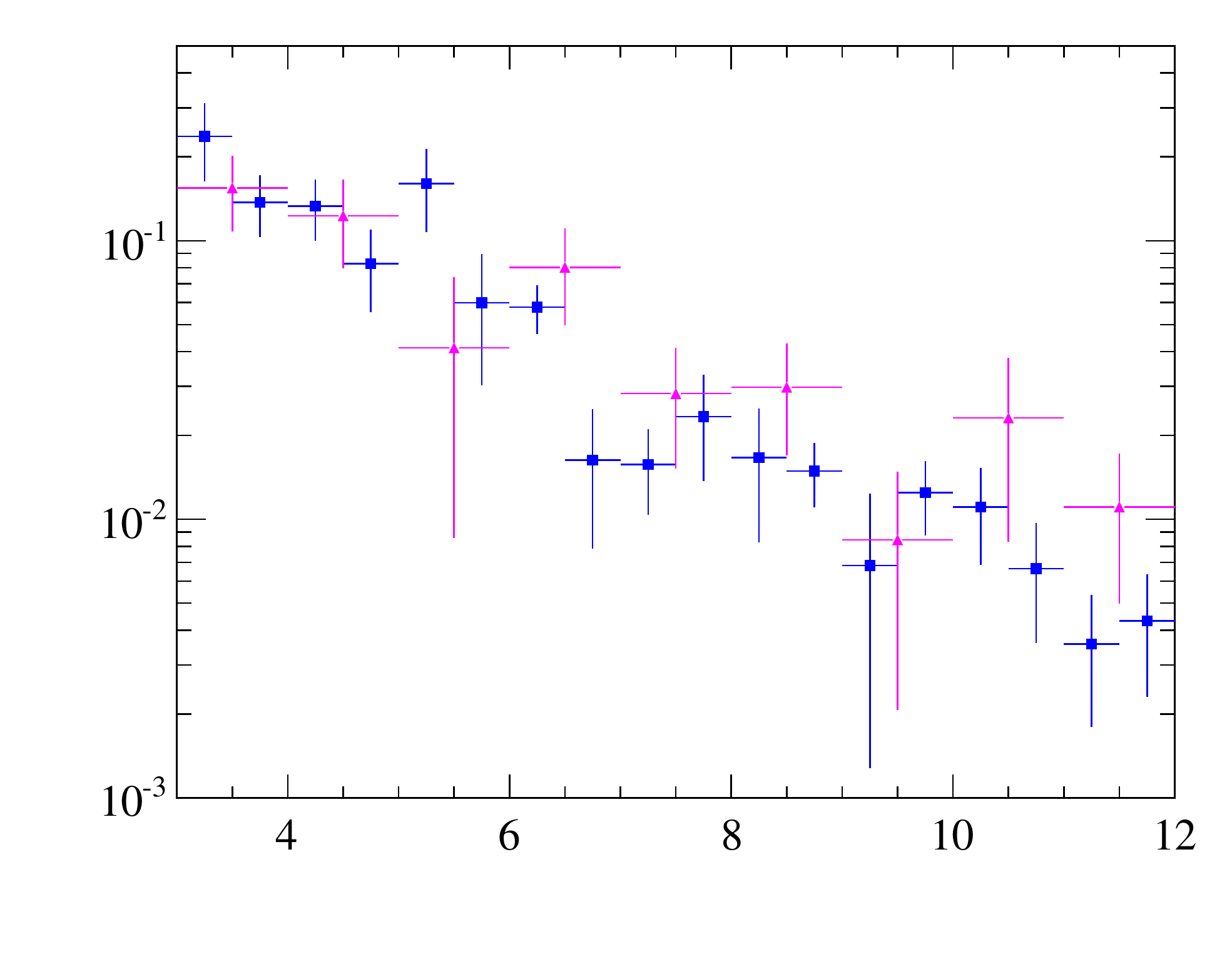}     
    }
    \put(35 ,1)  { $p^{\mathrm{T}}_{\mathrm{C}}$}
    \put(110,1)  { $p^{\mathrm{T}}_{\mathrm{C}}$}
    \put(58 ,1)  { $\left[ \mathrm{GeV}/c\right]$}
    \put(133,1)  { $\left[ \mathrm{GeV}/c\right]$}
    \put(-2, 28)  { \small 
      \begin{sideways}%
        $\tfrac{\mathrm{d}\ln\sigma^*}{{\mathrm{d}}p^{\mathrm{T}}_{\mathrm{C}}}%
        ~\left[ \tfrac{1}{250~\mathrm{MeV}/c}\right]$
      \end{sideways}%
    }
    \put(73,28)  { \small 
      \begin{sideways}%
        $\tfrac{\mathrm{d}\ln\sigma^*}{{\mathrm{d}}p^{\mathrm{T}}_{\mathrm{C}}}%
        ~\left[ \tfrac{1}{250~\mathrm{MeV}/c}\right]$
      \end{sideways}%
    }
     \put( 55,45){ \small 
       LHCb
    }
    \put(130,45){ \small 
      LHCb
    }
    \put( 65,52)   { a) }
    \put(140,52)   { b) }
    \put(12,20){ \tiny
      $\begin{array}{cl}
        {\color{red}     \text{\ding{108}} } & \DzDz \\
        {\color{blue}    \text{\ding{110}} } & \DzDp \\
        {\color{RootSix} \text{\ding{115}} } & \DzDs  
       \end{array}$
    } 
    \put(87,20){ \tiny
      $\begin{array}{cl}
        {\color{blue}    \text{\ding{110}} } & \DpDp \\
        {\color{RootSix} \text{\ding{115}} } & \DpDs 
       \end{array}$
    } 
  \end{picture}
  \caption { \small
    Transverse momentum  spectra of charm hadrons
    from \CC:
    a)~\DzDz, 
       \DzDp,
       \DzDs~and 
    b)~\DpDp~and  
       \DpDs~.
  }
  \label{fig:pt_CC_pt3}
\end{figure}

\begin{figure}[htb]
  \setlength{\unitlength}{1mm}
  \centering
  \begin{picture}(150,60)
    \put(0,0){
      \includegraphics*[width=75mm,height=60mm,%
      ]{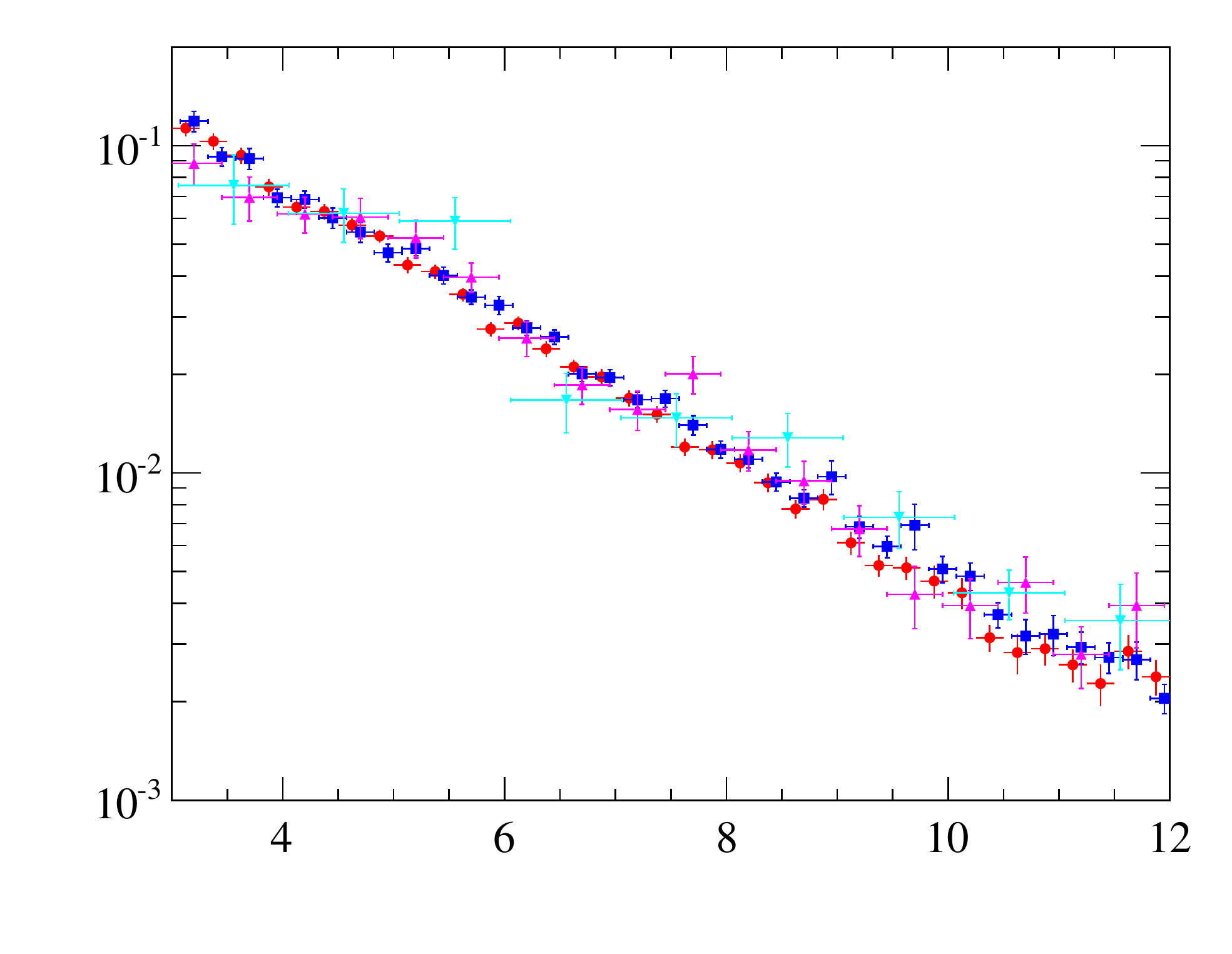}
    }
    \put(75,0){
      \includegraphics*[width=75mm,height=60mm,%
      ]{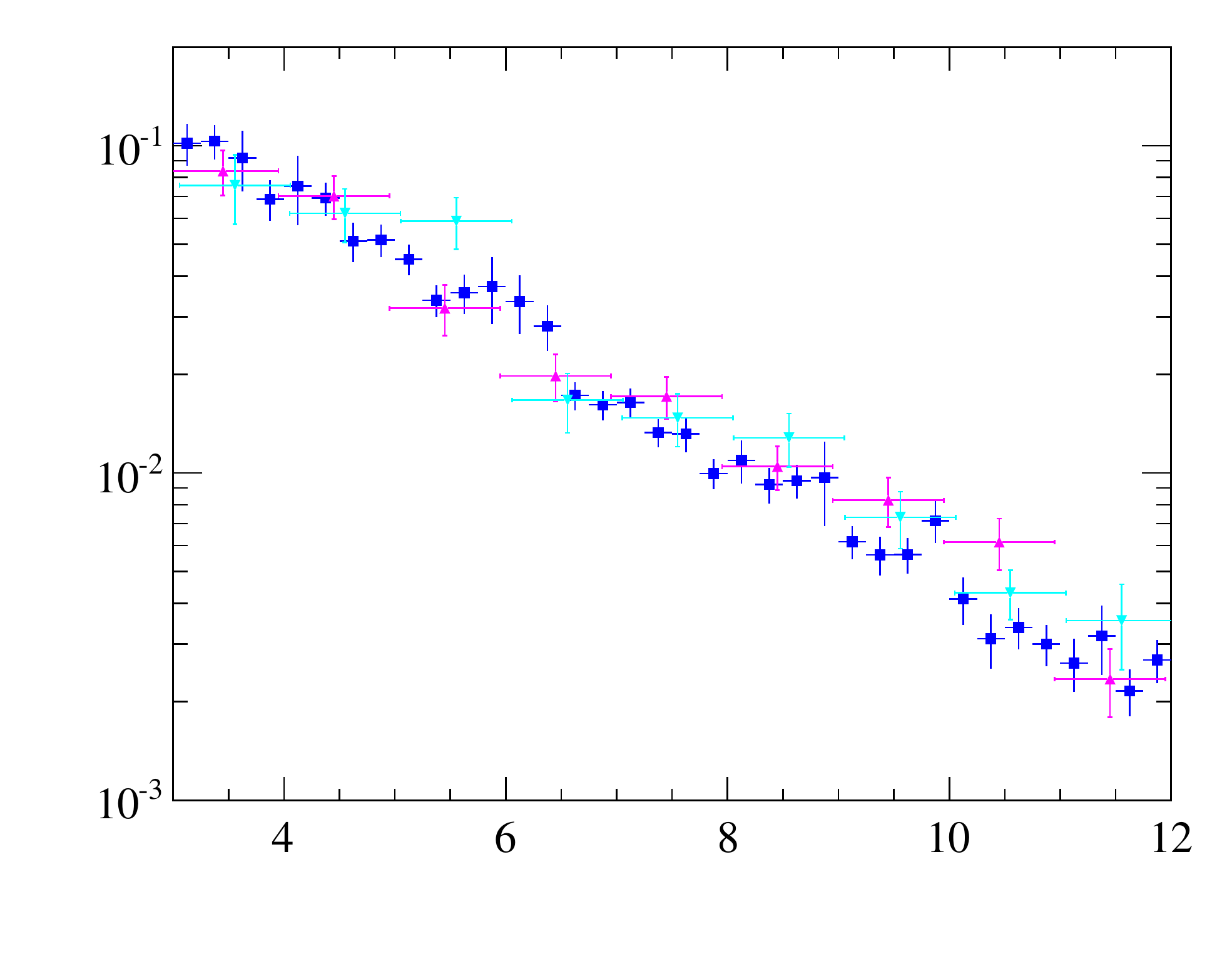}
    }
    \put(35 ,1)  { $p^{\mathrm{T}}_{\mathrm{C}}$}
    \put(110,1)  { $p^{\mathrm{T}}_{\mathrm{C}}$}
    \put(58 ,1)  { $\left[ \mathrm{GeV}/c\right]$}
    \put(133,1)  { $\left[ \mathrm{GeV}/c\right]$}
    \put(-2, 28)  { \small 
      \begin{sideways}%
        $\tfrac{\mathrm{d}\ln\sigma^*}{\mathrm{d}\mathrm{p}_{\mathrm{T}}^{\mathrm{C}}}%
        ~\left[ \tfrac{1}{250~\mathrm{MeV}/c}\right]$
      \end{sideways}%
    }
    \put(73,28)  { \small 
      \begin{sideways}%
        $\tfrac{\mathrm{d}\ln\sigma^*}{\mathrm{d}\mathrm{p}_{\mathrm{T}}^{\mathrm{C}}}%
        ~\left[ \tfrac{1}{250~\mathrm{MeV}/c}\right]$
      \end{sideways}%
    }
    \put( 55,45){ \small 
      LHCb
    }
    \put(130,45){ \small 
      LHCb
    }
    \put( 65,52)   { a) }
    \put(140,52)   { b) }
    \put(12,20){ \tiny
      $\begin{array}{cl}
        {\color{red}     \text{\ding{108}} } & \DzDzb  \\
        {\color{blue}    \text{\ding{110}} } & \DzDpb  \\
        {\color{RootSix} \text{\ding{115}} } & \DzDsb  \\
        {\color{RootSeven}    \text{\ding{116}} } & \DzLcb  
       \end{array}$
    } 
    \put(87,20){ \tiny
      $\begin{array}{cl}
        {\color{blue}    \text{\ding{110}} } & \DpDpb  \\
        {\color{RootSix} \text{\ding{115}} } & \DpDsb  \\
        {\color{RootSeven}    \text{\ding{116}} } & \DpLcb  
       \end{array}$
    } 
  \end{picture}
  \caption { \small
    Transverse momentum spectra of charm hadrons
    from \CCbar:
    a)~\DzDzb, 
       \DzDpb, 
       \DzDsb~and 
       \DzLcb;
    b)~\DpDpb,
       \DpDsb~and
       \DpLcb.
  }
  \label{fig:pt_CCbar_pt3}
\end{figure}

\begin{figure}[htb]
  \setlength{\unitlength}{1mm}
  \centering
  \begin{picture}(120,160)
    \put(0,0){
      \includegraphics*[height=160mm,width=130mm,%
      ]{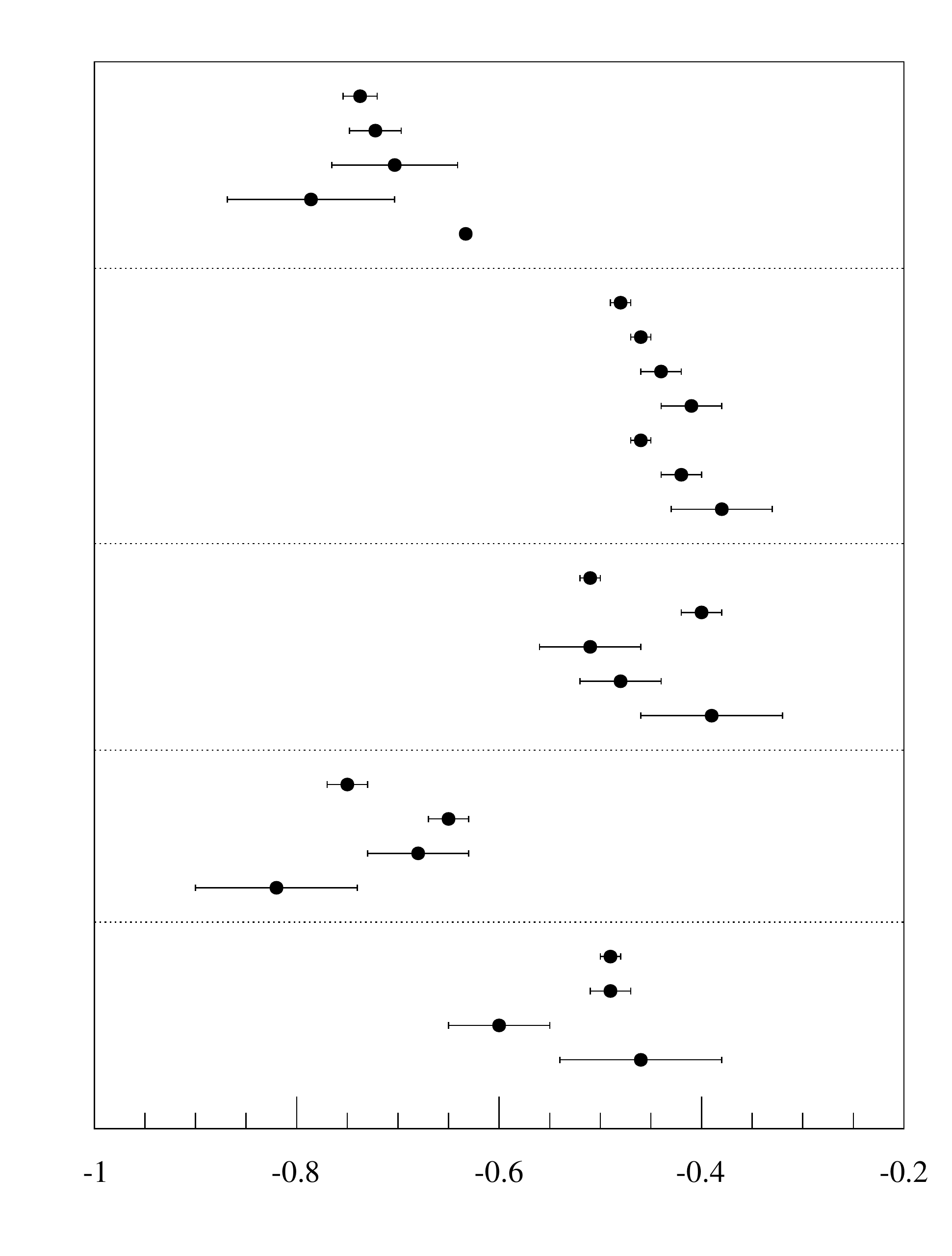}
    }
    \put(68,138) { \small
      $\left.
        \begin{array}{l}
        \Dz \\
        \Dp \\
        \Ds \\
        \Lc \\
        \jpsi
        \end{array}\right\} \text{prompt}$
     }
    \put(15,107) { \small
      $\left.
        \begin{array}{l}
        \DzDzb \\
        \DzDpb \\
        \DzDsb \\
        \DzLcb \\
        \DpDpb \\
        \DpDsb \\
        \DpLcb 
        \end{array}\right\}\CCbar$
     }
    \put(15,76) { \small
      $\left.
        \begin{array}{l}
        \DzDz \\
        \DzDp \\
        \DzDs \\
        \DpDp \\
        \DpDs 
        \end{array}\right\}\CC$
     }
    \put(75,53) { \small
      $\left.
       \begin{array}{l}
        \psiDz \\
        \psiDp \\
        \psiDs \\
        \psiLc 
        \end{array}{}\right\} p^{\mathrm{T}}_{\mathrm{C}}$
     }
    \put(13,30) { \small
      $\left.
       \begin{array}{l}
        \psiDz \\
        \psiDp \\
        \psiDs \\
        \psiLc 
        \end{array}\right\} p^{\mathrm{T}}_{\jpsi}$
     }
    \put(60  ,2)  { $p^{\mathrm{T}}$-slope}
    \put(100 ,2)  { $\left[ \dfrac{1}{\mathrm{GeV}/c}\right]$}
    \put(110,145){  \small 
      LHCb
    }
  \end{picture}
  \caption { \small
    Slope parameters of the transverse momentum 
    spectra for prompt charm particles~\cite{LHCb-CONF-2010-013}
    and charm particles from \psiC, \CCbar~and \CC~production. 
  }
  \label{fig:pt_spectra_cmp}
\end{figure}

The transverse momentum spectra of the \jpsi~meson in 
\psiC~events are similar for all species of open charm hadrons. 
The shape of 
the transverse momentum spectra of open charm hadrons 
also appears to be the same for all species.  The $p^{\mathrm{T}}_{\jpsi}$~spectra 
are harder than the corresponding spectrum 
of prompt \jpsi, while the $p^{\mathrm{T}}$-spectra for open charm hadrons
seem to be well compatible in shape with the spectra for prompt charm production.
To allow a more quantitative comparison, each spectrum is fitted
in the region $3<p^{\mathrm{T}}<12~\mathrm{GeV}/c$~with an exponential 
function. 
The results are summarized in Table~\ref{tab:pt_slopes_pt3}
and Fig.~\ref{fig:pt_spectra_cmp}.
They agree reasonably well within the uncertainties. 
\begin{table}[t]
  \centering
  \caption{ \small
   Slope parameters of the transverse momentum 
   spectra in the \psiC~mode and for prompt 
   charm particles.
   These parameters are determined from fits to the spectra 
   in the region $3<p^{\mathrm{T}}<12~\mathrm{GeV}/c$.
  }
  \label{tab:pt_slopes_pt3}
  \vspace*{3mm}
  \setlength{\extrarowheight}{1.5pt}
  \begin{tabular*}{0.75\textwidth}{@{\hspace{10mm}}c@{\extracolsep{\fill}}cc@{\hspace{10mm}}}
    \multirow{2}{*}{Mode} &  
    \multicolumn{2}{c}{$p^{\mathrm{T}}$-slope ~~$\left[ \frac{1}{\mathrm{GeV}/c}\right]$}
    \\
    & \jpsi
    & $\mathrm{C}$
    \\
    \hline 
    \psiDz
    & $-0.49\pm0.01$
    & $-0.75\pm0.02$
    \\
    \psiDp
    & $-0.49\pm0.02$
    & $-0.65\pm0.02$
    \\
    \psiDs
    & $-0.60\pm0.05$
    & $-0.68\pm0.05$
    \\
    \psiLc
    & $-0.46\pm0.08$
    & $-0.82\pm0.08$
    \\
    \hline 
    \jpsi
    &  $-0.633 \pm 0.003 $
    & 
    \\
    \Dz 
    & 
    &  $-0.77  \pm 0.03 $
    \\
    \Dp
    &
    &  $-0.70  \pm 0.03 $
    \\
    \Ds
    &
    &  $-0.57  \pm 0.13 $
    \\
    \Lc
    &
    &  $-0.79  \pm 0.08 $
    \\
  \end{tabular*}
\end{table}
The transverse momentum spectra of charm hadrons from 
\CC~and \CCbar~events are presented in 
Figs.~\ref{fig:pt_CC_pt3} and~\ref{fig:pt_CCbar_pt3}. 
The~fitted slope parameters of an exponential function 
are summarized 
in Table~\ref{tab:dd_pt_slopes_pt3} 
and Fig.~\ref{fig:pt_spectra_cmp}.
The $p^{\mathrm{T}}$-slopes, 
though similar for \CCbar~and \CC~events,
are significantly 
different from those for both single prompt charm particles 
and those found in \psiC~events. 

\begin{table}[t]
  \centering
  \caption{
    Slope parameters of transverse momentum spectra  
    for the \CC~and \CCbar~modes.
  } \label{tab:dd_pt_slopes_pt3}
  \vspace*{3mm}
  \setlength{\extrarowheight}{1.5pt}
  \begin{tabular*}{0.60\textwidth}{@{\hspace{10mm}}l@{\extracolsep{\fill}}c@{\hspace{10mm}}}
    Mode &  
    $p^{\mathrm{T}}$-slope ~~$\left[ \frac{1}{\mathrm{GeV}/c}\right]$
    \\
    \hline 
    \DzDz
    &  $-0.51 \pm 0.02 $ 
    \\
    \DzDzb
    &  $-0.48 \pm 0.01 $ 
    \\
    \DzDp
    &  $-0.40 \pm 0.02 $ 
    \\
    \DzDpb
    &  $-0.46 \pm 0.01 $ 
    \\
    \DzDs 
    &  $-0.51 \pm 0.05 $ 
    \\
    \DzDsb 
    &  $-0.44 \pm 0.02 $ 
    \\
    \DzLcb 
    &  $-0.41 \pm 0.03 $ 
    \\
    \DpDp 
    &  $-0.48 \pm 0.04 $ 
    \\
    \DpDpb 
    &  $-0.46 \pm 0.01 $ 
    \\
    \DpDs 
    &  $-0.39 \pm 0.07 $ 
    \\
    \DpDsb 
    &  $-0.42 \pm 0.02 $ 
    \\
    \DpLcb 
    &  $-0.38 \pm 0.05 $ 
    \\
  \end{tabular*}
\end{table}

\begin{figure}[htb]
  \setlength{\unitlength}{1mm}
  \centering
  \begin{picture}(150,60)
    \put(0,0){
      \includegraphics*[width=75mm,height=60mm,%
      ]{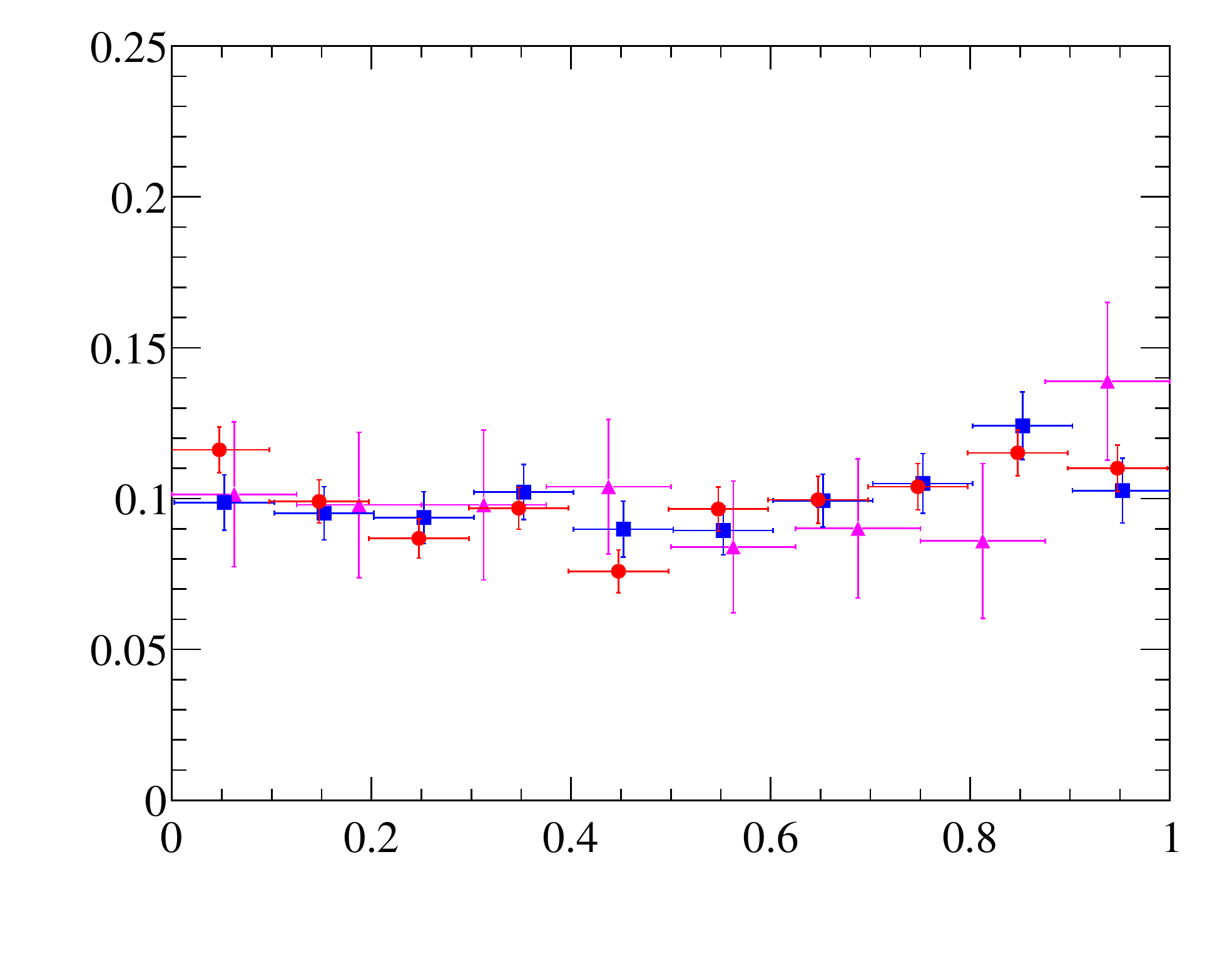}
    }
    \put(75,0){
      \includegraphics*[width=75mm,height=60mm,%
      ]{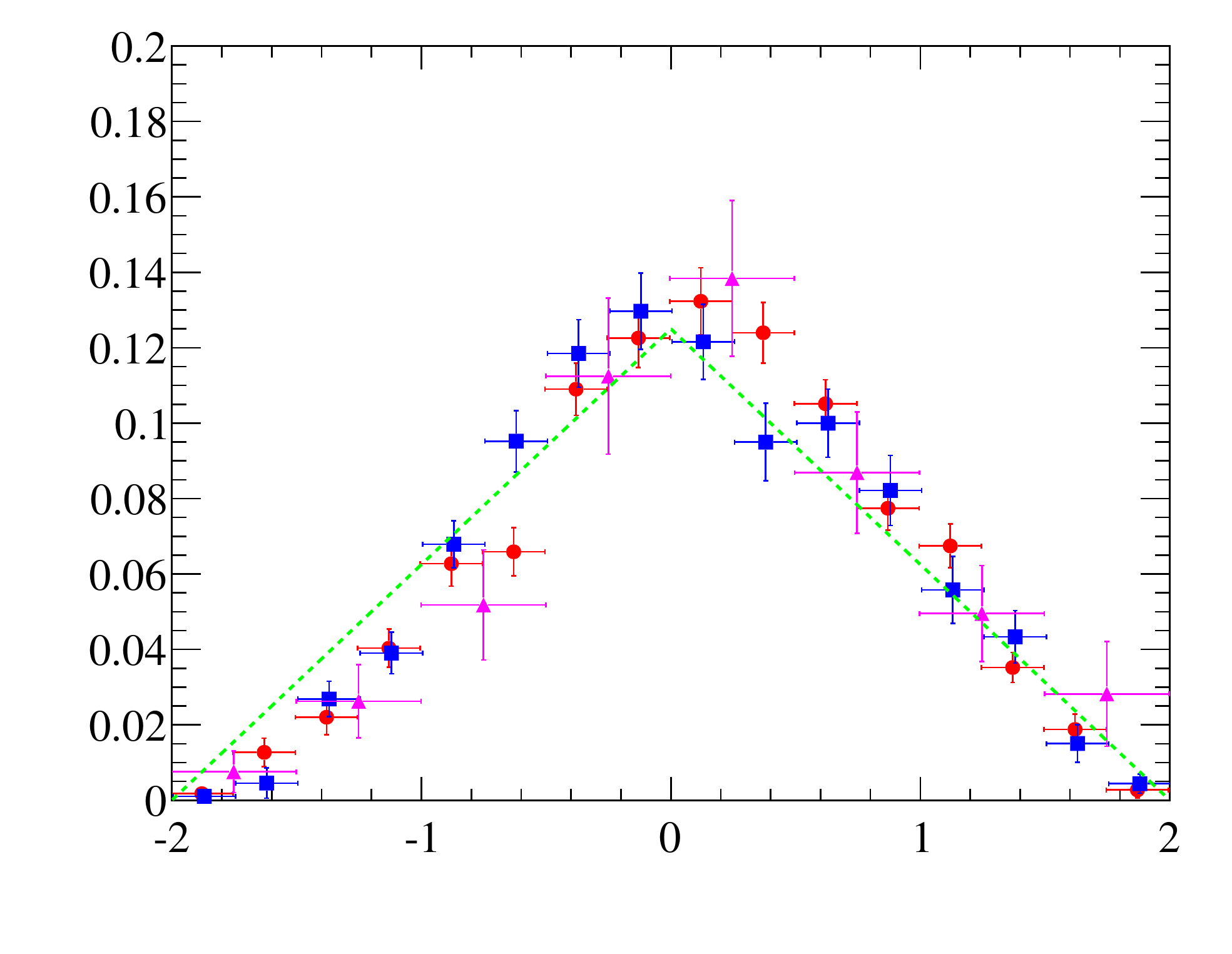}
    }
    \put(35, 1)   { $\left|\Delta\phi\right|/\pi$}
    \put(112,1)   { $\Delta y$}
    \put(-2 , 38 )  { \small 
      \begin{sideways}%
        $\tfrac{\mathrm{d}\ln\sigma^*}{\mathrm{d}\left| \Delta\phi \right|}%
        ~\left[ \tfrac{\pi}{0.1}\right]$ 
      \end{sideways}%
    }
    \put(73, 35 )  { \small 
      \begin{sideways}%
        $\tfrac{\mathrm{d}\ln\sigma^*}{\mathrm{d}\Delta y}%
        ~\left[ \tfrac{1}{0.125}\right]$
      \end{sideways}%
    }
    \put( 55,45){ \small 
      LHCb
    }
    \put(130,45){ \small 
      LHCb
    }
    \put( 65,52)   { a) }
    \put(140,52)   { b) }
    \put(12,50){ \tiny
      $\begin{array}{cl}
        {\color{red}     \text{\ding{108}} } & \psiDz  \\
        {\color{blue}    \text{\ding{110}} } & \psiDp  \\
        {\color{RootSix} \text{\ding{115}} } & \psiDs  
       \end{array}$
    } 
    \put(87,50){ \tiny
      $\begin{array}{cl}
        {\color{red}     \text{\ding{108}} } & \psiDz  \\
        {\color{blue}    \text{\ding{110}} } & \psiDp  \\
        {\color{RootSix} \text{\ding{115}} } & \psiDs  
       \end{array}$
    } 
  \end{picture}
  \caption { \small 
    Distributions of the difference in azimuthal angle~(a)
    and rapidity~(b) 
    for  
    \psiDz,  
    \psiDp~and \psiDs~events.
    The dashed line shows the expected 
    distribution for uncorrelated events.
  }
  \label{fig:psid_dphi_pt3}
\end{figure}

The correlations in azimuthal angle and rapidity between the two charm
hadrons have also been studied by measuring the distributions of
$\Delta\phi$ and $\Delta y$, where $\Delta\phi$~and $\Delta y$~are the
differences in azimuthal angle and rapidity between the two hadrons.  
These distributions for the charm hadrons in \psiC~events 
are shown in Fig.~\ref{fig:psid_dphi_pt3}. 
No significant azimuthal correlation is
observed. 
The $\Delta y$~distribution is compared to the triangular shape that
is expected if the rapidity distribution 
for single charm hadrons is flat and if there are no correlations.

\begin{figure}[htb]
  \setlength{\unitlength}{1mm}
  \centering
  \begin{picture}(150,60)
    \put(0,0){
      \includegraphics*[width=75mm,height=60mm,%
      ]{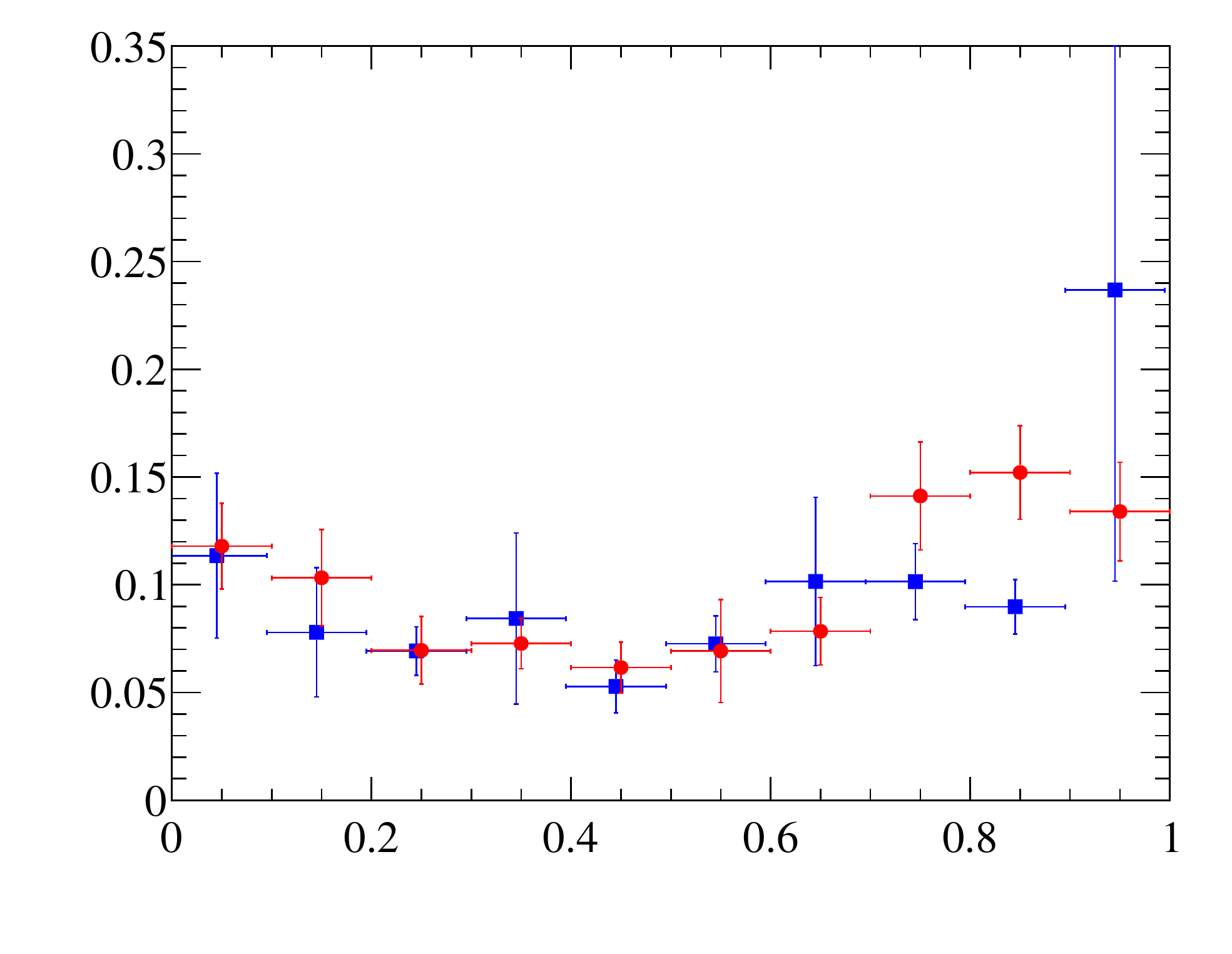}
    }
    \put(75,0){
      \includegraphics*[width=75mm,height=60mm,%
      ]{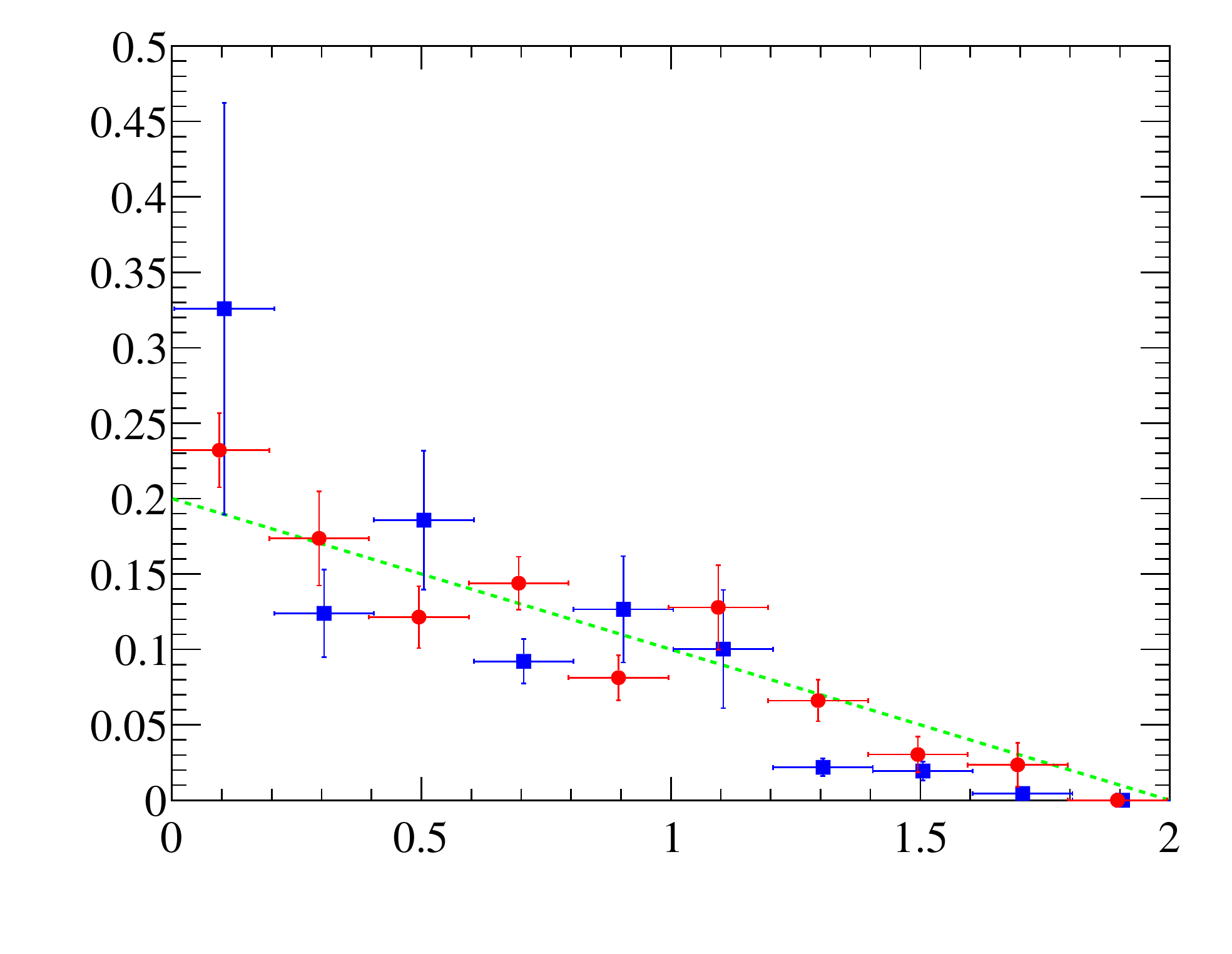}
    }
    \put(35, 1)   { $\left|\Delta\phi\right|/\pi$}
    \put(112,1)   { $\left|\Delta y\right|$}
    \put(-2 , 38 )  { \small 
      \begin{sideways}%
        $\tfrac{\mathrm{d}\ln\sigma^*}{\mathrm{d}\left| \Delta\phi \right|}%
        ~\left[ \tfrac{\pi}{0.1}\right]$ 
      \end{sideways}%
    }
    \put(73, 37 )  { \small 
      \begin{sideways}%
        $\tfrac{\mathrm{d}\ln\sigma^*}{\mathrm{d}\left|\Delta y\right|}%
        ~\left[ \tfrac{1}{0.1}\right]$
      \end{sideways}%
    }
    \put( 12,45){ \small 
      LHCb
    }
    \put(130,45){ \small 
      LHCb
    }
    \put( 13,52)   { a) }
    \put(140,52)   { b) }
    \put(50,50){ \tiny
      $\begin{array}{cl}
        {\color{red}     \text{\ding{108}} } & \DzDz  \\
        {\color{blue}    \text{\ding{110}} } & \DzDp  
       \end{array}$
    } 
    \put(90,50){ \tiny
      $\begin{array}{cl}
        {\color{red}     \text{\ding{108}} } & \DzDz  \\
        {\color{blue}    \text{\ding{110}} } & \DzDp  
       \end{array}$
    } 
  \end{picture}
  \caption { \small 
    Distributions of the difference in azimuthal angle~(a)
    and rapidity~(b) 
    for  
    \DzDz~and \DzDp~events.
    The dashed line shows the expected 
    distribution for uncorrelated events.
  }
  \label{fig:dd_dphi_CC_pt3}
\end{figure}

\begin{figure}[htb]
  \setlength{\unitlength}{1mm}
  \centering
  \begin{picture}(150,60)
    \put(0,0){
      \includegraphics*[width=75mm,height=60mm,%
      ]{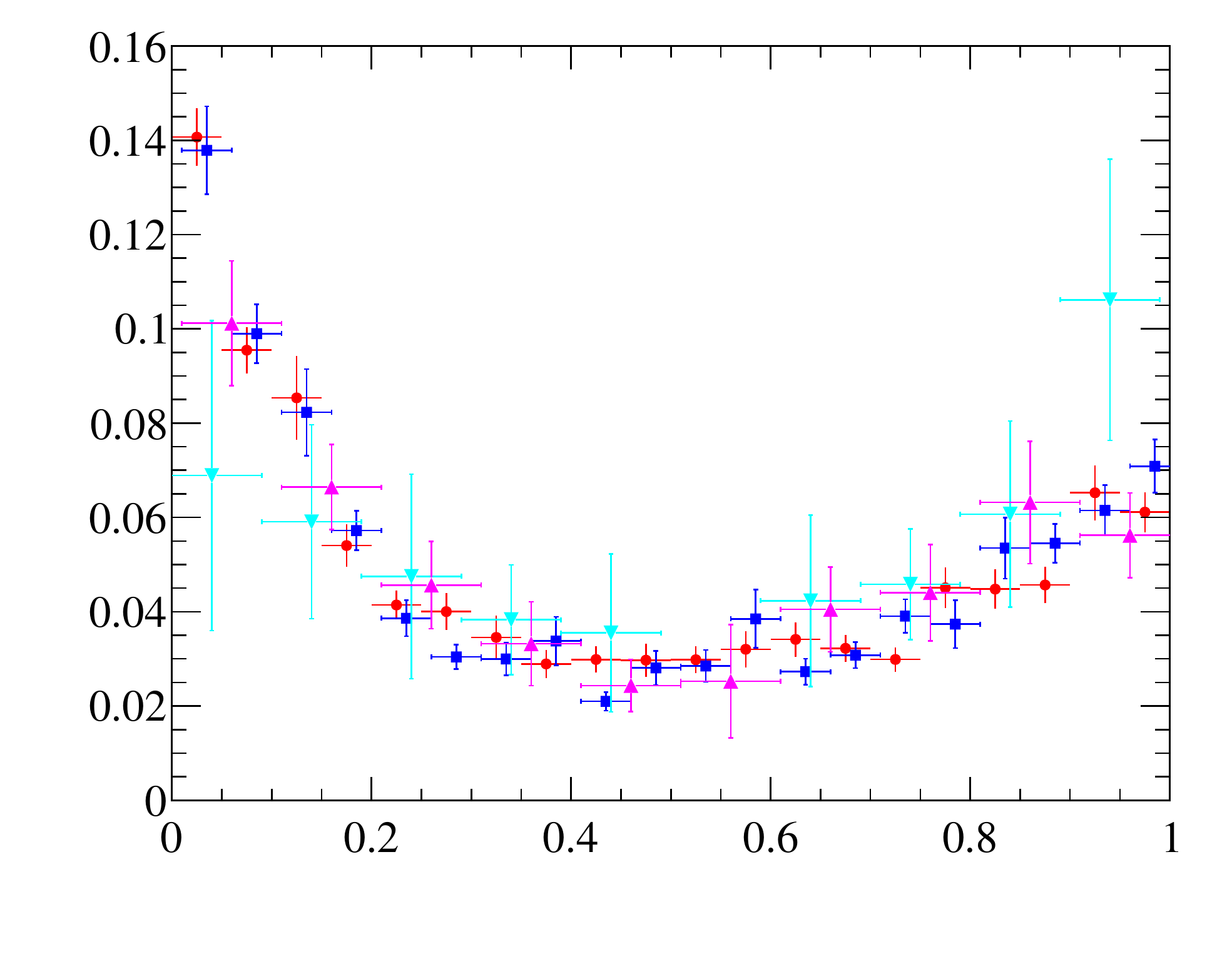}
    }
    \put(75,0){
      \includegraphics*[width=75mm,height=60mm,%
      ]{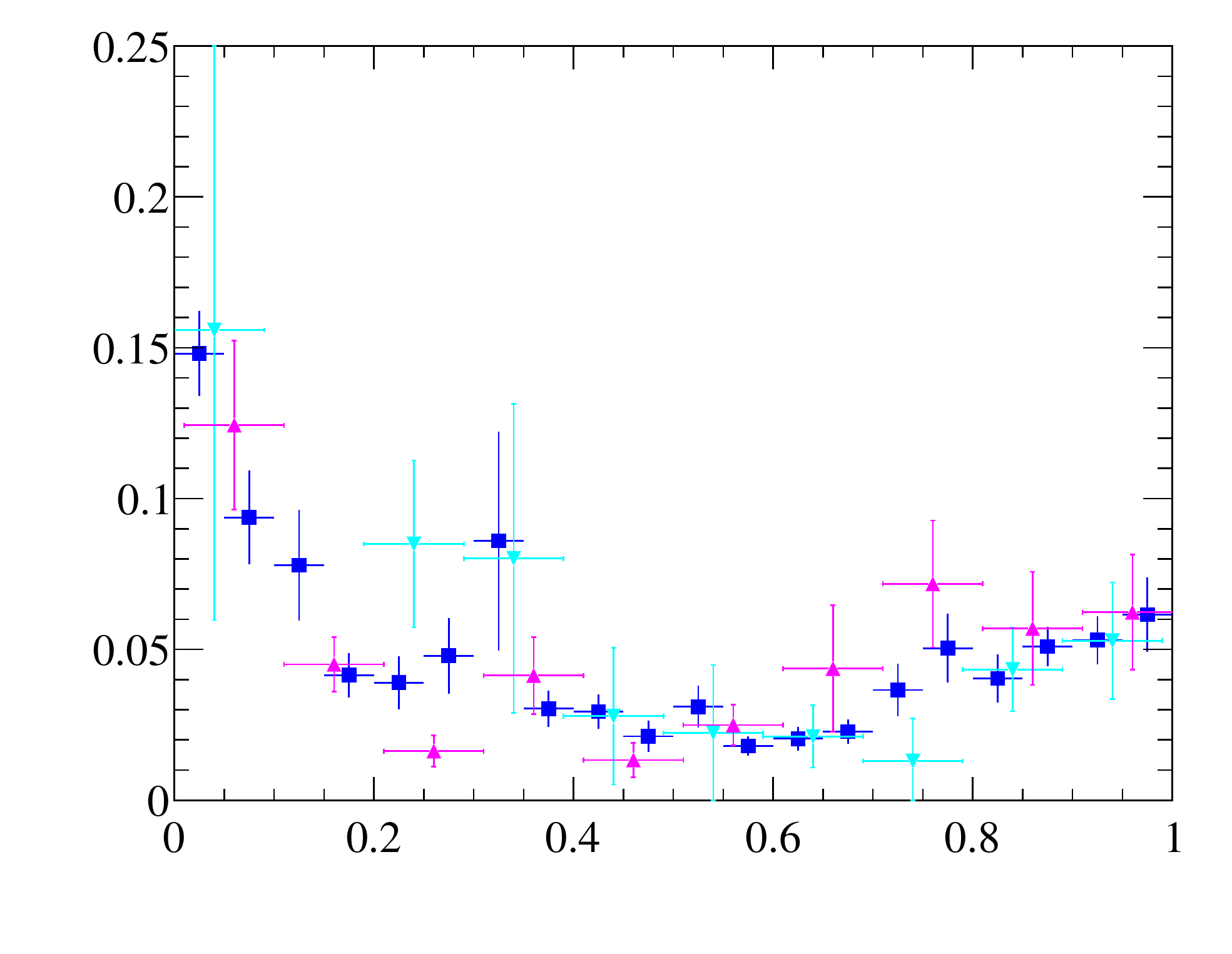}
    }
    \put( 35,1)   { $\left|\Delta\phi\right|/\pi$}
    \put(112,1)   { $\left|\Delta\phi\right|/\pi$}
    \put(0, 37)  { \small 
      \begin{sideways}%
        $\tfrac{\mathrm{d}\ln\sigma^*}{\mathrm{d}\left| \Delta\phi \right|}%
        ~\left[ \tfrac{\pi}{0.05}\right]$
      \end{sideways}%
    }
    \put(75, 37)  { \small 
      \begin{sideways}%
        $\tfrac{\mathrm{d}\ln\sigma^*}{\mathrm{d}\left| \Delta\phi \right|}%
        ~\left[ \tfrac{\pi}{0.05}\right]$
      \end{sideways}%
    }
    \put( 55,45){ \small 
      LHCb
    }
    \put(130,45){ \small 
      LHCb
    }
    \put( 63,52)   { a) }
    \put(140,52)   { b) }
    \put(17,45){ \tiny
      $\begin{array}{cl}
        {\color{red}       \text{\ding{108}} } & \DzDzb  \\
        {\color{blue}      \text{\ding{110}} } & \DzDpb  \\
        {\color{RootSix}   \text{\ding{115}} } & \DzDsb  \\  
        {\color{RootSeven} \text{\ding{116}} } & \DzLcb  
       \end{array}$
    } 
    \put(92,45){ \tiny
      $\begin{array}{cl}
        {\color{blue}      \text{\ding{110}} } & \DpDpb  \\
        {\color{RootSix}   \text{\ding{115}} } & \DpDsb  \\  
        {\color{RootSeven} \text{\ding{116}} } & \DpLcb  
       \end{array}$
    } 
  \end{picture}
  \caption { \small 
    Distributions of the difference in azimuthal angle
    for \CCbar events:
    a)~\DzDzb, 
       \DzDpb, 
       \DzDsb~and 
       \DzLcb; 
    b)~\DpDpb,  
       \DpDsb~and 
       \DpLcb.
  }
  \label{fig:dd_dphi_CCbar_pt3}
\end{figure}

\begin{figure}[htb]
  \setlength{\unitlength}{1mm}
  \centering
  \begin{picture}(150,60)
    \put(0,0){
      \includegraphics*[width=75mm,height=60mm,%
      ]{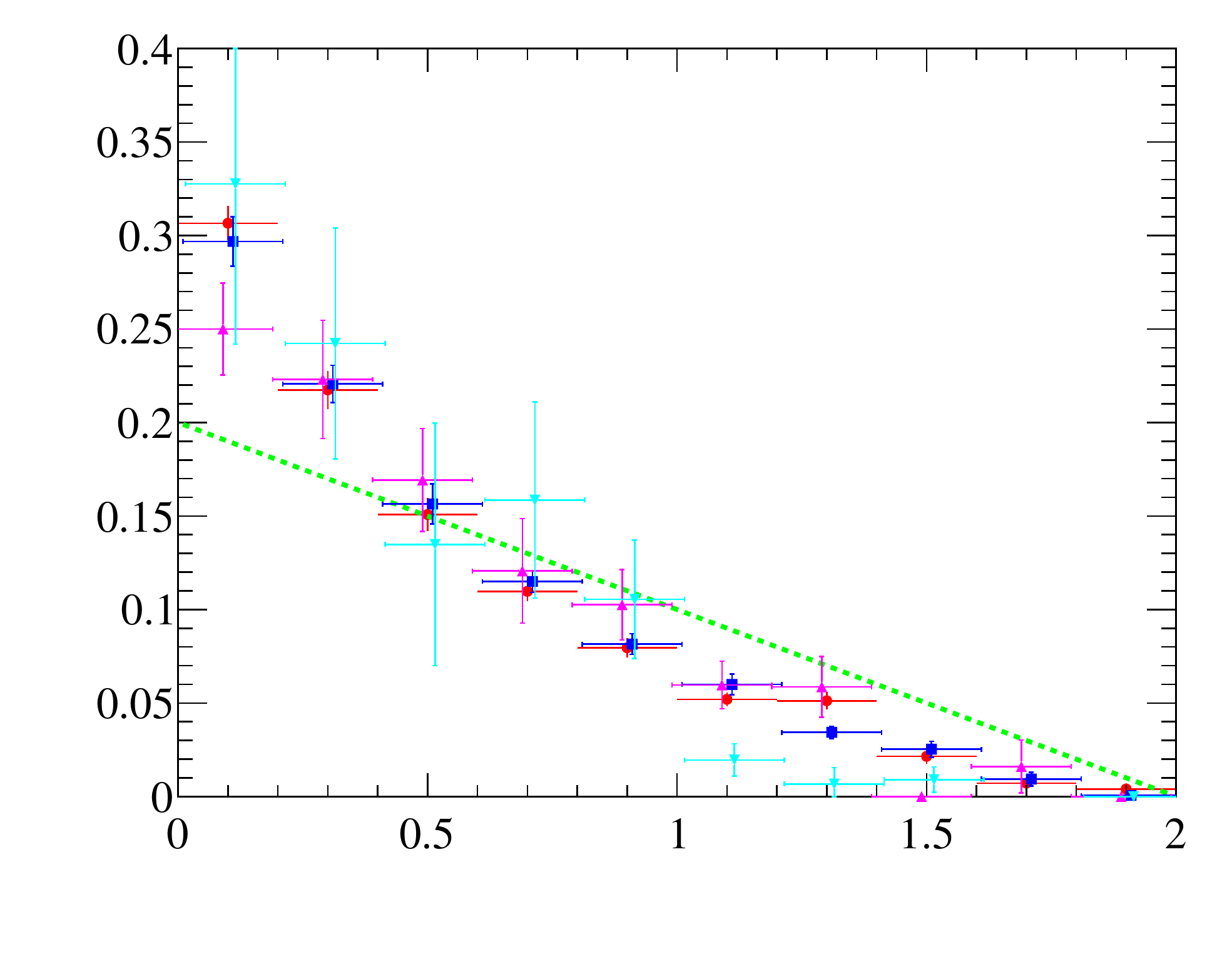}
    }
    \put(75,0){
      \includegraphics*[width=75mm,height=60mm,%
      ]{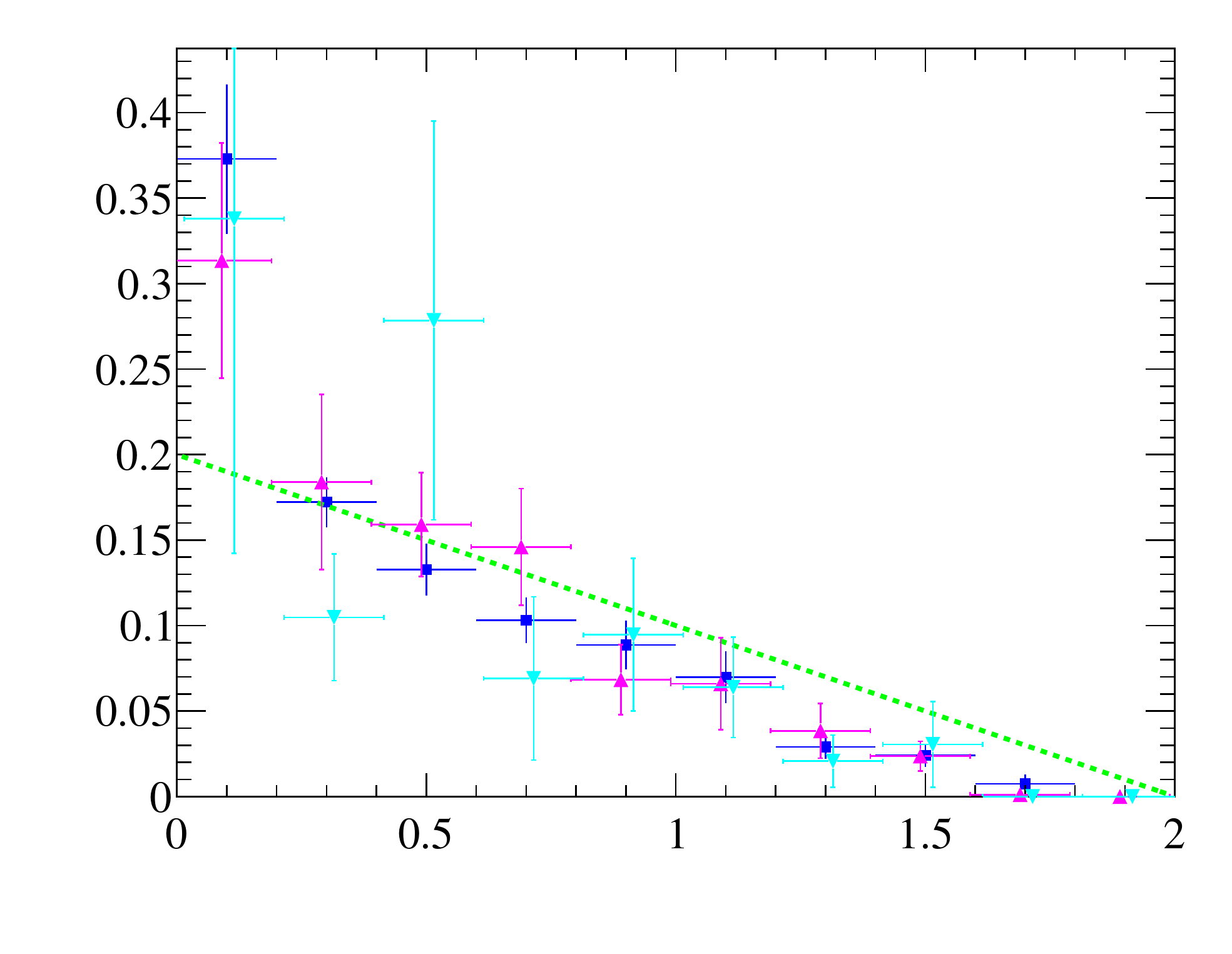}
    }
    \put( 37,3)   { $\left| \Delta y \right|$}
    \put(112,3)   { $\left| \Delta y \right|$}
    \put(-2 , 37 )  { \small 
      \begin{sideways}%
        $\tfrac{\mathrm{d}\ln\sigma^*}{\mathrm{d} \left| \Delta y \right|}%
        ~\left[ \tfrac{1}{0.2}\right]$
      \end{sideways}%
    }
    \put(73 , 37 )  { \small 
      \begin{sideways}%
        $\tfrac{\mathrm{d}\ln\sigma^*}{\mathrm{d} \left| \Delta y \right|}%
        ~\left[ \tfrac{1}{0.2}\right]$
      \end{sideways}%
    }
    \put( 55,45){ \small 
      LHCb
    }
    \put(130,45){ \small 
      LHCb
    }
    \put( 65,52)   { a) }
    \put(140,52)   { b) }
    \put(55,30){ \tiny
      $\begin{array}{cl}
        {\color{red}       \text{\ding{108}} } & \DzDzb  \\
        {\color{blue}      \text{\ding{110}} } & \DzDpb  \\
        {\color{RootSix}   \text{\ding{115}} } & \DzDsb  \\  
        {\color{RootSeven} \text{\ding{116}} } & \DzLcb  
       \end{array}$
    } 
    \put(130,30){ \tiny
      $\begin{array}{cl}
        {\color{blue}      \text{\ding{110}} } & \DpDpb  \\
        {\color{RootSix}   \text{\ding{115}} } & \DpDsb  \\  
        {\color{RootSeven} \text{\ding{116}} } & \DpLcb  
       \end{array}$
    } 
  \end{picture}
  \caption { \small 
    Distributions of the difference in rapidity
    for  \CCbar events: 
    a)~\DzDzb,  
       \DzDpb, 
       \DzDsb~and
       \DzLcb; 
    b)~\DpDpb, 
       \DpDsb~and
       \DpLcb. 
    The dashed line shows the expected 
    distribution for uncorrelated events.
  }
  \label{fig:dd_ccbar_dy_pt3}
\end{figure}

The azimuthal and rapidity 
correlations for \CC~and \CCbar~events are shown in
Figs.~\ref{fig:dd_dphi_CC_pt3}, 
\ref{fig:dd_dphi_CCbar_pt3}, and~\ref{fig:dd_ccbar_dy_pt3}. 
In the \CC~case the $\Delta\phi$~distribution is 
reasonably 
consistent with a flat distribution.
In contrast, for \CCbar~events a clear enhancement is seen 
for $\Delta\phi$~distributions 
at small $\left|\Delta \phi\right|$. 
This is consistent with $\mathrm{c}\bar{\mathrm{c}}$
production via the gluon 
splitting mechanism~\cite{Norrbin:2000zc}.
The \CCbar~events suggest some enhancement at small 
$\left|\Delta y\right|$, while 
the \CC~sample shows no clear difference from 
the triangular shape given the present statistics.

\begin{figure}[htb]
  \setlength{\unitlength}{1mm}
  \centering
  \begin{picture}(150,60)
    \put(0,0){
      \includegraphics*[width=75mm,height=60mm,%
      ]{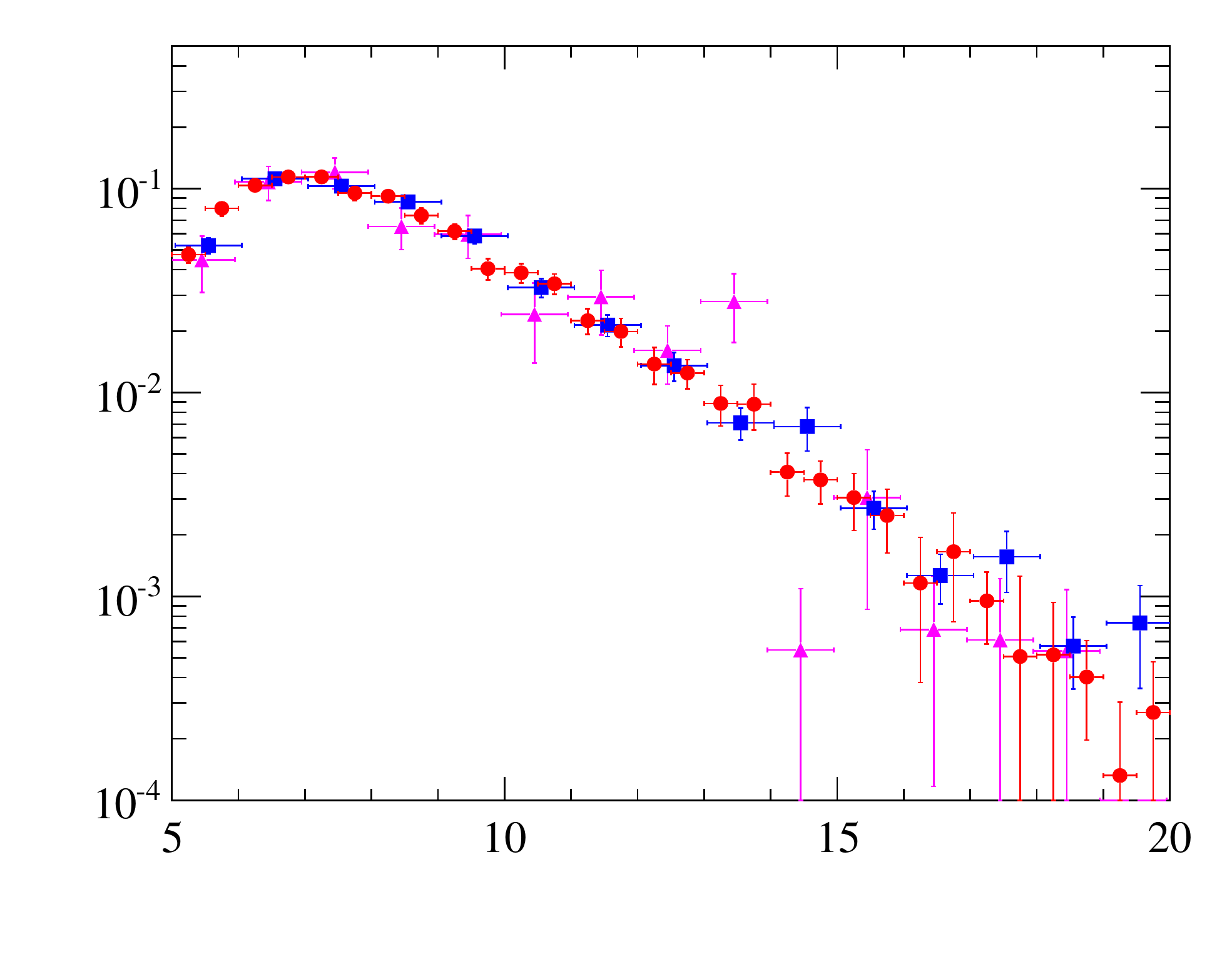}
    }
    \put(75,0){
      \includegraphics*[width=75mm,height=60mm,%
      ]{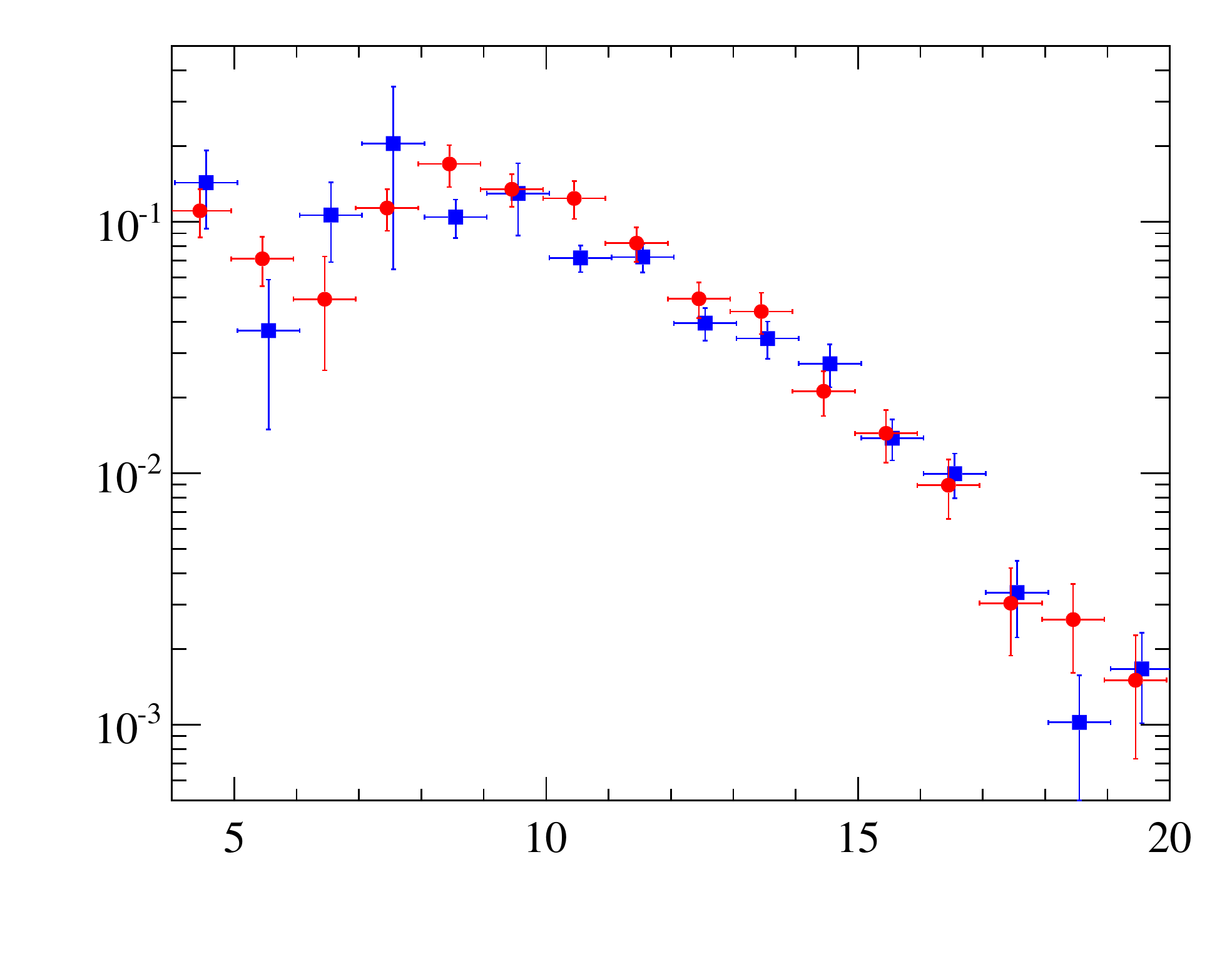}
    }
    \put( 35,1)   { $m_{\psiC}$}
    \put(110,1)   { $m_{\CC}$}
    \put( 58,1)   { $\left[ \mathrm{GeV}/c^2\right]$}
    \put(133,1)   { $\left[ \mathrm{GeV}/c^2\right]$}
    \put(-2 , 25 )  { \small 
      \begin{sideways}%
        $\tfrac{\mathrm{d}\ln\sigma^*}{\mathrm{d}m_{\psiC}}%
        ~\left[ \tfrac{1}{500~\mathrm{MeV}/c^2}\right]$
      \end{sideways}%
    }
    \put(73, 25 )  { \small 
      \begin{sideways}%
        $\tfrac{\mathrm{d}\ln\sigma^*}{\mathrm{d}m_{\CC}}%
        ~\left[ \tfrac{1}{500~\mathrm{MeV}/c^2}\right]$
      \end{sideways}%
    }
    \put( 55,45){ \small 
      LHCb
    }
    \put(130,45){ \small 
      LHCb
    }
    \put( 65,52)   { a) }
    \put(140,52)   { b) }
    \put(12,20){ \tiny
      $\begin{array}{cl}
        {\color{red}     \text{\ding{108}} } & \psiDz  \\
        {\color{blue}    \text{\ding{110}} } & \psiDp  \\
        {\color{RootSix} \text{\ding{115}} } & \psiDs  
       \end{array}$
    } 
    \put(87,20){ \tiny
      $\begin{array}{cl}
        {\color{red}     \text{\ding{108}} } & \DzDz \\
        {\color{blue}    \text{\ding{110}} } & \DzDp 
       \end{array}$
    } 
  \end{picture}
  \caption { \small 
    a)~Invariant mass spectra for \psiDz,  
    \psiDp~and 
    \psiDs~events.
    b)~Invariant mass spectra for 
    \DzDz~and \DzDp~events. 
  }
  \label{fig:psid_mass_pt2}
\end{figure}

Finally, the invariant mass distributions 
of the pairs of charm hadrons in 
these events have been
studied. The mass spectra for \psiC~and \CC~events 
are shown in Fig.~\ref{fig:psid_mass_pt2}.
The spectra appear to be independent of 
the type of the open charm hadron.

\begin{figure}[htb]
  \setlength{\unitlength}{1mm}
  \centering
  \begin{picture}(150,60)
    \put(0,0){
      \includegraphics*[width=75mm,height=60mm,%
      ]{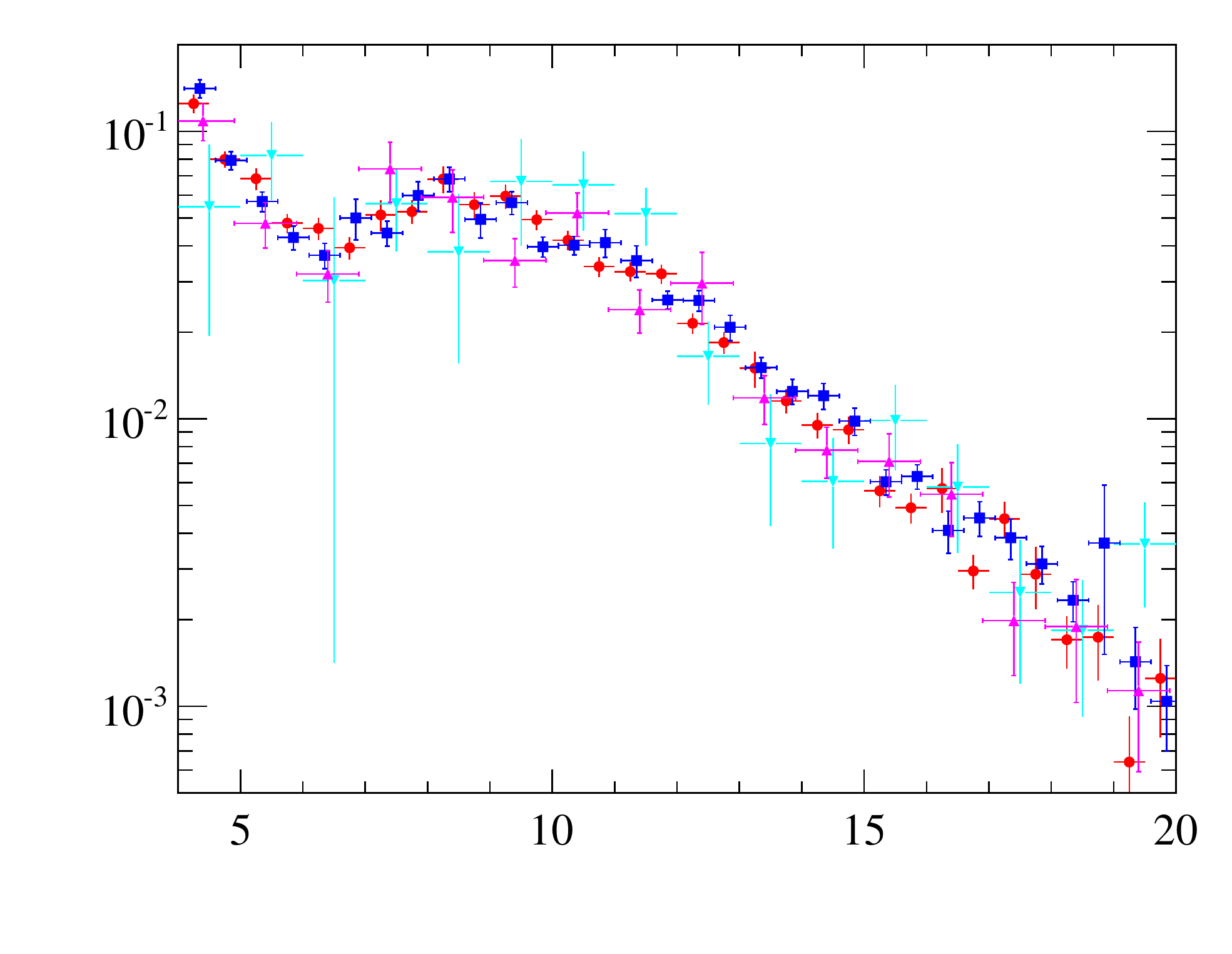}
    }
    \put(75,0){
      \includegraphics*[width=75mm,height=60mm,%
      ]{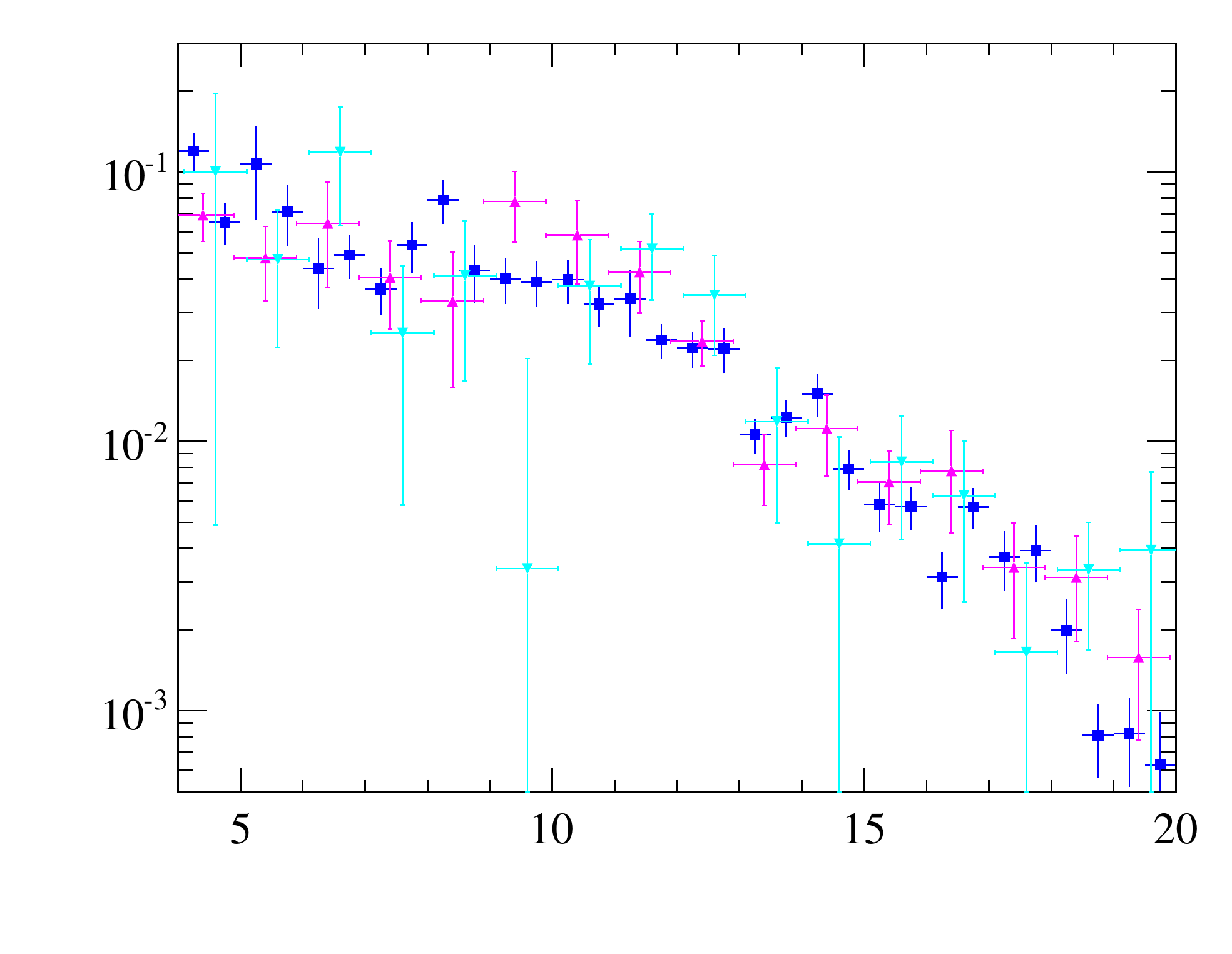}
    }
    \put( 35,3)   { \large $m_{\CCbar}$}
    \put(110,3)   { \large $m_{\CCbar}$}
    \put( 58,1)   { $\left[ \mathrm{GeV}/c^2\right]$}
    \put(133,1)   { $\left[ \mathrm{GeV}/c^2\right]$}
    \put(-2 , 25 )  { \small 
      \begin{sideways}%
        $\tfrac{\mathrm{d}\ln\sigma^*}{\mathrm{d}m_{\CCbar}}%
        ~\left[ \tfrac{1}{500~\mathrm{MeV}/c^2}\right]$
      \end{sideways}%
    }
    \put(73 , 25 )  { \small 
      \begin{sideways}%
        $\tfrac{\mathrm{d}\ln\sigma^*}{\mathrm{d}m_{\CCbar}}%
        ~\left[ \tfrac{1}{500~\mathrm{MeV}/c^2}\right]$
      \end{sideways}%
    }
    \put( 55,45){ \small 
      LHCb
    }
    \put(130,45){ \small 
      LHCb
    }
    \put( 65,52)   { a) }
    \put(140,52)   { b) }
    \put(12,20){ \tiny
      $\begin{array}{cl}
        {\color{red}        \text{\ding{108}} } & \DzDzb  \\
        {\color{blue}       \text{\ding{110}} } & \DzDpb  \\
        {\color{RootSix}    \text{\ding{115}} } & \DzDsb  \\  
        {\color{RootSeven}  \text{\ding{116}} } & \DzLcb  
       \end{array}$
    } 
    \put(87,20){ \tiny
      $\begin{array}{cl}
        {\color{blue}       \text{\ding{110}} } & \DpDpb  \\
        {\color{RootSix}    \text{\ding{115}} } & \DpDsb  \\  
        {\color{RootSeven}  \text{\ding{116}} } & \DpLcb  
      \end{array}$
    } 
  \end{picture}
  \caption { \small 
    Invariant mass spectra 
    for  \CCbar~events: 
    a)~\DzDzb,  
       \DzDpb,
       \DzDsb~and 
       \DzLcb;
    b)~\DpDpb, 
       \DpDsb~and
       \DpLcb. 
  }
  \label{fig:ccbar_mass_pt3}
\end{figure}

The invariant mass spectra for \CCbar~events 
are shown in Fig.~\ref{fig:ccbar_mass_pt3}.
Again, the spectra are similar and independent of 
the type of the open charm meson. The enhancement at small invariant
mass is most likely due to 
the gluon splitting process~\cite{Norrbin:2000zc}. 
For the region of invariant masses 
above $6~\mathrm{GeV}/c^2$ the spectra 
are similar 
for \CCbar~and \CC~events.

\clearpage 
%

\section{Conclusion}
The production of  \jpsi~mesons accompanied by open charm, 
and pairs of open charm hadrons 
has been observed in pp~collisions at $\sqrt{\mathrm{s}}=7~{\mathrm{TeV}}$.
This is the first observation of these phenomena in hadronic
collisions. Signals with a statistical significance in excess 
of five standard deviations have been observed for four \psiC modes:
\psiDz, \psiDp, \psiDs and \psiLc,
for six \CC~modes: 
\DzDz, \DzDp, \DzDs, 
\DzLc, \DpDp, and \DpDs, 
and for seven \CCbar~channels:
\DzDzb, \DzDpb, \DzDsb, 
\DzLcb, \DpDpb, \DpDsb and \DpLcb.

The cross-sections and the properties of these events have been
studied. The predictions from gluon-gluon
fusion~\cite{Berezhnoy:1998aa,Baranov:2006dh,Lansberg:2008gk}
are significantly smaller than the observed
cross-sections. Better agreement is found with the DPS 
model~\cite{Kom:2011bd, Baranov:2011ch, Novoselov:2011ff,Luszczak:2011zp}
if the
effective cross-section inferred from the Tevatron data is used. 
The absence of significant azimuthal or rapidity correlations provides 
support for  this hypothesis. 

The transverse momentum spectra for these events have also been
studied. The transverse momentum spectra for \jpsi~from \psiC~events
are significantly harder than those observed in prompt \jpsi
production. On the other hand the spectra for open charm mesons 
in \psiC~events appear to be similar to those observed for 
prompt charm hadrons. 
Similar  transverse momentum spectra for \CC~and 
\CCbar~events are observed.  However, the expectation of similar
transverse momentum  spectra for \CCbar~events and prompt charm 
events appears to be invalid.

For \CCbar~events significant rapidity and azimuthal correlations are observed.
These, as well as the invariant mass spectra for \CCbar~events, suggest
a sizeable contribution from the gluon splitting process 
to charm quark production~\cite{Norrbin:2000zc}. 



\section*{Acknowledgements}

\noindent 
We thank  M.~H.~Seymour and A.~Si\'odmok for the points raised 
in Ref.~\cite{Seymour:2013sya}, and have revised the paper 
to address these issues.
We would like 
to thank 
J.-P.~Lansberg, 
A.K.~Likhoded and  
A.~Szczurek for many fruitful discussions.
We express our gratitude to our colleagues in the CERN accelerator
departments for the excellent performance of the LHC. We thank the
technical and administrative staff at CERN and at the LHCb institutes,
and acknowledge support from the National Agencies: CAPES, CNPq,
FAPERJ and FINEP (Brazil); CERN; NSFC (China); CNRS/IN2P3 (France);
BMBF, DFG, HGF and MPG (Germany); SFI (Ireland); INFN (Italy); FOM and
NWO (The Netherlands); SCSR (Poland); ANCS (Romania); MinES of Russia and
Rosatom (Russia); MICINN, XuntaGal and GENCAT (Spain); SNSF and SER
(Switzerland); NAS Ukraine (Ukraine); STFC (United Kingdom); NSF
(USA). We also acknowledge the support received from the ERC under FP7
and the Region Auvergne.

\appendix
%

\section*{Appendix: Contribution from sea charm quarks} 
\label{instrinsic}
Estimates for the expected cross-section in the LHCb fiducial region due to
the sea charm quarks from the interacting protons have been made as follows. 
The LHCb rapidity window $2<y<4$ corresponds to a $x$ range for
the additional charm quarks of 
\begin{equation}
\frac{ 2  m^{\mathrm{T}}_{\mathrm{c}}}{ \sqrt{s} } \sinh{2} < x < 
\frac{ 2 m^{\mathrm{T}}_{\mathrm{c}}}{ \sqrt{s} } \sinh{4}
\end{equation}
where $m^{\mathrm{T}}_{\mathrm{c}}$ is the transverse mass of the charm quark.
Assuming the extra charm mesons are distributed over
$p^{\mathrm{T}}$~in a similar way to the inclusive charm
mesons measured in~\cite{LHCb-CONF-2010-013,Aaij:2011jh} one can take
\begin{equation} 
m^{\mathrm{T}}_{\mathrm{c}} \approx m_{\mathrm{c}} \oplus 2~\mathrm{GeV}/c^2,
\end{equation}
 where 
$2~\mathrm{GeV}/c$ is the mean transverse momentum of charm quarks produced. 
This leads to the $x$ range of $0.0026<x<0.02$.
Integration of Alekhin's LO parton distribution 
functions~\cite{Alekhin:2002fv}  
over this $x$ range gives $0.25$~additional charm quarks per event. 
In this calculation the parton density functions are taken at the scale~$\mu \approx m^{\mathrm{T}}_{\jpsi}$. 
The cross-sections of $\jpsi$ plus open charm mesons can then be
estimated using the probabilities for the $c$-quark transition to
different mesons given in~\cite{LHCb-CONF-2010-013,Aaij:2011jh}. 
Similarly cross-sections for double open charm 
production can be estimated. 
Taking $\mu \approx m^{\mathrm{T}}_{\mathrm{D}}$ and 
integrating Alekhin's LO parton density functions~\cite{Alekhin:2002fv}  
on gets approximately 0.17~additional charm quarks per event. 
This calculation assumes that  all extra charm quarks from protons hadronize 
to open charm states visible in the detector. The real cross-sections 
may be smaller, but the ratio of different open charm states is expected 
to remain the same. The integrated parton density functions provide 
no information about  the $p^{\mathrm{T}}$~distribution of charm quarks. 
Under the assumption that the $p^{\mathrm{T}}$-spectrum coincides with the 
distribution of prompt charm particles measured at LHCb~\cite{LHCb-CONF-2010-013}, 
the cross-sections in the LHCb fiducial range are
calculated (see last column of Table~\ref{tab:intro_theo}).


\clearpage 
\bibliographystyle{LHCb}
\bibliography{main,local}

\ifx\mcitethebibliography\mciteundefinedmacro
\PackageError{LHCb.bst}{mciteplus.sty has not been loaded}
{This bibstyle requires the use of the mciteplus package.}\fi
\providecommand{\href}[2]{#2}
\begin{mcitethebibliography}{10}
\mciteSetBstSublistMode{n}
\mciteSetBstMaxWidthForm{subitem}{\alph{mcitesubitemcount})}
\mciteSetBstSublistLabelBeginEnd{\mcitemaxwidthsubitemform\space}
{\relax}{\relax}

\bibitem{LHCb-CONF-2010-013}
\lhcb~collaboration, \ifthenelse{\boolean{articletitles}}{{\it {Prompt charm
  production in $\mathrm{pp}$~collisions at $\sqrt{s}=7~\mathrm{TeV}$}}, }{}
  \href{http://cdsweb.cern.ch/search?p=LHCb-CONF-2010-013&f=reportnumber&action_search=Search&c=LHCb+Reports&c=LHCb+Conference+Proceedings&c=LHCb+Conference+Contributions&c=LHCb+Notes&c=LHCb+Theses&c=LHCb+Papers}
  {LHCb-CONF-2010-013}\relax
\mciteBstWouldAddEndPuncttrue
\mciteSetBstMidEndSepPunct{\mcitedefaultmidpunct}
{\mcitedefaultendpunct}{\mcitedefaultseppunct}\relax
\EndOfBibitem
\bibitem{Aaij:2011jh}
\lhcb~collaboration, R.~Aaij {\em et~al.},
  \ifthenelse{\boolean{articletitles}}{{\it {Measurement of \jpsi~production in
  $\mathrm{pp}$~collisions at $\sqrt{s}=7~\mathrm{TeV}$}},
  }{}\href{http://dx.doi.org/10.1140/epjc/s10052-011-1645-y}{Eur.\ Phys.\ J.\
  {\bf C71} (2011) 1645}, \href{http://arxiv.org/abs/1103.0423}{{\tt
  arXiv:1103.0423}}\relax
\mciteBstWouldAddEndPuncttrue
\mciteSetBstMidEndSepPunct{\mcitedefaultmidpunct}
{\mcitedefaultendpunct}{\mcitedefaultseppunct}\relax
\EndOfBibitem
\bibitem{lansbergetc}
S.~J. Brodsky and J.-P. Lansberg, \ifthenelse{\boolean{articletitles}}{{\it
  {Heavy-quarkonium production in high energy proton-proton collisions at
  RHIC}}, }{}\href{http://dx.doi.org/10.1103/PhysRevD.81.051502}{Phys.\ Rev.\
  {\bf D81} (2010) 051502}, \href{http://arxiv.org/abs/hep-ph/0908.0754}{{\tt
  arXiv:hep-ph/0908.0754}}\relax
\mciteBstWouldAddEndPuncttrue
\mciteSetBstMidEndSepPunct{\mcitedefaultmidpunct}
{\mcitedefaultendpunct}{\mcitedefaultseppunct}\relax
\EndOfBibitem
\bibitem{Kom:2011bd}
C.~H. Kom, A.~Kulesza, and W.~J. Stirling,
  \ifthenelse{\boolean{articletitles}}{{\it {Pair production of \jpsi~as a
  probe of double parton scattering at LHCb}}, }{}Phys.\ Rev.\ Lett.\  {\bf
  107} (2011) 082002, \href{http://arxiv.org/abs/1105.4186}{{\tt
  arXiv:1105.4186}}\relax
\mciteBstWouldAddEndPuncttrue
\mciteSetBstMidEndSepPunct{\mcitedefaultmidpunct}
{\mcitedefaultendpunct}{\mcitedefaultseppunct}\relax
\EndOfBibitem
\bibitem{Baranov:2011ch}
S.~P. Baranov, A.~M. Snigirev, and N.~P. Zotov,
  \ifthenelse{\boolean{articletitles}}{{\it {Double heavy meson production
  through double parton scattering in hadronic collisions}},
  }{}\href{http://dx.doi.org/10.1016/j.physletb.2011.09.106}{Phys.\ Lett.\
  {\bf B705} (2011) 116}, \href{http://arxiv.org/abs/1105.6276}{{\tt
  arXiv:1105.6276}}\relax
\mciteBstWouldAddEndPuncttrue
\mciteSetBstMidEndSepPunct{\mcitedefaultmidpunct}
{\mcitedefaultendpunct}{\mcitedefaultseppunct}\relax
\EndOfBibitem
\bibitem{Novoselov:2011ff}
A.~Novoselov, \ifthenelse{\boolean{articletitles}}{{\it {Double parton
  scattering as a source of quarkonia pairs in LHCb}},
  }{}\href{http://arxiv.org/abs/1106.2184}{{\tt arXiv:1106.2184}}\relax
\mciteBstWouldAddEndPuncttrue
\mciteSetBstMidEndSepPunct{\mcitedefaultmidpunct}
{\mcitedefaultendpunct}{\mcitedefaultseppunct}\relax
\EndOfBibitem
\bibitem{Luszczak:2011zp}
M.~Luszczak, R.~Maciula, and A.~Szczurek,
  \ifthenelse{\boolean{articletitles}}{{\it {Production of two
  $\mathrm{c}{\bar{\mathrm{c}}}$~pairs in double-parton scattering}},
  }{}\href{http://arxiv.org/abs/1111.3255}{{\tt arXiv:1111.3255}}\relax
\mciteBstWouldAddEndPuncttrue
\mciteSetBstMidEndSepPunct{\mcitedefaultmidpunct}
{\mcitedefaultendpunct}{\mcitedefaultseppunct}\relax
\EndOfBibitem
\bibitem{Brodsky1980451}
S.~Brodsky, P.~Hoyer, C.~Peterson, and N.~Sakai,
  \ifthenelse{\boolean{articletitles}}{{\it The intrinsic charm of the proton},
  }{}\href{http://dx.doi.org/10.1016/0370-2693(80)90364-0}{Phys.\ Lett.\  {\bf
  B93} (1980) 451}\relax
\mciteBstWouldAddEndPuncttrue
\mciteSetBstMidEndSepPunct{\mcitedefaultmidpunct}
{\mcitedefaultendpunct}{\mcitedefaultseppunct}\relax
\EndOfBibitem
\bibitem{Aoki:1986uq}
WA75~collaboration, S.~Aoki {\em et~al.},
  \ifthenelse{\boolean{articletitles}}{{\it {The double associated production
  of charmed particles by the interaction of 350~$\mathrm{GeV}/c$~$\Ppi^-$
  mesons with emulsion nuclei}},
  }{}\href{http://dx.doi.org/10.1016/0370-2693(87)90097-9}{Phys.\ Lett.\  {\bf
  B187} (1987) 185}\relax
\mciteBstWouldAddEndPuncttrue
\mciteSetBstMidEndSepPunct{\mcitedefaultmidpunct}
{\mcitedefaultendpunct}{\mcitedefaultseppunct}\relax
\EndOfBibitem
\bibitem{Kartvelishvili:1984ur}
V.~G. Kartvelishvili and S.~M. Esakiya,
  \ifthenelse{\boolean{articletitles}}{{\it {On hadron induced production of
  \jpsi~meson pairs}}, }{}Yad.\ Fiz.\  {\bf 38} (1983) 722\relax
\mciteBstWouldAddEndPuncttrue
\mciteSetBstMidEndSepPunct{\mcitedefaultmidpunct}
{\mcitedefaultendpunct}{\mcitedefaultseppunct}\relax
\EndOfBibitem
\bibitem{Humpert:1983yj}
B.~Humpert and P.~Mery, \ifthenelse{\boolean{articletitles}}{{\it
  {\Ppsi{}\Ppsi~production at collider energies}},
  }{}\href{http://dx.doi.org/10.1007/BF01577721}{Z.\ Phys.\  {\bf C20} (1983)
  83}\relax
\mciteBstWouldAddEndPuncttrue
\mciteSetBstMidEndSepPunct{\mcitedefaultmidpunct}
{\mcitedefaultendpunct}{\mcitedefaultseppunct}\relax
\EndOfBibitem
\bibitem{Berezhnoy:2011xy}
A.~V. Berezhnoy, A.~K. Likhoded, A.~V. Luchinsky, and A.~A. Novoselov,
  \ifthenelse{\boolean{articletitles}}{{\it {Double \jpsi~meson Production at
  LHC and 4c-tetraquark state}},
  }{}\href{http://dx.doi.org/10.1103/PhysRevD.84.094023}{Phys.\ Rev.\  {\bf
  D84} (2011) 094023}, \href{http://arxiv.org/abs/1101.5881}{{\tt
  arXiv:1101.5881}}\relax
\mciteBstWouldAddEndPuncttrue
\mciteSetBstMidEndSepPunct{\mcitedefaultmidpunct}
{\mcitedefaultendpunct}{\mcitedefaultseppunct}\relax
\EndOfBibitem
\bibitem{Aaij:2011yc}
\lhcb~collaboration, R.~Aaij {\em et~al.},
  \ifthenelse{\boolean{articletitles}}{{\it {Observation of \jpsi~pair
  production in $\mathrm{pp}$~collisions at $\sqrt{s}=7~\mathrm{TeV}$}},
  }{}\href{http://dx.doi.org/10.1016/j.physletb.2011.12.015}{Phys.\ Lett.\
  {\bf B707} (2012) 52}, \href{http://arxiv.org/abs/1109.0963}{{\tt
  arXiv:1109.0963}}\relax
\mciteBstWouldAddEndPuncttrue
\mciteSetBstMidEndSepPunct{\mcitedefaultmidpunct}
{\mcitedefaultendpunct}{\mcitedefaultseppunct}\relax
\EndOfBibitem
\bibitem{Berezhnoy:1998aa}
A.~V. Berezhnoy, V.~V. Kiselev, A.~K. Likhoded, and A.~I. Onishchenko,
  \ifthenelse{\boolean{articletitles}}{{\it {Doubly charmed baryon production
  in hadronic experiments}},
  }{}\href{http://dx.doi.org/10.1103/PhysRevD.57.4385}{Phys.\ Rev.\  {\bf D57}
  (1998) 4385}, \href{http://arxiv.org/abs/hep-ph/9710339}{{\tt
  arXiv:hep-ph/9710339}}\relax
\mciteBstWouldAddEndPuncttrue
\mciteSetBstMidEndSepPunct{\mcitedefaultmidpunct}
{\mcitedefaultendpunct}{\mcitedefaultseppunct}\relax
\EndOfBibitem
\bibitem{Baranov:2006dh}
S.~P. Baranov, \ifthenelse{\boolean{articletitles}}{{\it {Topics in associated
  \mbox{$\jpsi{}\!+\!\mathrm{c}{}{\bar{\mathrm{c}}}$}~production at modern
  colliders}}, }{}\href{http://dx.doi.org/10.1103/PhysRevD.73.074021}{Phys.\
  Rev.\  {\bf D73} (2006) 074021}\relax
\mciteBstWouldAddEndPuncttrue
\mciteSetBstMidEndSepPunct{\mcitedefaultmidpunct}
{\mcitedefaultendpunct}{\mcitedefaultseppunct}\relax
\EndOfBibitem
\bibitem{Abe:1997xk}
\cdf~collaboration, F.~Abe {\em et~al.},
  \ifthenelse{\boolean{articletitles}}{{\it {Double parton scattering in
  \mbox{$\bar{\mathrm{p}}\mathrm{p}$}~collisions at $\sqrt{s}=
  1.8~\mathrm{TeV}$}},
  }{}\href{http://dx.doi.org/10.1103/PhysRevD.56.3811}{Phys.\ Rev.\  {\bf D56}
  (1997) 3811}\relax
\mciteBstWouldAddEndPuncttrue
\mciteSetBstMidEndSepPunct{\mcitedefaultmidpunct}
{\mcitedefaultendpunct}{\mcitedefaultseppunct}\relax
\EndOfBibitem
\bibitem{Blok:2011bu}
B.~Blok, Y.~Dokshitser, L.~Frankfurt, and M.~Strikman,
  \ifthenelse{\boolean{articletitles}}{{\it {pQCD physics of multiparton
  interactions}}, }{}\href{http://arxiv.org/abs/1106.5533}{{\tt
  arXiv:1106.5533}}\relax
\mciteBstWouldAddEndPuncttrue
\mciteSetBstMidEndSepPunct{\mcitedefaultmidpunct}
{\mcitedefaultendpunct}{\mcitedefaultseppunct}\relax
\EndOfBibitem
\bibitem{Lansberg:2008gk}
J.-P. Lansberg, \ifthenelse{\boolean{articletitles}}{{\it {On the mechanisms of
  heavy-quarkonium hadroproduction}},
  }{}\href{http://dx.doi.org/10.1140/epjc/s10052-008-0826-9}{Eur.\ Phys.\ J.\
  {\bf C61} (2009) 693}, \href{http://arxiv.org/abs/0811.4005}{{\tt
  arXiv:0811.4005}}\relax
\mciteBstWouldAddEndPuncttrue
\mciteSetBstMidEndSepPunct{\mcitedefaultmidpunct}
{\mcitedefaultendpunct}{\mcitedefaultseppunct}\relax
\EndOfBibitem
\bibitem{Alves:2008zz}
LHCb collaboration, A.~A. Alves~Jr. {\em et~al.},
  \ifthenelse{\boolean{articletitles}}{{\it {The \lhcb detector at the LHC}},
  }{}\href{http://dx.doi.org/10.1088/1748-0221/3/08/S08005}{JINST {\bf 3}
  (2008) S08005}\relax
\mciteBstWouldAddEndPuncttrue
\mciteSetBstMidEndSepPunct{\mcitedefaultmidpunct}
{\mcitedefaultendpunct}{\mcitedefaultseppunct}\relax
\EndOfBibitem
\bibitem{LHCb-PUB-2011-016}
V.~Gligorov, C.~Thomas, and M.~Williams,
  \ifthenelse{\boolean{articletitles}}{{\it {The HLT inclusive $B$ triggers}},
  }{}
  \href{http://cdsweb.cern.ch/search?p=LHCb-PUB-2011-016&f=reportnumber&action_search=Search&c=LHCb+Reports&c=LHCb+Conference+Proceedings&c=LHCb+Conference+Contributions&c=LHCb+Notes&c=LHCb+Theses&c=LHCb+Papers}
  {LHCb-PUB-2011-016}\relax
\mciteBstWouldAddEndPuncttrue
\mciteSetBstMidEndSepPunct{\mcitedefaultmidpunct}
{\mcitedefaultendpunct}{\mcitedefaultseppunct}\relax
\EndOfBibitem
\bibitem{Sjostrand:2006za}
T.~Sj\"{o}strand, S.~Mrenna, and P.~Skands,
  \ifthenelse{\boolean{articletitles}}{{\it {PYTHIA 6.4 Physics and manual}},
  }{}\href{http://dx.doi.org/10.1088/1126-6708/2006/05/026}{JHEP {\bf 05}
  (2006) 026}, \href{http://arxiv.org/abs/hep-ph/0603175}{{\tt
  arXiv:hep-ph/0603175}}\relax
\mciteBstWouldAddEndPuncttrue
\mciteSetBstMidEndSepPunct{\mcitedefaultmidpunct}
{\mcitedefaultendpunct}{\mcitedefaultseppunct}\relax
\EndOfBibitem
\bibitem{LHCb-PROC-2011-005}
I.~Belyaev {\em et~al.}, \ifthenelse{\boolean{articletitles}}{{\it {Handling of
  the generation of primary events in \gauss, the \lhcb simulation framework}},
  }{}\href{http://dx.doi.org/10.1109/NSSMIC.2010.5873949}{Nuclear Science
  Symposium Conference Record (NSS/MIC) {\bf IEEE} (2010) 1155}\relax
\mciteBstWouldAddEndPuncttrue
\mciteSetBstMidEndSepPunct{\mcitedefaultmidpunct}
{\mcitedefaultendpunct}{\mcitedefaultseppunct}\relax
\EndOfBibitem
\bibitem{Lange:2001uf}
D.~J. Lange, \ifthenelse{\boolean{articletitles}}{{\it {The EvtGen particle
  decay simulation package}},
  }{}\href{http://dx.doi.org/10.1016/S0168-9002(01)00089-4}{Nucl.\ Instrum.\
  Meth.\  {\bf A462} (2001) 152}\relax
\mciteBstWouldAddEndPuncttrue
\mciteSetBstMidEndSepPunct{\mcitedefaultmidpunct}
{\mcitedefaultendpunct}{\mcitedefaultseppunct}\relax
\EndOfBibitem
\bibitem{Agostinelli:2002hh}
GEANT4 collaboration, S.~Agostinelli {\em et~al.},
  \ifthenelse{\boolean{articletitles}}{{\it {GEANT4: A simulation toolkit}},
  }{}\href{http://dx.doi.org/10.1016/S0168-9002(03)01368-8}{Nucl.\ Instrum.\
  Meth.\  {\bf A506} (2003) 250}\relax
\mciteBstWouldAddEndPuncttrue
\mciteSetBstMidEndSepPunct{\mcitedefaultmidpunct}
{\mcitedefaultendpunct}{\mcitedefaultseppunct}\relax
\EndOfBibitem
\bibitem{Hulsbergen:2005pu}
W.~D. Hulsbergen, \ifthenelse{\boolean{articletitles}}{{\it {Decay chain
  fitting with a Kalman filter}},
  }{}\href{http://dx.doi.org/10.1016/j.nima.2005.06.078}{Nucl.\ Instrum.\
  Meth.\  {\bf A552} (2005) 566},
  \href{http://arxiv.org/abs/physics/0503191}{{\tt
  arXiv:physics/0503191}}\relax
\mciteBstWouldAddEndPuncttrue
\mciteSetBstMidEndSepPunct{\mcitedefaultmidpunct}
{\mcitedefaultendpunct}{\mcitedefaultseppunct}\relax
\EndOfBibitem
\bibitem{Skwarnicki:1986xj}
T.~Skwarnicki, {\em {A study of the radiative cascade transitions between the
  Upsilon-prime and Upsilon resonances}}, PhD thesis, Institute of Nuclear
  Physics, Krakow, 1986,
  {\href{http://inspirehep.net/record/230779/files/230779.pdf}{DESY-F31-86-02}}\relax
\mciteBstWouldAddEndPuncttrue
\mciteSetBstMidEndSepPunct{\mcitedefaultmidpunct}
{\mcitedefaultendpunct}{\mcitedefaultseppunct}\relax
\EndOfBibitem
\bibitem{Lees:2011gw}
\babar~collaboration, J.-P. Lees {\em et~al.},
  \ifthenelse{\boolean{articletitles}}{{\it {Branching fraction measurements of
  the color-suppressed decays $\bar{\mathrm{B}}^0 \to
  \mathrm{D}^{\left(*\right)0} \Ppi^0$, $\mathrm{D}^{\left(*\right)0} \Peta$,
  $\mathrm{D}^{\left(*\right)0} \Pomega$, and $\mathrm{D}^{\left(*\right)0}
  \Peta^{\prime}$ and measurement of the polarization in the decay
  $\bar{\mathrm{B}}^0 \to \mathrm{D}^{*0} \Pomega$}},
  }{}\href{http://dx.doi.org/10.1103/PhysRevD.84.112007}{Phys.\ Rev.\  {\bf
  D84} (2011) 112007}, \href{http://arxiv.org/abs/1107.5751}{{\tt
  arXiv:1107.5751}}\relax
\mciteBstWouldAddEndPuncttrue
\mciteSetBstMidEndSepPunct{\mcitedefaultmidpunct}
{\mcitedefaultendpunct}{\mcitedefaultseppunct}\relax
\EndOfBibitem
\bibitem{Bickel:1983}
P.~J. Bickel and L.~Breiman, \ifthenelse{\boolean{articletitles}}{{\it Sums of
  functions of nearest neighbor distances, moment bounds, limit theorems and a
  goodness of fit test}, }{}The Annals of Probability {\bf 11} (1983), no.~1
  185\relax
\mciteBstWouldAddEndPuncttrue
\mciteSetBstMidEndSepPunct{\mcitedefaultmidpunct}
{\mcitedefaultendpunct}{\mcitedefaultseppunct}\relax
\EndOfBibitem
\bibitem{Williams:2010vh}
M.~Williams, \ifthenelse{\boolean{articletitles}}{{\it {How good are your fits?
  Unbinned multivariate goodness-of-fit tests in high energy physics}},
  }{}\href{http://dx.doi.org/10.1088/1748-0221/5/09/P09004}{JINST {\bf 5}
  (2010) P09004}, \href{http://arxiv.org/abs/1006.3019}{{\tt
  arXiv:1006.3019}}\relax
\mciteBstWouldAddEndPuncttrue
\mciteSetBstMidEndSepPunct{\mcitedefaultmidpunct}
{\mcitedefaultendpunct}{\mcitedefaultseppunct}\relax
\EndOfBibitem
\bibitem{Powell:2011zz}
A.~Powell, \ifthenelse{\boolean{articletitles}}{{\it {Reconstruction and PID
  performance of the LHCb RICH detectors}},
  }{}\href{http://dx.doi.org/10.1016/j.nima.2010.09.048}{Nucl.\ Instrum.\
  Meth.\  {\bf A639} (2011) 260}\relax
\mciteBstWouldAddEndPuncttrue
\mciteSetBstMidEndSepPunct{\mcitedefaultmidpunct}
{\mcitedefaultendpunct}{\mcitedefaultseppunct}\relax
\EndOfBibitem
\bibitem{LHCb-PROC-2011-008}
A.~Powell {\em et~al.}, \ifthenelse{\boolean{articletitles}}{{\it {Particle
  identification at LHCb}}, }{}PoS {\bf ICHEP2010} (2010) 020,
  \href{https://cdsweb.cern.ch/record/1322666?ln=en}{LHCb-PROC-2011-008}\relax
\mciteBstWouldAddEndPuncttrue
\mciteSetBstMidEndSepPunct{\mcitedefaultmidpunct}
{\mcitedefaultendpunct}{\mcitedefaultseppunct}\relax
\EndOfBibitem
\bibitem{Aaij:2010nx}
\lhcb~collaboration, R.~Aaij {\em et~al.},
  \ifthenelse{\boolean{articletitles}}{{\it {Prompt
  $\mathrm{K}^{0}_{\mathrm{S}}$ production in $\mathrm{pp}$~collisions at
  $\sqrt{s}=0.9~\mathrm{TeV}$}},
  }{}\href{http://dx.doi.org/10.1016/j.physletb.2010.08.055}{Phys.\ Lett.\
  {\bf B693} (2010) 69}, \href{http://arxiv.org/abs/1008.3105}{{\tt
  arXiv:1008.3105}}\relax
\mciteBstWouldAddEndPuncttrue
\mciteSetBstMidEndSepPunct{\mcitedefaultmidpunct}
{\mcitedefaultendpunct}{\mcitedefaultseppunct}\relax
\EndOfBibitem
\bibitem{Pivk:2004ty}
M.~Pivk and F.~R. Le~Diberder, \ifthenelse{\boolean{articletitles}}{{\it
  {\sPlot{}: A Statistical tool to unfold data distributions}},
  }{}\href{http://dx.doi.org/10.1016/j.nima.2005.08.106}{Nucl.\ Instrum.\
  Meth.\  {\bf A555} (2005) 356},
  \href{http://arxiv.org/abs/physics/0402083}{{\tt
  arXiv:physics/0402083}}\relax
\mciteBstWouldAddEndPuncttrue
\mciteSetBstMidEndSepPunct{\mcitedefaultmidpunct}
{\mcitedefaultendpunct}{\mcitedefaultseppunct}\relax
\EndOfBibitem
\bibitem{Nakamura:2010zzi}
Particle Data Group, K.~Nakamura {\em et~al.},
  \ifthenelse{\boolean{articletitles}}{{\it {Review of particle physics}},
  }{}\href{http://dx.doi.org/10.1088/0954-3899/37/7A/075021}{J.\ Phys.\  {\bf
  G37} (2010) 075021}\relax
\mciteBstWouldAddEndPuncttrue
\mciteSetBstMidEndSepPunct{\mcitedefaultmidpunct}
{\mcitedefaultendpunct}{\mcitedefaultseppunct}\relax
\EndOfBibitem
\bibitem{vandermeer}
S.~van~der Meer, \ifthenelse{\boolean{articletitles}}{{\it {Calibration of the
  effective beam height in the ISR}}, }{}
  \href{https://cdsweb.cern.ch/record/296752} {ISR-PO/68-31, 1968}\relax
\mciteBstWouldAddEndPuncttrue
\mciteSetBstMidEndSepPunct{\mcitedefaultmidpunct}
{\mcitedefaultendpunct}{\mcitedefaultseppunct}\relax
\EndOfBibitem
\bibitem{Ferroluzzi:2005em}
M.~Ferro-Luzzi, \ifthenelse{\boolean{articletitles}}{{\it {Proposal for an
  absolute luminosity determination in colliding beam experiments using vertex
  detection of beam-gas interactions}},
  }{}\href{http://dx.doi.org/10.1016/j.nima.2005.07.010}{Nucl.\ Instrum.\
  Meth.\  {\bf A553} (2005) 388}\relax
\mciteBstWouldAddEndPuncttrue
\mciteSetBstMidEndSepPunct{\mcitedefaultmidpunct}
{\mcitedefaultendpunct}{\mcitedefaultseppunct}\relax
\EndOfBibitem
\bibitem{LHCb-PAPER-2011-015}
LHCb collaboration, R.~Aaij {\em et~al.},
  \ifthenelse{\boolean{articletitles}}{{\it {Absolute luminosity measurements
  with the LHCb detector at the LHC}},
  }{}\href{http://dx.doi.org/10.1088/1748-0221/7/01/P01010}{JINST {\bf 7}
  (2012) P01010}, \href{http://arxiv.org/abs/1110.2866}{{\tt
  arXiv:1110.2866}}\relax
\mciteBstWouldAddEndPuncttrue
\mciteSetBstMidEndSepPunct{\mcitedefaultmidpunct}
{\mcitedefaultendpunct}{\mcitedefaultseppunct}\relax
\EndOfBibitem
\bibitem{Norrbin:2000zc}
{Norrbin, E. and Sj\"{o}strand, T.}, \ifthenelse{\boolean{articletitles}}{{\it
  {Production and hadronization of heavy quarks}},
  }{}\href{http://dx.doi.org/10.1007/s100520000460}{Eur.\ Phys.\ J.\  {\bf C17}
  (2000) 137}, \href{http://arxiv.org/abs/hep-ph/0005110}{{\tt
  arXiv:hep-ph/0005110}}\relax
\mciteBstWouldAddEndPuncttrue
\mciteSetBstMidEndSepPunct{\mcitedefaultmidpunct}
{\mcitedefaultendpunct}{\mcitedefaultseppunct}\relax
\EndOfBibitem
\bibitem{Seymour:2013sya}
M.~H. Seymour and A.~Si\'odmok, \ifthenelse{\boolean{articletitles}}{{\it
  {Extracting $\upsigma_{\mathrm{effective}}$ from the LHCb double-charm
  measurement}}, }{}\href{http://arxiv.org/abs/1308.6749}{{\tt
  arXiv:1308.6749}}\relax
\mciteBstWouldAddEndPuncttrue
\mciteSetBstMidEndSepPunct{\mcitedefaultmidpunct}
{\mcitedefaultendpunct}{\mcitedefaultseppunct}\relax
\EndOfBibitem
\bibitem{Alekhin:2002fv}
S.~Alekhin, \ifthenelse{\boolean{articletitles}}{{\it {Parton distributions
  from deep-inelastic scattering data}},
  }{}\href{http://dx.doi.org/10.1103/PhysRevD.68.014002}{Phys.\ Rev.\  {\bf
  D68} (2003) 014002}, \href{http://arxiv.org/abs/hep-ph/0211096}{{\tt
  arXiv:hep-ph/0211096}}\relax
\mciteBstWouldAddEndPuncttrue
\mciteSetBstMidEndSepPunct{\mcitedefaultmidpunct}
{\mcitedefaultendpunct}{\mcitedefaultseppunct}\relax
\EndOfBibitem
\end{mcitethebibliography}

\end{document}